\documentclass[blockstyle,preprint,nocopyrightspace]{sigplanconf}

\usepackage{amssymb,amsmath,amsthm}
\usepackage{latexsym}
\usepackage{graphicx}
\usepackage{textcomp}
\usepackage{txfonts} %% for various points-to symbols. use pxfonts for 5-tip stars
\usepackage{verbatim} %% for comment environment
\usepackage{xspace} %% for comment environment
\usepackage{cancel} % for crossing Owns 
\usepackage{mathpartir}
\usepackage{soul}
\usepackage{xcolor}
\usepackage{multirow}
\usepackage{stmaryrd} 
\usepackage{url}
\usepackage{float} 
\usepackage{framed}
\usepackage{hyperref}
\usepackage{bbm}
\usepackage{alltt}
\usepackage{verbdef}
\usepackage{titlesec}
\usepackage{parskip}

\titlespacing*{\section}{0pt}{*1}{*1}
\titlespacing*{\subsection}{0pt}{*0.7}{*0.5}
\titlespacing*{\paragraph}{0pt}{*0.5}{*0.5}

\renewcommand{\topfraction}{0.95}

\renewcommand{\floatpagefraction}{0.9}

\usepackage[T1]{fontenc}

\DeclareFontFamily{OML}{zavm}{\skewchar\font=127 }
\DeclareFontShape{OML}{zavm}{m}{n}{<-> s*[.80] zavmr7t}{}
\DeclareFontShape{OML}{zavm}{b}{n}{<-> s*[.80] zavmb7t}{}
\DeclareFontShape{OML}{zavm}{m}{it}{<-> s*[.80] zavmri7m}{}
\DeclareFontShape{OML}{zavm}{b}{it}{<-> s*[.80] zavmbi7m}{}
\DeclareFontShape{OML}{zavm}{m}{sl}{<->ssub * zavm/m/it}{}
\DeclareFontShape{OML}{zavm}{bx}{it}{<->ssub * zavm/b/it}{}
\DeclareFontShape{OML}{zavm}{b}{sl}{<->ssub * zavm/b/it}{}
\DeclareFontShape{OML}{zavm}{bx}{sl}{<->ssub * zavm/b/sl}{}
\DeclareMathAlphabet{\mathsf}{OML}{zavm}{m}{n} % not `n'

\newcommand{\ifext}[2]{\ifdefined\extflag{#1}\else{#1}\fi}

\newcommand{\jpts}{\mathrel{\overset{j}{\mapsto}}}
\newcommand{\spts}{\mathrel{\overset{s}{\mapsto}}}
\newcommand{\opts}{\mathrel{\overset{o}{\mapsto}}}
\newcommand{\lsep}{\ast}
\newcommand{\lk}{\mathit{lk}}
\newcommand{\angled}[1]{\langle {#1} \rangle}

\newcommand{\tfr}[1]{t^{#1}_{\text{fresh}}}
\newcommand{\ssep}{\circledast}
\newcommand{\hpts}{\mapsto}
\newcommand{\dotcup}{\ensuremath{\mathaccent\cdot\cup}}
\newcommand{\Dotcup}[1]{\ensuremath{\mathaccent\cdot{\bigcup}}_{#1}} 
\newcommand{\hunion}{\mathbin{\dotcup}}
\newcommand{\Hunion}[1]{\mathbin{\Dotcup{#1}}}
\newcommand{\wand}{\mathop{-\!\!\!\!-\!\!\!\ast}}

\newcommand{\pfun}{\rightharpoonup}
\newcommand{\hpriv}{\mathsf{pv}}
\newcommand{\hfc}{\mathsf{fc}}
\newcommand{\hps}{\ell}

\newcommand{\eqm}{=_{\!/_{\!\text{mset}}}}
\newcommand{\htb}{\mathsf{tb}}
\newcommand{\sent}{\mathit{snt}}
\newcommand{\hlock}{\mathsf{lk}}

\newcommand{\taupop}{{\mathit{pop}}}
\newcommand{\alphapush}{{\mathit{push}}}

\newcommand{\nat}{\mathsf{nat}}
\newcommand{\ap}{\text{\emph{a}}_p}
\newcommand{\AAArray}{\mathsf{Arr}}
\newcommand{\Array}[4]{\AAArray_{#2}(#1, #3, #4)}

\newcommand{\Pushed}[2]{\mathsf{Pushed}(#1, #2)}
\newcommand{\Popped}[2]{\mathsf{Popped}(#1, #2)}
\newcommand{\push}{\text{push}}
\newcommand{\pop}{\text{pop}}
\newcommand{\NoReqs}{\mathsf{NoReq}}
\newcommand{\noreqs}[1]{\NoReqs(#1)}
\newcommand{\Requested}{\mathsf{HasReq}}
\newcommand{\requested}[1]{\Requested(#1)}
\newcommand{\Locked}{\mathsf{Locked}}
\newcommand{\locked}[1]{\Locked(#1)}
\newcommand{\Ack}{\mathsf{Ack}}
\newcommand{\ack}[1]{\Ack(#1)}
\newcommand{\LHR}{\mathsf{LHR}}
\newcommand{\lhr}[1]{\LHR(#1)}

\newcommand{\reqtrans}{\text{\emph{req}}}
\newcommand{\helptrans}{\text{\emph{help}}}
\newcommand{\colltrans}{\text{\emph{coll}}}

\newcommand{\Prod}{\mathsf{Pr}}
\newcommand{\Cons}{\mathsf{Cn}}
\newcommand{\hlabel}{\ell}
\newcommand{\llist}{\mathsf{list}}

\newcommand{\complete}{\mathsf{complete}}
\newcommand{\cont}{\mathsf{stacklike}}
\newcommand{\garb}{\mathit{grb}}
\newcommand{\relentangle}{\bowtie}
\newcommand{\entangle}{\rtimes}

\newcommand{\Null}{\text{\texttt{null}}}

\renewcommand{\phi}{\varphi} 

\DeclareFontFamily{U}{mathb}{\hyphenchar\font45}
\DeclareFontShape{U}{mathb}{m}{n}{
      <5> <6> <7> <8> <9> <10> gen * mathb
      <10.95> mathb10 <12> <14.4> <17.28> <20.74> <24.88> mathb12
}{}
\DeclareSymbolFont{mathb}{U}{mathb}{m}{n}
\DeclareMathSymbol{\blacktriangleleft} {2}{mathb}{"9E}% name to be checked
\DeclareMathSymbol{\blacktriangleright}{2}{mathb}{"9F}% name to be checked

\newcommand{\self}{\emph{self}\xspace}
\newcommand{\other}{\emph{other}\xspace}
\newcommand{\joint}{\emph{joint}\xspace}

\newcommand{\mathself}{\!\mathop{s}}
\newcommand{\mathother}{\!\mathop{o}}
\newcommand{\mathjoint}{\!\mathop{j}}
\newcommand{\zip}{\circ}

\newcommand{\id}{\mathsf{id}}

\newcommand{\hempty}{\mathsf{empty}}

\newcommand{\Histso}{\hookrightarrow}
\newcommand{\histso}[2]{{#1} \Histso ({#2})}

\newcommand{\ldot}{\mathord{.}\,}

\newcommand{\aand}{\wedge}
\newcommand{\oor}{\vee}

\newcommand{\eqdef}{\mathrel{\:\widehat{=}\:}}

\newcommand{\zig}{\triangleleft}
\newcommand{\zag}{\triangleright}

\newcommand{\acon}{\mathcal{A}}
\newcommand{\econ}{\mathcal{E}}
\newcommand{\ucon}{\mathcal{U}}
\newcommand{\vcon}{\mathcal{V}}
\newcommand{\wcon}{\mathcal{W}}
\newcommand{\privcon}{\mathcal{P}}
\newcommand{\pscon}{\mathcal{S}}
\newcommand{\tbcon}{\mathcal{T}}
\newcommand{\fccon}{\mathcal{F}}
\newcommand{\lcon}{\mathcal{L}}

\newcommand{\cmd}[1]{\text{\texttt{#1}}\xspace}
\newcommand{\Cmd}[2]{\text{\texttt{{#1}(}}{#2}\text{\texttt{)}}\xspace}

\newcommand{\sstar}{\circledast}

\newcommand{\nil}{\mathsf{nil}}

\newcommand{\size}[1]{|~{#1}~|}
\newcommand{\act}[1]{\textsf{\small{#1}}}
\newcommand{\num}[1]{\({\text{{\scriptsize{#1}}}}\)}
\newcommand{\Num}[1]{{\text{{\scriptsize{#1}}}}}
\newcommand{\esc}[1]{\text{\texttt{\small{#1}}}}
\newcommand{\eesc}[1]{\text{\texttt{\small{#1}}}}
\newcommand{\Unit}{\({\text{\textsf{Unit}}}\)}

\newcommand{\cmdLock}[1]{\cmd{lock}}

\newcommand{\lockOwn}{\mathsf{Own}}
\newcommand{\lockNown}{\cancel\lockOwn}
\newcommand{\lcl}{{\mathsf{S}}}%L
\newcommand{\env}{{\mathsf{O}}}%E

\newcommand{\mX}[1]{\mathsf{m}_{#1}}
\newcommand{\mL}{\mX{\lcl}}
\newcommand{\mE}{\mX{\env}}
\newcommand{\mS}{\mX{\lcl}}
\newcommand{\mO}{\mX{\env}}

\newcommand{\tX}[1]{\mathsf{t}_{#1}}

\newcommand{\tS}{\tX{\lcl}}
\newcommand{\tO}{\tX{\env}}

\newcommand{\g}{\mathsf{g}}
\newcommand{\gX}[1]{\g_{#1}}
\newcommand{\gL}{\gX{\lcl}}
\newcommand{\gE}{\gX{\env}}
\newcommand{\gS}{\gX{\lcl}}
\newcommand{\gO}{\gX{\env}}
\newcommand{\gp}{\gX{p}}
\newcommand{\gd}{\gX{\Delta}}
\newcommand{\gall}{\gX{\text{all}}}

\newcommand{\h}{h}

\newcommand{\Hist}[1]{\mathsf{hist}~{#1}}
\newcommand{\hist}{\tau}
\newcommand{\histL}{\hist_\lcl}
\newcommand{\histE}{\hist_\env}
\newcommand{\histS}{\histL}
\newcommand{\histO}{\histE}
\newcommand{\histd}{\hist_{\Delta}}

\newcommand{\hL}{h_\lcl}
\newcommand{\hE}{h_\env}
\newcommand{\hS}{h_\lcl}
\newcommand{\hO}{h_\env}
\newcommand{\Inv}{\mathit{Inv}} 

\newcommand{\specK}[1]{\ensuremath{\textcolor{blue}{#1}}}
\newcommand{\spec}[1]{\specK{\left\{{#1}\right\}}}
\newcommand{\drspec}[1]{\specK{\langle{#1}\rangle}}
\newcommand{\sspec}[1]{\specK{\{{#1}\}}}

\newcommand{\prog}[4]{\sspec{#1}\, #2 \, \sspec{#3}@{#4}}%{$\mathsf{prog}\ #1\ #2\ #3$}
\newcommand{\stconc}[4]{\prog{#1}{#2}{#3}{#4}}
\newcommand{\stconcTy}[5]{\stconc{#1}{#2 : #3}{#4}{#5}}
\newcommand{\pcmS}{\mathbb{U}}
\newcommand{\pcmF}{\bullet}
\newcommand{\join}{\pcmF}
\newcommand{\jjoin}[1]{\ensuremath{\bigodot}_{#1=1}^n} 
\newcommand{\pcmU}{\mathbbm{1}}
\newcommand{\fspec}[1]{{#1}^{\sharp}}
\newcommand{\None}{\mathsf{None}}
\newcommand{\Some}[1]{\mathsf{Some}\ {#1}}
\newcommand{\Req}[2]{\mathsf{Req}\ {#1}~{#2}}
\newcommand{\Resp}[1]{\mathsf{Resp}\ {#1}}
\newcommand{\Done}{\mathsf{Init}}

\newcommand{\flatten}[1]{\lfloor #1 \rfloor}

\theoremstyle{remark}
\newtheorem{example}{Example}[section]
\theoremstyle{definition} 
\newtheorem{definition}{Definition}[section]
\newtheorem{theorem}{Theorem}[section]
\newtheorem{lemma}[theorem]{Lemma}

{\unskip\nobreak\hskip 1em plus 1fil\nobreak$\square$
\parfillskip=0pt%
\endtrivlist}
\theoremstyle{plain}

\newcommand{\pre}{\sqsubseteq}
\newcommand{\last}[1]{\mathsf{last}({#1})}

\newcommand{\inat}[3]{{#2}[{#3}] = {#1}}

\newcommand{\tani}[3]{{#2}[{#3}] = {#1}}

\newcommand{\etc}{\emph{etc}}
\newcommand{\ie}{\emph{i.e.}\xspace}

\newcommand{\eg}{\emph{e.g.}\xspace}

\newcommand{\etal}{\emph{et~al.}\xspace}

\newcommand{\dom}[1]{\mathsf{dom}(#1)}

\newcommand{\wrt}{\emph{wrt.}\xspace}

\newcommand{\result}{\mathsf{res}}
\newcommand{\state}[3]{[{#1}~|~{#2}~|~{#3}]}

\newcommand{\sep}{\textasteriskcentered}

\newcommand{\res}{\result}

\newcommand{\set}[1]{\{{#1}\}}
\newcommand{\mset}[1]{\{\!\{{#1}\}\!\}}

\newcommand{\gs}{\mathsf{g}_{\mathsf{S}}}

\newcommand{\True}{\mathsf{true}}
\newcommand{\False}{\mathsf{false}}

\definecolor{shadecolor}{gray}{1.00}
\definecolor{ddarkgray}{gray}{0.75}
\definecolor{darkgray}{gray}{0.30}
\definecolor{light-gray}{gray}{0.85}

\makeatletter
\newcommand*{\textalltt}{}
\DeclareRobustCommand*{\textalltt}{%
  \begingroup
    \let\do\@makeother
    \dospecials
    \catcode`\\=\z@
    \catcode`\{=\@ne
    \catcode`\}=\tw@
    \verbatim@font\@noligs
    \@vobeyspaces
    \frenchspacing
    \@textalltt
}
\newcommand*{\@textalltt}[1]{%
    #1%
  \endgroup
}
\makeatother

\newcommand{\code}[1]{\texttt{\small{#1}}}
\newcommand{\tid}{\mathit{tid}}

\newcounter{tags}

\newcommand{\mytitle}{Specifying and Verifying Concurrent
  Algorithms\\
with Histories and Subjectivity}

\begin{document} 

%\special{papersize=8.5in,11in}
\setlength{\pdfpageheight}{\paperheight}
\setlength{\pdfpagewidth}{\paperwidth}

% \subtitle{Extended version}
%\subtitle{\large{\color{red}\today}}

%\titlebanner{In submission} % These are ignored unless
%\preprintfooter{} % 'preprint' option specified.

% Commented out for the double-blind review

% \authorinfo{Ilya Sergey}
%            {IMDEA Software Institute}
%            {ilya.sergey@imdea.org}
% %
% \authorinfo{Aleksandar Nanevski}
%            {IMDEA Software Institute}
%            {aleks.nanevski@imdea.org}
% %
% \authorinfo{Anindya Banerjee}
%            {IMDEA Software Institute}
%            {anindya.banerjee@imdea.org}

% \title{\mytitle\vspace{-10pt}}

\ifext{
\authorinfo{Ilya Sergey\vspace{-2pt}}
           {IMDEA Software Institute}
           {ilya.sergey@imdea.org}
\authorinfo{Aleksandar Nanevski\vspace{-2pt}}
           {IMDEA Software Institute}
           {aleks.nanevski@imdea.org}
\authorinfo{Anindya Banerjee\vspace{-2pt}}
           {IMDEA Software Institute}
           {anindya.banerjee@imdea.org}

\title{\mytitle}
%\subtitle{Extended version\vspace{-5pt}}
}{
\title{\mytitle\vspace{-21pt}}
\authorinfo{}{}{} 
} 

\maketitle 

\begin{abstract}

We present a lightweight approach to Hoare-style specifications for
fine-grained concurrency, based on a notion of \emph{time-stamped
  histories} that abstractly capture atomic changes in the program
state.
Our key observation is that histories form a \emph{partial commutative
  monoid}, a structure fundamental for representation of concurrent
resources.
This insight provides us with a unifying mechanism that allows us to
treat histories just like heaps in separation logic. For example, both
are subject to the same assertion logic and inference rules (\eg, the
frame rule). Moreover, the notion of ownership transfer, which usually
applies to heaps, has an equivalent in histories. It can be used to
formally represent helping---an important design pattern for
concurrent algorithms whereby one thread can execute code on behalf of
another.
Specifications in terms of histories naturally abstract granularity,
in the sense that sophisticated fine-grained algorithms can be given
the same specifications as their simplified coarse-grained
counterparts, making them equally convenient for client-side
reasoning.
We illustrate our approach on a number of examples and validate all of
them in~Coq.
  
\end{abstract}

% \category{D.1.1}{Programming Techniques}{Applicative (Functional) Programming}
% %
% \category{F.3.2}{Logics and Meanings of Programs}{Semantics of
%  Programming Languages --- Program analysis, Operational semantics}

% do nothing

% \terms
% Languages, Theory, Verification

\section{Introduction}
\label{sec:intro}

For sequential programs and data structures, Hoare-style
specifications (or specs) in the form of pre- and postconditions are %uniformly
%accepted as 
a declarative way to express a program's behavior. For example, an
abstract specification of stack operations can be given as follows:
\[
\tag{\arabic{tags}}\refstepcounter{tags}\label{eq:seqstack-spec}
{\small
\hspace{-5pt}
\begin{array}{r@{\ }c@{\ }l}
\spec{~s \hpts \mathit{xs}~} &\!\Cmd{push}{x}\! & \spec{~ s \hpts
  x::\mathit{xs}~}
\\[3pt]
\spec{~s \hpts \mathit{xs}~} & \!\!\Cmd{pop}{}\!\! &\spec{\!\!\!
\begin{array}{l}
  \res = \None \aand \mathit{xs} = \nil \aand s \hpts \nil \oor \hbox{}\\
  \res = \Some x \aand \exists \mathit{xs'},~\mathit{xs} = x \!::\!\mathit{xs'} \aand
  s \hpts \mathit{xs'}
\end{array}
\!\!\!\!}
\end{array}
}
\]
where $s$ is an ``abstract pointer'' to the data structure's logical
contents, and the logical variable $\mathit{xs}$ is universally
quantified over the spec.
The result $\res$ of $\cmd{pop}$ is either $\Some x$, if $x$ was on
the top of the stack, or $\None$ if the stack was empty.
The spec (\ref{eq:seqstack-spec}) is usually accepted as canonical for
stacks: it hides the details of method implementation, but exposes
what's important about the method behavior, so that a verification of
a stack \emph{client} doesn't need to explore the implementations of
$\cmd{push}$ and $\cmd{pop}$.

The situation is much more complicated in the case of concurrent data
structures. In the concurrent setting, (\ref{eq:seqstack-spec}) is of
little use, as the interference of the threads executing concurrently
may invalidate the assertions about the stack. For example, a call to
$\cmd{pop}$ may encounter an empty stack, and decide to return
$\None$, but by the time it returns, the stack may be filled by the
other threads, thus invalidating the postcondition of $\cmd{pop}$ in
(\ref{eq:seqstack-spec}). To soundly reason about concurrent data
structures, one has to devise specs that are \emph{stable} (\ie,
invariant under interference), but this may require trade-offs.

\ifext{
\enlargethispage{\baselineskip}
}{}

For instance, a few recent
proposals~\cite{Turon-al:ICFP13,Svendsen-al:ESOP13} rely on the
following spec, which restricts the stack elements to satisfy a fixed
client-chosen predicate~$P$:
\[
\tag{\arabic{tags}}\refstepcounter{tags}\label{eq:stack-weak}
{\small
\begin{array}{r@{\ }c@{\ }l}
  \spec{~~P(x)~~} &\Cmd{push}{x} & \spec{~~ \mathsf{true}~~}
  \\ 
  \spec{~~ \mathsf{true}~~} & \Cmd{pop}{} &\spec{~~ \res = \Some x
  \implies P(x) ~~}
\end{array}
}
\]
Specification (\ref{eq:stack-weak}) is stable, but it isn't canonical,
as it doesn't capture the LIFO element management policy. It holds of
any other container structure, such as queues.

%Another recent
%proposal~\cite{Svendsen-al:ESOP13,Svendsen-Birkedal:ESOP14} involves
%parametrizing the spec of the data structure \wrt~auxiliary callback
%code~\cite{Jacobs-Piessens:POPL11} that describes how the abstract
%contents of the stack changes upon execution of the methods. The
%clients can choose the callbacks that makes the spec stable for their
%particular use, thus achieving canonicity, but at the expense of
%polluting the specs and proofs of the data structure implementation
%with client-side information. Moreover, the parametrization requires
%impredicative quantification over callbacks and leads to rather
%indirect specs and non-elementary model~theory.

Reasoning about concurrent data structures is further complicated by
the fact that their implementations are often \emph{fine-grained}.
Striving for better performance, they avoid explicit locking, and
implement sophisticated synchronization patterns that deliberately
rely on interference.
For reasoning purposes, however, it is desirable that the clients can
perceive such fine-grained implementations as if they were
\emph{coarse-grained}; that is, as if the effects of their methods
take place \emph{atomically}, at singular points in time.  The
standard correctness criteria of
\emph{linearizability}~\cite{Herlihy-WingTOPLAS90} establishes that a
fine-grained data structure implementation \emph{contextually refines}
a coarse-grained one~\cite{Filipovic-al:TCS10}. One can make use of a
refined, fine-grained, implementation for efficiency in programming,
but then soundly replace it with a more abstract coarse-grained
implementation, to simplify the reasoning about clients.
% therefore
%taking advantage of a \emph{granularity abstraction}.

Semantically, one program linearizes to another if the
\emph{histories} of the first program (\ie, the sequence of actions it
executed) can be transformed, in a suitable sense, into the histories
of the second. Thus, histories are an essential ingredient in
specifying fine-grained concurrent data structures. However, while a
number of logical methods exist for establishing the linearizability
relation between two programs, for a class of
structures~\cite{OHearn-al:PODC10,Vafeiadis-al:PPOPP06,Elmas-al:TACAS10,Vafeiadis:PhD,Liang-Feng:PLDI13},
in general, it's a non-trivial property to prove and use. First, in a
setting that employs Hoare-style reasoning, showing that a
fine-grained structure refines a coarse-grained one is not an end in
itself. One still needs to ascribe a stable spec to the coarse-grained
version~\cite{Turon-al:ICFP13,Liang-Feng:PLDI13}.
Second, the standard notion of linearizability doesn't directly
account for modern programming features, such as ownership transfer of
state between threads, pointer aliasing, and higher-order
procedures. Theoretical extensions required to support these features
are a subject of active ongoing
research~\cite{Cerone-al:ICALP14,Gotsman-Yang:CONCUR12}.
Finally, being a relation on \emph{two} programs, deriving
linearizability by means of logical inference inherently requires a
\emph{relational program
  logic}~\cite{Turon-al:ICFP13,Liang-Feng:PLDI13}, even though the
spec one is ultimately interested in (\eg,~\eqref{eq:stack-weak} for a
concurrent stack) may be expressed using a Hoare triple that operates
over a \emph{single} program.

\ifext{
\enlargethispage{\baselineskip}
}{}

In this paper, we propose a novel method to specify and verify
fine-grained programs \emph{as well as} provide a form of granularity
abstraction, by directly reasoning about histories in the specs of an
elementary Hoare logic. We propose using \emph{timestamped}
histories, which carry information about the atomic changes in the
abstract state of the program, indexed by discrete time stamps, and
tracking the history of a program as a form of auxiliary state.

Histories can help abstract the granularity of a program as
follows. We consider a program \emph{logically} atomic (irrespective
of the physical granularity of its implementation), if its history is
a singleton history $t \hpts a$, containing only an abstract action
$a$ time-stamped with $t$. This spec provides an abstraction that the
effect $a$ of the program takes place at a singular point in time $t$,
as if the program were coarse-grained, thus achieving exactly the main
goal of linearizability, without needing contextual refinement.
%
%(but see Section~\ref{sec:related} for comparison
%with~\cite{ArrozPincho-al:ECOOP14,Svendsen-al:ESOP13,Svendsen-Birkedal:ESOP14}).
%
Client-side proofs can be developed out of such a spec, while ignoring
the details of a potentially fine-grained implementation. The user can
select the desired level of granularity, by choosing the actions $a$
to use in the histories.
%
%%
%Our starting point is to allow the Hoare-style preconditions and
%postconditions to specify the program behavior in terms of an
%algebraic structure of \emph{time-stamped histories} (or just
%\emph{histories}), which carry information about the atomic changes in
%the abstract state of a concurrent data structure, indexed by discrete
%time stamps.  
%
%Specifically, we consider a program atomic (in the abstract,
%irrespective of the granularity of its implementation) if it
%increments its history by a singleton history $t \hpts a$, containing
%a single abstract action $a$, time-stamped with $t$.  This notion of
%atomicity in the abstract doesn't rely on linearizability or
%contextual refinement, but achieves a similar goal to them, when it
%comes to Hoare-style reasoning. In particular, history-based specs
%provide the abstraction that the effects of the program take place at
%singular points in time, determined by the time stamps, as if the
%program were coarse-grained.  Client-side proofs can be developed out
%of such specs, while ignoring the details of the potentially
%fine-grained implementation.
%
While using histories in Hoare logic specs is a simple and natural
idea, and has been employed
before~\cite{Fu-al:CONCUR10,Gotsman-al:ESOP13}, in our paper it comes
with two additional novel observations. 

First, timestamped histories are technically very similar to heaps,
as both satisfy the algebraic properties of a \emph{partial
  commutative monoid} (PCM). A PCM is a set $\pcmS$ with an
associative and commutative \emph{join} operation $\pcmF$ and unit
element $\pcmU$. Both heaps and histories form a PCM with disjoint
union and empty heap/history as the unit. Also, a singleton history $t
\hpts a$ is very similar to the singleton heap $x \hpts v$ containing
only the pointer $x$ with value $v$. 
%
%They both are \emph{prime} elements of their respective PCMs, \ie,
%neither is representable as a disjoint union of two non-units.
We emphasize the connection by using the same notation for both.

The common PCM structure makes it possible to reuse for histories the
ideas and results developed for heaps in the work on separation
logic~\cite{Calcagno-al:LICS07}. In particular, in this paper, we make
both heaps and histories subject to the same assertion logic and the
same rules of inference (\eg, the frame rule). Moreover, concepts such
as ownership transfer, that have been developed for heaps, apply to
histories as well. For example, in Section~\ref{sec:flatco}, we use
ownership transfer on histories to formalize the important design
pattern of \emph{helping}~\cite{Hendler-al:SPAA10}, whereby a
concurrent thread may execute a task on behalf of other threads. That
helping corresponds to a kind of ownership transfer (though not on
histories, but on auxiliary commands) has been noticed
before~\cite{Turon-al:POPL13,Liang-Feng:PLDI13}.  However, commands
don't form a PCM, while histories do -- a fact that makes our
development simple and uniform.

Second, we argue that precise history-based specs have to
differentiate between the actions that have been performed by the
specified thread, from the actions that have been performed by the
thread's concurrent environment. Thus, our specs will range over
\emph{two} different history-typed variables, capturing the
timestamped actions of the specified thread (\emph{self}) and its
environment (\emph{other}), respectively. This split between self and
other will provide us with a novel and very direct way of relating the
functional behavior of a program to the interference of its concurrent
environment, leading to specs that have a similar canonical ``feel''
in the concurrent setting, as the specs (\ref{eq:seqstack-spec}) have
in the sequential one. 

The self/other dichotomy required of histories is a special case of
the more general specification pattern of \emph{subjectivity},
observed in the recent related work on Subjective and Fine-grained
Concurrent Separation Logic
(FCSL)~\cite{LeyWild-Nanevski:POPL13,Nanevski-al:ESOP14}. That work
generalized Concurrent Separation Logic (CSL)~\cite{OHearn:TCS07}
%,and its modular rules for framing and parallel composition, 
to apply not only to heaps, but to any abstract notion of state (real
or auxiliary) satisfying the PCM properties.
We thus reuse FCSL~\cite{Nanevski-al:ESOP14} off-the-shelf, and
instantiate it with histories, \emph{without any additions to the
  logic or its meta-theory}. Surprisingly, the FCSL style of auxiliary
state is sufficient to enable expressive history-based,
granularity-abstracting specs, and proofs of realistic fine-grained
algorithms, including those with helping. We show how a number of
well-known algorithms can be proved logically atomic, and illustrate
how the atomic specs facilitate client-side reasoning. We consider an
atomic pair snapshot data
structure~\cite{Qadeer-al:TR09,Liang-Feng:PLDI13}
(Section~\ref{sec:overview}), Treiber stack~\cite{Treiber:TR} along
with its clients (Section~\ref{sec:examples}), and Hendler \etal's
flat combining algorithm~\cite{Hendler-al:SPAA10}, a highly
non-trivial example employing higher-order functions and helping
(Section~\ref{sec:flatco}). All our proofs, including the theory of
histories, have been checked mechanically in Coq.\footnote{Available
  at \url{http://ilyasergey.net/other/fcsl-histories.zip}.}

\section{Overview: specifying snapshots with histories}
\label{sec:overview}

In this section, we illustrate history-based specifications by applying
them to the fine-grained \emph{atomic pair snapshot} data
structure~\cite{Qadeer-al:TR09,Liang-Feng:PLDI13}.
This data structure contains a pair of pointers, $x$ and $y$, pointing
to tuples $(c_x, v_x)$ and $(c_y, v_y)$, respectively. The components
$c_x$ and $c_y$ of type $A$ represent the accessible contents of $x$
and $y$, that may be read and updated by the client. The components
$v_x$ and $v_y$ are $\mathsf{nat}$s, encoding ``version numbers'' for
$x$ and $y$. They are internal to the structure and not directly
accessible by the client.

The structure exports three methods: \code{readPair}, \code{writeX},
and \code{writeY}. \code{readPair} is the main method, and the focus of the section. It returns the
\emph{snapshot} of the data structure, \ie, the accessible contents of
$x$ and $y$ as they appear together at the moment of the
call. However, while $x$ and $y$ are being read by \code{readPair},
other threads may change them, by invoking \code{writeX} or
\code{writeY}. Thus, a na\"{i}ve implementation of \code{readPair} which
first reads $x$, then $y$, and returns the pair $(c_x, c_y)$ does not
guarantee that $c_x$ and $c_y$ ever appeared together in the
structure. One may have \code{readPair} first lock $x$ and $y$ to
ensure exclusive access, but here we consider a fine-grained
implementation which relies on the version numbers to ensure that
\code{readPair} returns a valid snapshot.

The idea is that $\esc{writeX}(\esc{cx})$ (and symmetrically,
$\esc{writeY}(\esc{cy})$), changes the logical contents of $x$ to
$\esc{cx}$, while incrementing the internal version number,
\emph{simultaneously}. Since the operation involves changes to the
contents of a single pointer, in this paper we assume that it can be
performed atomically (\eg, by some kind of read-modify-write
operation~\cite[\S5.6]{Herlihy-Shavit:08}). We also assume atomic
operations \code{readX} and \code{readY} for reading from $x$ and $y$
respectively. Then the implementation of \code{readPair}
(Figure~\ref{fig:readpair}) reads from $x$ and $y$ in succession, but
makes a check (line~5) to compare the version numbers for $x$ obtained
before and after the read of $y$. In case $x$'s version has changed,
the procedure is restarted.

\begin{figure}[t]
\centering 
\small
\begin{tabular}{l@{\ \ \ }l}
\begin{minipage}[l]{4.3cm}
\begin{alltt}
\num{1}  readPair(): \(A {\times} A\)  \{
\num{2}    (cx, vx) <- \act{readX}();
\num{3}    (cy, _)  <- \act{readY}();
\num{4}    (_, tx)  <- \act{readX}();
\num{5}    \textbf{if} vx == tx 
\num{6}    \textbf{then return} (cx, cy);
\num{7}    \textbf{else return} readPair();\}
\end{alltt} 
\end{minipage}
& 
%
%      \(\sspec{x {\spts} \histS}\)
%
%\begin{minipage}[l]{3.8cm}
%\begin{alltt}
%writeX(cx: \(A\)): \Unit \{ 
%  \act{writeAndIncX}(cx); 
%\}%
%
%writeY(cy: \(A\)): \Unit \{
%  \act{writeAndIncY}(cy); 
%\}
%\end{alltt}
%\end{minipage}
%
\end{tabular}
\caption{Main method of the atomic pair snapshot data structure.}
\label{fig:readpair}
\end{figure}

We want to specify and prove that such an implementation of
\code{readPair} is correct; that is, if it returns a pair $(c_x,
c_y)$, then $c_x$ and $c_y$ occurred simultaneously in the structure.
To do so, we use histories as auxiliary state of every method of the
structure. Histories, ranged over by $\hist$, are finite maps from the
natural numbers to pairs of elements of some type $S$; \ie,
$\Hist{S}\ {\eqdef}\ \mathsf{nat} \pfun S \times S$. The natural
numbers represent the moments in time, and the pairs represent the
change of state. Thus, a singleton history $t \hpts (s_1, s_2)$
encodes an atomic change from abstract state $s_1$ to abstract state
$s_2$ at the time moment $t$. We will only consider \emph{continuous}
histories, for which $t \hpts (s_1, s_2)$ and $t+1 \hpts (s_3, s_4)$
implies $s_2 = s_3$. We use the following abbreviations to work with
histories:
\[
\tag{\arabic{tags}}\refstepcounter{tags}\label{eq:hist-not}
\begin{array}{lcl}
{\hist}[t] & \eqdef & s,~\text{such that}~\exists s',~\hist(t) = (s', s) \\ 
\hist \le t & \eqdef & \forall t' \in \dom{\hist}, t' \le t\\
\hist \pre \hist' & \eqdef & \mbox{$\hist$ is a subset of $\hist'$}
\end{array}
\]

Similarly to heaps, histories form a PCM under the operation $\hunion$
of disjoint union, with the $\hempty$ history as the unit. The type
$S$ can be chosen arbitrarily, depending on the application, to
capture whichever logical aspects of the actual physical state are of
interest.
For the snapshot structure, we take $S = A \times A \times \nat$. That
is, the entries in the histories for pair snapshot will be of the form
\vspace{-2pt}
\[
\tag{\arabic{tags}}\refstepcounter{tags}\label{eq:entry}
t \hpts
(\angled{c_x, c_y, v_x}, \angled{c'_x, c'_y, v'_x}).\] 
\ifext{
\enlargethispage{\baselineskip}
}{}
The entry encodes that at time moment $t$, the contents of $x$, $y$,
and the version of $x$ have changed from $(c_x, c_y, v_x)$ to $(c'_x,
c'_y, v'_x)$. We ignore $v_y$, as it doesn't factor in the
implementation of \code{readPair}.
%
%\an{I think we should just drop vy
%  from the story completely. That way we don't waste space explaining
%  it, and then saying it does not matter.}
%
%\is{We'd rather not, as it makes the program assymetric and quite
%  ugly, especially, taking into account that the original program
%  from~\cite{Liang-Feng:PLDI13} dealt with a pair snapshot over an
%  arbitrary number of cells (pick any two).}  \an{Well, the whole
%  thing is already assymetric since the histories don't keep track of
%  y, and thus the specs for readX and readY are completely
%  different. But ok, it is not a big deal.}
%

All the threads working over the pair snapshot structure respect a
protocol on histories consisting of the following three properties. We
explain in Section~\ref{sec:background} how these are formally
specified and enforced, but for now simply assume them. They will be
important in the proof outline for \code{readPair}.
\begin{itemize}
\item [$(i)$] Whenever a thread modifies $x$ or $y$ (\eg, by calling
  \code{writeX} or \code{writeY}), its history gets augmented by an
  entry such as (\ref{eq:entry}), where the timestamp $t$ is chosen
  afresh. Thus, histories only grow, and only by adding valid
  snapshots.
\item [$(ii)$] Whenever the contents of $x$ is changed in a history,
  its version number changes too. In contrapositive form, if
  $\tau[t_1] = \angled{c_1, -, v}$ and $\tau[t_2] = \angled{c_2, -,
    v}$, then $c_1 = c_2$.
\item [$(iii)$] Version numbers in a history grow monotonically. That
  is, if $\tau[t_1] = \angled{-, -, v_1}$ and $\tau[t_2] = \angled{-,
    -, v_2}$ and $t_1 \le t_2$, then $v_1 \le v_2$.
\end{itemize}
\paragraph{Specification.}
We now describe an FCSL spec for \code{readPair} and explain how it
captures that its result is a valid snapshot of $x$ and $y$.
%
%\[
%\tag{\arabic{tags}}\refstepcounter{tags}\label{eq:pair-spec-read}
%{\scriptsize
%\begin{array}{r@{\ }c@{\ }l}
%\spec{\!\!\!\!\begin{array}{l}
%  \exists \histO\ldot
%  \hps \spts \hempty \aand \hbox{}\\ 
%  \quad \hps \opts \histO \aand \hist \pre \histO  
%  \end{array}
%\!\!\!\!} &\Cmd{readPair}{} & 
%\spec{\!\!\!\!\begin{array}{l}
%  \exists \histO\ t\ldot
%    \hps \spts \hempty \aand \hps \opts \histO \aand \hbox{} \\ 
%    \quad \hist \pre \histO \aand \tau \le t \aand \hbox{}\\
%    \quad \inat{\angled{\res.1, \res.2, -}}{\histO}{t}
%  \end{array}
%\!\!\!\!}
%\end{array}
%}
%\]
%
%
{\small
\begin{gather*}
\spec{\exists \histO\ldot \hps \spts \hempty \aand \hps \opts \histO
  \aand \hist \pre \histO}
\\[-4pt]
\Cmd{readPair}{} 
\tag{\normalsize \arabic{tags}}\refstepcounter{tags}\label{eq:pair-spec-read}
\\[-3pt]
\spec{\!\!\!\begin{array}{l}
  \exists \histO\ t\ldot \hps \spts \hempty \aand \hps \opts \histO \aand \hist \pre \histO \aand \hbox{}\\
          \qquad\ \ \tau \le t \aand \inat{\angled{\res.1, \res.2, -}}{\histO}{t}
    \end{array}\!\!\!}
\end{gather*}
}
\!First, note the label $\hps$, which serves as an ``abstract
pointer'' that differentiates the instance of the pair snapshot
structure from any other structure that may exist in the program. In
particular, $\hps$ identifies the histories of concern to
\code{readPair}. Each thread keeps track of two such histories: the
self-history, describing the operations that the thread itself has
executed, and the other-history, describing the operations executed by
all the other threads combined. They are captured by the assertions
$\hps \spts \tau$ and $\hps \opts \tau$, respectively.

Thus, the precondition in (\ref{eq:pair-spec-read}) requires that
\code{readPair} starts with the empty self-history, \ie, the calling
thread has not performed any updates to $x$ or $y$. We show in
Section~\ref{sec:background} that the frame rule can be used to relax
the requirement, so that \code{readPair} can be invoked by threads
with an arbitrary self history.
The precondition allows an arbitrary initial other-history
$\histO$. As $\histO$ is bound locally in the precondition, and we
need to relate to it in the postcondition, we use the logical variable
$\hist$, and a conjunct $\hist \pre \histO$ to ``name'' it. The
conjunct uses inclusion (instead of equality). Inclusion makes the
precondition stable under growth of $\histO$ due to interfering
threads, according to $(i)$.
%However, we can't achieve the naming in the usual way, by conjoining
%$\hist = \histO$ to the precondition. Such an equation is unstable,
%because the environment threads can augment the history
%$\histO$. Fortunately, conjoining $\hist \pre \histO$ suffices.

The postcondition states that \code{readPair} does not perform any
changes to $x$ and $y$; it's a \emph{pure} method, thus its
self-history remains empty.
%
%\an{Should I say that according to our criterion of atomicity from
%  Section~\ref{sec:intro}, this makes \code{readPair} an atomic
%  program.}
%
The main novelty of the specification is that the postcondition
directly relates the result of \code{readPair} to the interference of
the environment, \ie, to the value of $\histO$. Referring to $\histO$
may look odd at first, but it's appropriate, and precisely specifies
what \code{readPair} returns. In particular, the postcondition says
that $\inat{\angled{\res.1, \res.2, -}}{\histO}{t}$, \ie, that the
components of the returned pair $\res$ appear in the environment
history. Since according to the property $(i)$ above, the histories
only store valid snapshots, the resulting pair must be a valid
snapshot too. In other words, \code{readPair} behaves as if it read
$x$ and $y$ atomically, at time $t$.
Moreover, $\tau \le t$, \ie, the read occurred after \code{readPair}
was invoked.

The specification pattern whereby a logical variable $\tau$ names the
initial history of the environment is very common, so we streamline it
by introducing the following notation.
\[
\tag{\arabic{tags}}\refstepcounter{tags}\label{eq:histso}
{\small
\begin{array}{r@{\ }c@{\ }l}
\histso{\ell}{\histS, \histO, \hist} \eqdef \ell \spts \histS \aand \ell
\opts \histO \aand \hist \pre \histS \hunion \histO
 \end{array}
}
\]

\begin{figure}
\begin{align*}
{\small
\begin{array}{r@{\ \ \ }l@{\ }l}
 \Num{1} & \spec{~~\histso{\hps}{\hempty, -, \hist}~~} & 
  \\[1pt]
  \Num{2} &  ~\esc{readPair():}~ A \times A~\{ & 
  \\[1pt]
 \Num{3}  & \spec{~~\histso{\hps}{\hempty, -, \hist}~~} & 
  \\[1pt]
 \Num{4} &  ~\esc{(cx, vx) <- \act{readX}();}  & 
  \\[1pt]
 \Num{5} & \spec{
    \begin{array}{l@{\ }l}
     \histso{\hps}{\hempty, \hist_1, \hist} \aand \hist \le t_1 \aand 
     \inat{\angled{\esc{cx}, -, \esc{vx}}}{\hist_1}{t_1}
    \end{array}
  } &
  \\[1pt]
  \Num{6} &  ~\esc{(cy, \_) <- \act{readY}();} & 
  \\[1.5pt]
  \Num{7} & \spec{
    \def\arraystretch{1.2}
    \begin{array}{l}
    \histso{\hps}{\hempty, \hist_2, \hist} \aand \hist \le t_1 \le t_2 \aand \esc{vx} \le v \aand \hbox{} \\
    \inat{\angled{\esc{cx}, -, \esc{vx}}}{\hist_2}{t_1} \aand
    \inat{\angled{c, \esc{cy}, v}}{\hist_2}{t_2}   
    \end{array}
  } & 
  \\[1.5pt]
\Num{8} &  ~\esc{(\_, tx) <- \act{readX}();} & 
  \\[1.5pt]
  \vspace{1mm}
 \Num{9} & \spec{
    \def\arraystretch{1.2}
    \begin{array}{l@{\ }l}
      \histso{\hps}{\hempty, \hist_3, \hist} \aand \hist \le t_1 \le t_2 \le t_3 \aand \esc{vx}\le v \le \esc{tx} \aand \hbox{}\\
      \inat{\angled{\esc{cx}, -, \esc{vx}}}{\hist_3}{t_1}  \aand
      \inat{\angled{c, \esc{cy}, v}}{\hist_3}{t_2} \aand \inat{\angled{-, -, \esc{tx}}}{\hist_3}{t_3} \\
    \end{array}
  } & 
  \\[1.5pt]
  \Num{10} &  ~\esc{\textbf{if} vx == tx} \\
 \Num{11} & \quad\spec{
    \begin{array}{l@{\ }l}
      \histso{\hps}{\hempty, \hist_3, \hist} \aand \hist \le t_2
      \aand \esc{cx} = c \aand
      \inat{\angled{\esc{cx}, \esc{cy}, v}}{\hist_3}{t_2}
    \end{array}
  } & 
  \\[1.5pt]
  \Num{12} &  \quad~\esc{\textbf{then return}~(cx, cy);} 
  \\[1.5pt]
  \Num{13} & 
  \quad \spec{~~ 
  \exists \histO\ t\ldot \histso{\hps}{\hempty, \histO, \hist} \aand \hist \le t \aand \inat{\angled{\res.1, \res.2, -}}{\histO}{t} ~}
  &
 \\[1.5pt]
  \Num{14} &  ~\esc{\textbf{else return} readPair();\}}
  \\[1.5pt]
\Num{15} & \spec{~~
   \exists \histO\ t\ldot \histso{\hps}{\hempty, \histO, \hist} \aand \hist \le t \aand \inat{\angled{\res.1, \res.2, -}}{\histO}{t} ~}
\end{array}
}
\end{align*}
\caption{Proof outline for \code{readPair}. Note that $\hist \pre
  \histO$ is folded into the definition of $\histso{\hps}{\hempty,
    \histO, \hist}$.}
\label{fig:pair-proof}
\vspace{-2mm}
\end{figure}

\paragraph{Proof outline.}
Figure~\ref{fig:pair-proof} contains the proof outline for
\code{readPair}, which we discuss next. Lines 1 and 3 abbreviate the
precondition in (\ref{eq:pair-spec-read}). The \code{readX} method has
the following spec:
\[
\tag{\arabic{tags}}\refstepcounter{tags}\label{eq:readx-spec}
{\small
\hspace{-3.5mm}
\begin{array}{r@{\ }c@{\ }l}
\spec{\!\!\!
  \begin{array}{l}
    \histso{\hps}{\hempty, -, \hist}
  \end{array}
\!\!\!} &\Cmd{readX}{} & 
\spec{\!\!\! 
  \begin{array}{l}
    \exists \histO\ t\ldot \histso{\hps}{\hempty, \histO, \hist} \aand \hbox{}\\
    \quad \hist \le t \aand  \inat{\angled{\res.1, -, \res.2}}{\histO}{t}
  \end{array}
\!\!\!}
\end{array}
}
\]
Thus in line~5 of the proof outline, we infer the existence of the
history $\hist_1$ and time stamp $t_1 \geq \hist$, such that the
\code{cx} and \code{vx} appear in $\hist_1$ at the time
$t_1$. 
Similarly, \code{readY} has the spec:
%\
\[
\tag{\arabic{tags}}\refstepcounter{tags}\label{eq:ready-spec}
{\small
\begin{array}{r@{\ }c@{\ }l}
\spec{\!\!\!
  \begin{array}{l}
    \histso{\hps}{\hempty, -, \hist}
  \end{array}
\!\!\!} & \Cmd{readY}{} & 
\spec{\!\!\! 
  \begin{array}{l}
    \exists \histO\ t\ldot \histso{\hps}{\hempty, \histO, \hist} \aand \hbox{}\\
    \quad \hist \le t \aand \inat{\angled{-, \res.1, -}}{\histO}{t}
  \end{array}
\!\!\!}
\end{array}
}
\]
To obtain line~7, instantiate $\hist$ with $\hist_1$ in the spec of
\code{readY}. This derives the existence of $\tau_2$, $t_2$, $c$ and
$v$, such that $\histso{\hps}{\hempty, \hist_2, \hist_1}$, $\hist_1
\le t_2$, and $\inat{\angled{c, \esc{cy}, v}}{\hist_2}{t_2}$. Because
$t_1 \in \mathsf{dom}(\hist_1)$, it must be that $t_1 \le
t_2$. Moreover, because $\tau \pre \tau_1 \pre \tau_2$, we further
obtain $\histso{\hps}{\hempty, \hist_2, \hist}$, and $\hist \le t_2$,
and lifting from line 5, $\inat{\angled{\esc{cx}, -,
    \esc{vx}}}{\hist_2}{t_1}$. Because $t_1, t_2$ appear in the same
history $\hist_2$, with versions $\esc{vx}$ and $v$, respectively, by
property $(iii)$, $\esc{vx} \le v$. Similarly, instantiating $\hist$
in the spec of \code{readX} with $\hist_2$, and invoking $(iii)$,
derives line~9 of the proof outline, and in particular $\esc{vx} \le v
\le \esc{tx}$.

From this property, if $\esc{vx} = \esc{tx}$ in the conditional on
line~10, it must be that $\esc{vx} = v$, and thus by $(ii)$, $\esc{cx}
= c$. Substituting $c$ by $\esc{cx}$ in line~9 gives us
$\inat{\angled{\esc{cx}, \esc{cy}, v}}{\hist_3}{t_2}$, which, after
$(\esc{cx}, \esc{cy})$ are returned in $\res$, obtains the
postcondition of \code{readPair}. Otherwise, if $\esc{vx} \neq
\esc{tx}$ in the conditional 10, we perform the recursive call to
\code{readPair}. The precondition for the call is
$\histso{\hps}{\hempty, -, \hist}$, which is clearly met in line~9, so
the postcondition immediately follows.

\paragraph{Monolithic histories.}
We compare the spec (\ref{eq:pair-spec-read}) with an alternative spec
where the history is not split into self/other portions, but is kept
monolithically as a \emph{joint} (or shared) state. We use the
predicate $\hps \jpts \tau$ to specify such state:
\[
\tag{\arabic{tags}}\refstepcounter{tags}\label{eq:readpair-spec2}
{\small
\begin{array}{c}
\spec{\exists \histO\ldot \hps \jpts \histO \aand \hist \pre \histO}\\
\Cmd{readPair}{} \\
\spec{\exists \histO\ t\ldot \hps \jpts \histO \aand \hist \pre \histO \aand 
       \tau \le t \aand \inat{\angled{\res.1, \res.2, -}}{\histO}{t}}
\end{array}
}
\]
Note that the spec (\ref{eq:readpair-spec2}) imposes no restrictions
on the growth of $\histO$ (unlike (\ref{eq:pair-spec-read}) which
keeps the self history $\hempty$). Thus, (\ref{eq:readpair-spec2}) is
weaker than (\ref{eq:pair-spec-read}), as it allows more behaviors. In
particular, it can be ascribed to any program which, in addition to
calling \code{readPair}, also modifies $x$ and $y$. This substantiates
our claim from Section~\ref{sec:intro} that the self/other dichotomy
is required to prevent history-based specs from losing precision. We
provide further evidence for this claim in Section~\ref{sec:examples},
where we show that subjective specs for \emph{stacks} generalize the
sequential canonical ones~\eqref{eq:seqstack-spec}. The latter can be
derived from the former by restricting $\histO$ to be the empty
history. Such a restriction isn't possible if the history is kept
monolithic.

\section{Background: a review of FCSL}
\label{sec:background}
In this section we review the relevant aspects of the previous work on
Fine-grained Concurrent Separation Logic
(FCSL)~\cite{Nanevski-al:ESOP14}. We explain FCSL by showing how it
can be specialized to our novel contribution of specifying concurrent
objects by means of histories. FCSL has been previously implemented as
a shallow embedding in Coq; thus our assertions will freely use Coq's
higher-order logic and datatype definition mechanism whenever
required.

%
%We will focus on several main aspects of FCSL. First, we illustrate
%how FCSL allows encoding and enforcing invariant properties of a
%stateful structure (such as the properties $(i)-(iii)$ in
%Section~\ref{sec:overview}). Second, we illustrate how specifications
%in terms of histories are generalized in FCSL to working with
%arbitrary PCMs. The generality will allow us to compose histories with
%other specification-level structures in the future examples. Third, we
%illustrate how structures in FCSL can be composed into larger ones,
%and the inference rules that support reasoning about composed
%structures.

%\subsection{Basic connectives}
%\label{sec:basic-definitions}
FCSL is a Hoare logic, generalizing CSL, hence its assertions are
predicates on state. But unlike in CSL where state is a heap, in FCSL
state may consist of a number of labeled components, each of which may
represent state by a different type. If the type used by some label is
non-heap, then that label encodes auxiliary state, used for logical
specification, but erased at run time. For example, histories are an
auxiliary state identified by the label $\hps$ in the atomic snapshot
example. If we had a program which used two different atomic snapshot
structures, we may label these by $\hps_1$ and $\hps_2$, \etc.

\subsection{Subjectivity}
The state recorded in labels is further divided across another
orthogonal axis -- ownership. Each label identifies three different
chunks of state: self, joint and other portion. The self portion is
private to the specified thread, and can't be accessed by the other
threads. Dually, other is private to the environment threads, and
can't be accessed by the one being specified. Finally, the joint
section is shared and can be accessed by everyone. The self and other
portions of any given label have to belong to a common PCM, and are
often combined together by means of the $\join$ operation of that
PCM. Of course, different labels can use different PCMs.

The FCSL assertions reflect the division across these axes. We have
already illustrated the assertions $\hlabel\,{\spts}\,v$,
$\hlabel\,{\jpts}\,v$ and $\hlabel\,{\opts}\,v$, which identify the
self/joint/other component stored in the label $\hps$ of the
state. These three basic assertions can be combined by the usual
propositional connectives, such as $\aand$ and $\oor$, as we have
already shown in Section~\ref{sec:overview}. FCSL further provides two
connectives that generalize the \emph{separating conjunction} $\lsep$
from separation logic, along the two axes of state splitting. We next
illustrate the \emph{subjective separating conjunction} $\ssep$, and
defer the discussion of the \emph{resource separating conjunction}
$\lsep$ until additional technical material has been introduced. The
formal definitions of all the connectives can be found in %
\ifext{Appendix~\ref{sec:broccoli}.}{the Appendix of the long version
  of the paper.}

The subjective conjunction $\sstar$ is used to model the division of
state between concurrent threads upon forking and joining. In
particular, the parallel composition rule of FCSL is:
\[
\tag{\arabic{tags}}\refstepcounter{tags}\label{eq:parcom}
\begin{array}{c}
\stconc{p_1}{c_1}{q_1}{\ucon} \qquad \stconc{p_2}{c_2}{q_2}{\ucon}\\\hline
\stconc{p_1 \ssep p_2}{c_1 \parallel c_2}{q_1 \ssep q_2}{\ucon}
\end{array}
\]
Ignoring $\ucon$ and the result types of $c_1$ and $c_2$ for now, we
describe how $\ssep$ works. In this rule, it splits the pre-state of
$c_1 \parallel c_2$ into two parts, satisfying $p_1$ and $p_2$
respectively. The parts contain the same labels, and equal joint
portions, but the self and other portions are recombined to match the
thread-relative views of $c_1$ and $c_2$. 
Concretely, in the case of
%assertions with 
one label $\hlabel$, with a PCM $\mathbb U$ and values $a, b, c \in
\mathbb U$, we have the following illustrative implication.
\[
\tag{\arabic{tags}}\refstepcounter{tags}\label{sep-star-impl}
\hspace{-2mm}
{\small
\!\!\!\begin{array}{l}
\hlabel \spts a \join b \aand %\hlabel \jpts h \aand 
  \hlabel \opts c \implies 
  (\hlabel \spts a \aand %\hlabel \jpts h \aand 
   \hlabel \opts b \join c) \ssep 
  (\hlabel \spts b \aand %\hlabel \jpts h \aand 
   \hlabel \opts a \join c)
\end{array}\!\!\!}
\]
Thus, if before the fork, the self-state of the parent thread
contained $a \join b$, and the other-state contained $c$, then after
the fork, the children will have self-states $a$ and $b$, and the
other-states $b \join c$ and $a \join c$, respectively.
In the opposite direction:
\[\tag{\arabic{tags}}\refstepcounter{tags}\label{sep-star-inverse}
{\small
\begin{array}{l}
(\hlabel \spts a \aand \hlabel \opts c_1) \ssep
(\hlabel \spts b \aand \hlabel \opts c_2) \implies \hbox{}\\
\quad \exists c\ldot c_1 = b \join c \aand c_2 = a \join c \aand
\hlabel \spts a \join b \aand \hlabel \opts c
\end{array}
}\] 
That is, if the state can be subjectively split between two child
threads so that their other-views are $c_1$, $c_2$ (with self-views
$a$, $b$), then there exists a common $c$---the other-view of the
parent thread---such that $c_1 = b \join c$ and $c_2 = a \join c$.  In
this sense, the rule for parallel composition models the important
effect that upon a split, $c_1$ becomes an environment thread for
$c_2$, and vice-versa.

There are a few further equations that illustrate the interaction
between the different assertions. First, every label contains all
three of the self/joint/other components. Thus:
\[\tag{\arabic{tags}}\refstepcounter{tags}\label{spts-opts}
{\small
 \hlabel \spts a \iff \hlabel \spts a \aand \hlabel \jpts - \aand \hlabel \opts -
}\]
and similarly for $\hlabel \jpts a$ and $\hlabel \opts a$. Also:
%
%We commonly encounter assertions of the form $\hlabel \spts a \aand
%\hlabel \opts -$, where the other-value is existentially
%abstracted. These can be simplified into $\hlabel \spts a$, and in
%those cases, (\ref{sep-star-impl} and \ref{sep-star-inverse}) reduce
%into a bi-implication:
%We commonly encounter cases when $c$ is existentially abstracted (and
%hence the conjunct $\hlabel \opts c$ is omitted), in which case, we
%use the form:
\[
\tag{\arabic{tags}}\refstepcounter{tags}\label{biimpl}
{\small
%\begin{array}{r@{\ }c@{\ }l}
%\hlabel \spts a \join b \aand \hlabel \opts c & \implies & (\hlabel \spts a
%\aand \hlabel \opts c \join b) \ssep (\hlabel \spts b \aand \hlabel \opts c \join a)\\
\hlabel \spts a \join b \iff \hlabel \spts a \ssep \hlabel \spts b
%\end{array}
}\]
which is provable from (\ref{sep-star-impl}), (\ref{sep-star-inverse})
and (\ref{spts-opts}).

%Both (\ref{sep-star-impl}) and (\ref{biimpl}) generalize in the
%obvious way to $\sstar$-separated assertions with more than one label.

FCSL also provides a \emph{frame rule}, obtained as a special case of
parallel composition when $c_2$ is the idle thread, and $p_2 = q_2 =
r$ is a stable predicate, as usual in fine-grained
logics~\cite{Feng:POPL09,Vafeiadis:PhD,DinsdaleYoung-al:ECOOP10}.
\[
\tag{\arabic{tags}}\refstepcounter{tags}\label{eq:frame}
\begin{array}[m]{c}
  \stconc{p}{c}{q}{\ucon}\\\hline
  \stconc{p \ssep r}{c}{q \ssep r}{\ucon}
\end{array} \quad \mbox{$r$ stable under $\ucon$}
\]

We illustrate the frame rule by deriving from the \code{readPair}
spec~(\ref{eq:pair-spec-read}) a relaxed spec which allows
\code{readPair} to apply when the calling thread has non-trivial self
history $\histS$:
\[
\tag{\arabic{tags}}\refstepcounter{tags}\label{eq:pair-spec-read-framed}
{\small
\hspace{-3mm}
\begin{array}{r@{\ }c@{\ }l}
\spec{~ \histso{\hps}{\histS, -, \hist} ~} 
&\Cmd{readPair}{} & 
\spec{\!\!\!\!\begin{array}{l}
  \exists \histO\ t\ldot
    \histso{\hps}{\histS, \histO, \hist} \aand \tau \le t \aand \hbox{}\\
    \quad \inat{\angled{\res.1, \res.2, -}}{(\histS \hunion \histO)}{t}
  \end{array}
\!\!\!\!}
\end{array}
}
\]
Note that (\ref{eq:pair-spec-read-framed}), when compared to
(\ref{eq:pair-spec-read}), changes the self component from $\hempty$
to $\histS$, but also ${\histO}[t]$ changes into ${(\histS \hunion
  \histO)}[t]$. The latter accounts for the possibility that the
returned snapshot may have been recorded in $\histS$ as a consequence
of the thread itself changing $x$ or $y$, immediately before invoking
\code{readPair}.

The spec (\ref{eq:pair-spec-read-framed}) derives from
(\ref{eq:pair-spec-read}) by framing with the predicate $r = \hps
\spts \histS$. $r$ is trivially stable, as it describes self-state,
which is inaccessible to the interfering threads. We only show how to
weaken the framed postcondition of (\ref{eq:pair-spec-read}) to the
postcondition in (\ref{eq:pair-spec-read-framed}); the preconditions
can be strengthened similarly. Abbreviating $\hist \pre \histO \aand
\hist \le t \aand \inat{\angled{\res.1, \res.2, -}}{\histO}{t}$ by
$P(\histO)$, which is a label-free (\ie pure) assertion, and thus
commutes with $\ssep$, we get:
%\[
%{\small
%\begin{array}{l}
%(\hps \spts \hempty \aand \hps \opts \histO \aand \hist \pre \histO \aand \hist \le t \aand 
% \inat{X}{\histO}{t}) \ssep (\hps \spts \histS) \implies \hbox{}\\
%\quad\mbox{by (\ref{spts-opts}) and commuting out label-free assertions}\\
%(\hps \spts \hempty \aand \hps \opts \histO) \ssep (\hps \spts \histS \aand \hps \opts -) \aand \hbox{}\\
%\hspace{3cm}\hist \pre \histO \aand \hist \le t \aand  \inat{X}{\histO}{t} \implies \mbox{by (\ref{sep-star-inverse})}\\
%\exists \histO'\ldot \histO = \histS \hunion \histO' \aand 
%\hps \spts \histS \aand \hps \opts \histO' \aand \hist \pre \histO \aand \hist \le t \aand \inat{X}{\histO}{t} \implies \hbox{}\\
%\quad\mbox{by substituting $\histO$}\\
%\exists \histO'\ldot \histso{\hps}{\histS, \histO', \hist} \aand \hist \le t \aand \inat{X}{(\histS \hunion \histO')}{t}.
%\end{array}
%}
%\]
\[
{\small
\begin{array}{l}
(\hps \spts \hempty \aand \hps \opts \histO \aand P(\histO)) \ssep (\hps \spts \histS) \implies \mbox{by (\ref{spts-opts}) and $P$-pure}\\
(\hps \spts \hempty \aand \hps \opts \histO) \ssep (\hps \spts \histS \aand \hps \opts -) \aand P(\histO) \implies \mbox{by (\ref{sep-star-inverse})}\\
\exists \histO'\ldot \histO = \histS \hunion \histO' \aand 
\hps \spts \histS \aand \hps \opts \histO' \aand P(\histO) \implies \mbox{by substituting $\histO$}\\
\exists \histO'\ldot \histso{\hps}{\histS, \histO', \hist} \aand \hist \le t \aand \inat{\angled{\res.1, \res.2, -}}{(\histS \hunion \histO')}{t}.
\end{array}}
\]
Intuitively, the frame history $\histS$ is ``subtracted'' from the
other-history $\histO$ of (\ref{eq:pair-spec-read}), and moved to the
self-history in (\ref{eq:pair-spec-read-framed}). This illustrates one
important difference between the frame rule of FCSL and that of
CSL. In FCSL, the frame is always subtracted from the other component,
whereas in CSL the frame simply materializes out of nowhere. On the
flip side, CSL doesn't consider the other component, and can't easily
express a spec such as (\ref{eq:pair-spec-read}).
%At the same time, the result $\histO'$ of the subtraction is
%existentially quantified.

\subsection{Concurroids}
\label{sec:concurroids}
We now turn to the component $\ucon$ of the FCSL specs, which is
called \emph{concurroid}. Concurroids are responsible for enforcing
the invariants on the evolution of the state. For example, the
properties $(i)${--}$(iii)$ in Section~\ref{sec:overview} will be
enforced by defining an appropriate concurroid to govern the
pair-snapshot structure. Thus, concurroids formally represent
concurrent data structures, over which the programs operate.

A concurroid is (a form of) a state transition system (STS). It's a
quadruple $\ucon = ({L}, W, I, {E})$ where: (1) $L$ is a set of
labels, identifying different data structures; (2) $W$ is a set of
admissible states (alternatively, an FCSL assertion); (3) $I$ is the
set of \emph{internal transitions} on $W$; (4) $E$ is a set of pairs
$(\alpha, \rho)$, where $\alpha$ is a \emph{heap-acquiring} and $\rho$
is a \emph{heap-releasing} transition, collectively called
\emph{external} transitions. The internal transitions are relations on
states, describing how a state of the STS evolves in one atomic
step. The external transitions serve for transfer of state
ownership. The concurroids thus bound the moves of the concurrent
programs that operate on a data structure, and therefore represent a
structured form of rely/guarantee transitions from Rely/Guarantee
logics~\cite{Feng-al:ESOP07,Vafeiadis:PhD,Jones:IFIP83,Feng:POPL09,Vafeiadis-Parkinson:CONCUR07}. We
next illustrate concurroids by example.

\paragraph{Pair-snapshot concurroid.}
Given a label $\hps$, pointers $x$, $y$, and the type $A$ of the
accessible contents of $x$ and $y$, the concurroid for the
pair-snapshot structure is ${\pscon} = (\{\hps\}, W_{\pscon}, \{wr_x,
wr_y, \id\}, \emptyset)$.
The set of states $W_{\pscon}$ is described below. We assume that
$\histS, \histO$ are histories, $c_x, c_y\,{:}\,A$ and $v_x,
v_y\,{:}\,\mathsf{nat}$, and are implicitly existentially quantified.
\vspace{-5pt}
\[
{\small
\begin{array}{l@{\ }c@{\ }l}
W_{\pscon} & \eqdef & \hps \spts \histS \aand \hps \jpts (x \hpts (c_x, v_x) \hunion y \hpts (c_y, v_y)) \aand \hps \opts \histO \aand \hbox{}\\
    & & \mbox{$\histS$, $\histO$ satisfy $(ii)-(iii)$}, \mbox{$\histS \hunion \histO$ is continuous, and}\\
    & & \mbox{if $t = \last{\histS \hunion \histO}$, then $(\histS \hunion \histO)[t] = (c_x, c_y, v_x)$}
\end{array}
}
\]
A state in $W_{\pscon}$ consists of the auxiliary part, which are
histories in the self and other components, and concrete part, which
is a joint heap, storing pointers $x$ and $y$, with accessible
contents $c_x, c_y$, and version numbers $v_x, v_y$,
respectively.\footnote{Notice the overloading of the $\hpts$ notation
  for singleton heaps and histories.}
It requires several additional properties of the auxiliary
histories. First, the combined history $\histS \hunion \histO$ is
continuous; that is, adjacent timestamps have matching
states. Second, the last timestamp in $\histS \hunion \histO$
correctly reflects what's stored in $x$ and $y$. Finally, $W_{\pscon}$
also bakes in the properties $(ii)-(iii)$ required in the proof
outline of \code{readPair}.

The internal transitions $wr_x$ and $wr_y$ synchronize the changes to
$x$ and $y$ with histories. In both transitions, $\tfr{\histS \hunion
  \histO}$ is the smallest timestamp unused by $\histS$ and $\histO$.
\vspace{-5pt}
\[ 
\tag{\arabic{tags}}\refstepcounter{tags}\label{eq:pair-trans}
{\small 
  \begin{array}{l@{\ }c@{\ }l@{\ }c}
    wr_x & \eqdef &\hps \jpts (x \hpts (c_x, v_x) \hunion y
    \hpts (c_y, v_y))  \aand 
    \hps \spts \histL & \rightsquigarrow \\
    & &
    \hps \jpts (x \hpts (c'_x, v_x + 1) \hunion y \hpts (c_y, v_y) ~\aand \\
    && \hps \spts \histL \hunion \tfr{\histS \hunion \histO} \hpts (\angled{c_x, c_y, v_x},\angled{c'_x, c_y, v_x+1})
    \\
    wr_y & \eqdef & \hps \jpts (x \hpts (c_x, v_x) \hunion y
    \hpts (c_y, v_y)) \aand \hps \spts \histL 
    & {\rightsquigarrow} \\
    && \hps \jpts (x \hpts (c_x, v_x) \hunion y \hpts (c'_y, v_y
    +1) ~\aand \\ 
    && \hps \spts \histL \hunion \tfr{\histS \hunion \histO} \hpts (\angled{c_x, c_y, v_x},\angled{c_x, c'_y, v_x})
    \end{array}
}\] 
The first conjunct after $\rightsquigarrow$ in $wr_x$ (and $wr_y$ is
similar) allows that the version number of $x$ can only increase by 1
in an atomic step. The second conjunct shows that simultaneously with
the change of $x$, the snapshot of the changed state is committed to
the self-history of the invoking thread.
Together, $wr_x$ and $wr_y$ ensure that histories only grow, and only
by adding valid snapshots; \ie, precisely the property $(i)$ from
Section~\ref{sec:overview}.

$\ucon$ also contains the identity transition $\id$, whose presence
enables programs that don't modify the state at all. In the
pair-snapshot example, these are the \code{readX} and \code{readY}
actions, and the \code{readPair} method.  The pair-snapshot example
doesn't involve ownership transfer, so $\pscon$ has no external
transitions, but these will be important in the forthcoming examples.

\paragraph{Entanglement and private heaps.} 
Larger concurroids may be constructed out of smaller ones. A
particularly common construction is
\emph{entanglement}~\cite{Nanevski-al:ESOP14}. Given concurroids
$\ucon$ and $\vcon$, the entanglement $\ucon \entangle \vcon$ is a
concurroid whose state space is the Cartesian product $W_{\ucon}
\times W_{\vcon}$, and the transitions allow the $\ucon$ portion to
perform a $\ucon$ transition, while the $\vcon$ portion remains idle,
and vice-versa. Additionally, $\ucon$ and $\vcon$ portions can
communicate to \emph{transfer a heap} between themselves, by having
one take a heap-acquiring, and the other \emph{simultaneously} taking
a heap-releasing transition.

The most common is the entanglement with the concurroid $\privcon$ of
\emph{private heaps} (see
\ifext{Appendix~\ref{sec:conc-priv-heaps}}{the Appendix of the
  extended version}). Entangling with $\privcon$ lets the
concurroids temporarily move heaps to a private section, via the
communication discussed above, where threads may then perform the
customary operations of reading, writing, allocating, and deallocating
pointers, without interference.\footnote{Our Coq proofs actually use
  two different concurroids, one for reading/writing, another for
  allocation/deallocation, which we entangle to provide all four
  operations. For simplicity, here we assume a monolithic
  implementation.}
$\privcon$ comes with a dedicated label $\hpriv$. As an illustration,
the following assertion may describe one possible state in the state
space of the entanglement $\privcon \entangle \pscon$ with the
snapshot concurroid.
%
%\vspace{-5pt}
%
\[
{\small
\begin{array}{c}
\hpriv \spts (z \hpts 0) \lsep \hps \jpts (x \hpts (c_x, v_x) \hunion
y \hpts (c_y, v_y))
\end{array}
}
\]
The $\hps \jpts -$ portion describes the part of the state coming from
$\pscon$, which is joint, containing pointers $x$ and $y$, as
explained before. The $\hpriv \spts (z \hpts 0)$ describes the part of
the state coming from $\privcon$. In this particular case, it contains
a heap with a single pointer $z$. The heap is private, \ie, owned by
the self thread, so $z$ can't be modified by other threads.
Notice that the assertions about $\hpriv$ and $\hps$ are separated by
the resource separating conjunction $\lsep$, which splits the state
into portions with disjoint labels and heaps. In this particular case,
it signifies that the labels $\hpriv$ and $\hps$ are distinct, as are
the pointers $z$, $x$ and $y$.

%
%Moreover, heaps can be transfered to $\privcon$ from the other
%components of the entanglement, thus becoming private, or out of
%$\privcon$, when the private status of the heap is not desired
%anymore.

\newcommand{\inject}[1]{[#1]}

\subsection{Extending and hiding concurroids}
Concurroids represent concurrent data structures; thus it's important
to be able to introduce and eliminate them. FCSL provides two
programming constructors (both no-ops operationally), and
corresponding inference rules for that purpose. For completeness, we
introduce them here, but postpone the illustration until
Section~\ref{sec:examples}.

The injection rule shows that if a program is proved correct with
respect to a smaller concurroid $\ucon$, then it can be extended to
$\ucon \entangle \vcon$, without invalidating the proof.
\[
\tag{\arabic{tags}}\refstepcounter{tags}\label{eq:inject}
\begin{array}{c}
 \stconc{p}{c}{q}{\ucon}\\\hline
 \stconc{p \lsep r}{\inject c}{q \lsep r}{{\ucon} \entangle {\vcon}}
\end{array} \quad \mbox{$r \subseteq W_{\vcon}$ stable under $\vcon$}
\]
This is a form of framing rule, along the axis of adding new
resources. The operator $\lsep$ splits the state into portions with
disjoint labels, and the side-condition that $r \subseteq W_{\vcon}$
forces $r$ to remove the labels of the concurroid $\vcon$, so that $c$
is verified \wrt~the labels of $\ucon$. The program constructor
$\inject -$ is a coercion from $\ucon$ to ${\ucon} \entangle {\vcon}$.

Hiding is the ability to introduce a concurroid $\vcon$, \ie, install
it in a private heap, for the scope of a thread $c$. The children
forked by $c$ can interfere on $\vcon$'s state, respecting $\vcon$'s
transitions, but $\vcon$ is hidden from the environment of $c$. To the
environment, $\vcon$'s state changes look like changes of the private
heap of $c$. Upon termination of $c$, $\vcon$ is deinstalled.
{\small
\begin{gather*}
\begin{array}{c}
{\stconc{\hpriv\spts h \lsep p}{c}{\hpriv\spts h' \lsep q}{(\privcon
    \entangle {\ucon}) \entangle {\vcon}}} 
\\ \hline
{\stconc{\Psi\ g\ h \lsep (\Phi\,(g) \wand p)}{\mathsf{hide}_{\Phi, g}\ c}{\exists g'. \Psi\ g'\ h' \lsep (\Phi\,(g') \wand q)}{\privcon \entangle \ucon}}
\end{array}
\\
\tag{\normalsize \arabic{tags}}\refstepcounter{tags}\label{eq:hide-rule}
\mbox{where}\ \Psi\ g\ h = \exists k{:}\mathsf{heap}.\, \hpriv \spts {h \hunion k} \aand \Phi\,(g)\ \mbox{erases to}\ k
\end{gather*}
}
%\[
%\begin{array}[t]{c}
%  {\stconc{\hpriv\spts h_1 * r * \Phi\ g_1}{C}{\hpriv\spts h_2 * q * \Phi\ g_2}{P \entangle (U \apart V)}}\\\hline
%  {\stconc{\hpriv \spts {h_1 \hunion x_1} * r \aand \Phi g_1 \downarrow x_1}{\mathsf{hide}_{\Phi} C}{\hpriv \spts {h_2 \hunion x_2} * q \aand \Phi g_2 \downarrow x_2}{P \entangle U}}
%\end{array}
%\]
%\an{Here's where the point-free notation may look attractive. We can
%  define $\Psi$ simply as $\mathsf{this}\ h \lsep_\pi \Phi\
%  g\downarrow$ and save one existential.}  
%
\hspace{-8pt}
Since installing $\vcon$ consumes a chunk of private heap, the rule
requires the overall concurroid to support private heaps, \ie, to be
an entanglement of $\privcon$ with an arbitrary $\ucon$. In programs,
we use the coercion $\mathsf{hide}\ c$ to indicate the change from
$(\privcon \entangle {\ucon}) \entangle {\vcon}$ to $\privcon
\entangle \ucon$. If $\ucon$ is of no interest, one can take it to be
the empty concurroid $\econ$, which is a right unit for $\entangle$
(see \ifext{Appendix~\ref{sec:empty-concurroid}}{Appendix}).

The annotation $\Phi$ is a predicate; it describes an invariant that
holds within the scope of $\mathsf{hide}$, parametrized by an
argument. It's subject to a number of conditions (see
\ifext{Appendix~\ref{sec:phi-properties}}{Appendix}). $g$ is the
initial argument, so $\Phi(g)$ holds in the initial state into which
$\vcon$ is placed upon installation. The rule guarantees that the
ending state of $c$ satisfies $\exists g'\ldot \Phi(g')$. The
surrounding connectives $\lsep$ and $\wand$ merely mediate between
$\ucon$, $\vcon$, and the erasure of $\vcon$ to heaps. We explain the
precondition, and the postcondition is similar.

In the precondition, $*$ separates private heaps from $\ucon$, and
$\Psi$ requires that every state in $\Phi(g)$ obtains the same private
heap when the auxiliary fields are erased. $\wand$ is inherited from
separation logic. $\Phi(g) \wand p$ says that if the initial state
(which is in $W_{\ucon}$) is extended with a state from $\Phi(g)$
(which is in $W_{\vcon}$), then the result is a state satisfying
$p$. In other words, if a state satisfying $\Phi(g)$ is installed in
the initial state of $c$, while its heap footprint is removed from the
private heaps, then $c$'s precondition is satisfied.

\section{Treiber stack and its client}
\label{sec:examples}

In this section we illustrate how histories can be used to specify and
verify the fine-grained data structure of Treiber
stack~\cite{Treiber:TR}. We also show how the specs can be used by
clients, where they provide an abstraction that facilitates client
reasoning as if the structure were coarse-grained.
\begin{figure}
\centering
\small
\begin{tabular}{l@{}l}
\begin{minipage}[l]{3.9cm}
\begin{alltt}
\num{1} push(e : \(A\)): \Unit \{
\num{2}  p <- \act{alloc}();
\num{3}  \textbf{fix} loop() \{
\num{4}   p1 <- \act{readSentinel}();
\num{5}   \act{write}(p, (e, p1));
\num{6}   ok <- \act{tryPush}(p1, p);
\num{7}   \textbf{if} ok \textbf{then return} ();
\num{8}   \textbf{else} loop();\}();
\num{9} \}

\end{alltt} 
\end{minipage}
& 
\begin{minipage}[l]{4.2cm}
\begin{alltt}
\num{ 1} pop(): \act{option} \(A\) \{
\num{ 2}  p <- \act{readSentinel}();
\num{ 3}  \textbf{if} p == null 
\num{ 4}  \textbf{then return} \act{None};
\num{ 5}  \textbf{else} \{
\num{ 6}   (e,p1) <- \act{readNode}(p);
\num{ 7}   ok     <- \act{tryPop}(p,p1);
\num{ 8}   \textbf{if} ok 
\num{ 9}   \textbf{then return} \act{Some} e;
\num{10}   \textbf{else return} pop();\}\}
\end{alltt}
\end{minipage}
\end{tabular} 
\caption{Code of Treiber stack procedures.}
\label{fig:treiber-code}
\end{figure}

%\subsection{Specifying and verifying the Treiber stack}
%\label{sec:treib-stack-proc}
%
The Treiber stack works as follows. Physically, the stack is kept as a
singly-linked list in the heap, with a sentinel pointer $\sent$
pointing to the stack top $\esc{p1}$. \code{push(e)} allocates a node
$\esc{p}$ that's supposed to go to the top of stack, and attempts to
link the node into the stack, by changing the sentinel to
$\esc{p}$. Clearly, this operation shouldn't succeed if some
interfering thread has in the meantime changed the top by pushing or
popping elements. Thus \code{push} applies a CAS read-modify-write
operation~\cite{Herlihy-Shavit:08}, which atomically reads $\sent$,
compares its contents with $\esc{p1}$, and if the two are equal (\ie,
if the stack's top hasn't changed), writes $\esc{p}$ into $\sent$,
thus en-linking the new top. Otherwise, \code{push} is restarted.

$\code{pop()}$ behaves similarly. It reads the first node $\code{p}$, pointed
to by $\sent$, and obtains its value $\esc{e}$ and pointer $\esc{p1}$
to the next node. Then it tries to de-link $\code{p}$, by changing the
sentinel to $\esc{p1}$ using a CAS to identify interference. Note that
\code{pop} doesn't deallocate the de-linked node $\esc{p}$, which thus
remains in the data structure as garbage. This is by design, to
prevent the ABA problem~\cite[\S10]{Herlihy-Shavit:08}: if
$\esc{p}$ is deallocated, then some other \code{push} may allocate it
again, and place it back on top of the stack. A procedure that
observed $\esc{p}$ on top of the stack, but hasn't performed its CAS
yet may thus be fooled as follows. Its CAS may encounter $\esc{p}$
on top of the stack, and proceed as if the stack hadn't changed,
producing invalid results.

The described code of the Treiber stack operations is given in
Figure~\ref{fig:treiber-code}, where we used descriptive names for the
atomic operations. Instead of CAS, we used \code{tryPush} and
\code{tryPop}, and instead of pointer \code{read}, we used
\code{readSentinel} and \code{readNode}. The reason for the
descriptive names is that the atomic operations in FCSL operate not
only on concrete heap pointers, but on auxiliary state as well. In the
particular case of Treiber, the auxiliary state will be histories,
which \code{tryPush} and \code{tryPop} change in different ways, even
though they both operationally perform a CAS. Similarly,
\code{readSentinel} and \code{readNode} deduce different facts about
the histories, even though they both simply read from a pointer.

We elide here any further discussion on how the atomic operations are
specified and verified in FCSL (it can be found
in~\cite{Nanevski-al:ESOP14} and
\ifext{Appendix~\ref{sec:appactions}}{the Appendix}). Instead,
whenever needed, we simply state the Hoare specs for the atomics and
proceed to use them in proof outlines, as if the atomics were ordinary
procedures. Of course, our Coq files contain proofs that all such
Hoare triples are valid.

\newcommand{\tbinv}{I}

\paragraph{Treiber concurroid.}
Given a label $\htb$, the sentinel pointer $\sent$, and the type $A$
of the stack elements, the state space of the Treiber concurroid
$\tbcon$ is described as follows. Its auxiliary self/other components
are histories $\histS$ and $\histO$ that store mathematical sequences
$l$ corresponding to the logical contents of the stack at various
timestamps. The joint component contains a heap $h_s$ storing a
sentinel $\sent$ pointing to a linked list, a heap~$h$ implementing
the list, and a garbage section $\garb$ of de-linked nodes.
{\small
\hspace{-11pt}
\begin{align*}
W_{\tbcon} & \eqdef \exists \histS~\histO~h_s\ldot \htb \spts \histS \aand
\htb \opts \histO \aand \htb \jpts h_s \aand \tbinv~(\histS \hunion
\histO)~h_s\\[5pt]
\tag{\normalsize \arabic{tags}}\refstepcounter{tags}\label{eq:tb-states}
\tbinv~\hist~h_s & \eqdef
\exists p~h~\garb~l\ldot h_s = (\sent \hpts p) \hunion h \hunion \garb
\aand \llist(p, l, h) \aand\hbox{}\\[-2pt]
& ~~\complete(\hist) \aand \mathsf{continuous}(\hist) \aand \cont(\hist) \aand \inat{l}{\hist}{\last{\hist}}
\end{align*}
}
The auxiliary predicates are:
%
%\vspace{-10pt}
%
\[
%\tag{\arabic{tags}}\refstepcounter{tags}\label{eq:tb-preds}
\hspace{-5pt}
{\small
\begin{array}{r@{\ }c@{\ }l}
\llist(p, l, h) & \eqdef & 
p = \Null \aand l = \nil \aand h = \hempty ~\oor \\
& & \exists e\ p'\ l'\ h'\ldot l = e\!::\!l' \aand h = p\!\hpts\!(e, p') \hunion h' \aand
\llist(p',l',h')\\[5pt]
\complete(\hist) & \eqdef &
\exists l_0\ldot \hist(0) = (l_0, l_0) \aand
\forall t\ldot t < |\dom{\hist}| \Rightarrow t \in \dom{\hist}
\\[5pt]
\cont(\hist) & \eqdef &
\forall t \in \dom{\hist}\ldot t > 0 ~\Rightarrow \exists l~e\ldot \hist(t) = (l,e\!::\!l) \oor \hist(t)
= (e\!::\!l,e)
\end{array}
}
\]
In particular: (1) the overall history $\histS \hunion \histO$ is
complete, \ie no gaps exist between timestamps; (2) aside from the
initialization in timestamp $0$, the history only stores events
corresponding to pushing or popping, and (3) the last recorded state
in the history captures the current contents of the stack. For
simplicity, we disable reasoning about the structure's inherent memory
leak by not relating histories to $\garb$ in \eqref{eq:tb-states}.

The transitions of $\tbcon$ allow for popping and pushing only. 
\[ 
{ 
\small
\hspace{-3pt}
  \begin{array}{r@{\ }c@{\ }l@{\ }c}
    \taupop & \eqdef &\htb \jpts \sent \hpts p \hunion h \hunion
    \garb \aand \htb \spts \histS  ~\aand \\
    & & h = (p \hpts (e, p')\hunion h') \aand \llist(p, (e::l), h) &
    \rightsquigarrow 
    \\
    & &\htb \jpts \sent \hpts p' \hunion h' \hunion (p \hpts (e,
    p') \hunion \garb) \aand \\
    & & \htb \spts \histS \hunion \tfr{\histS \hunion \histO} \hpts
    (e::l, l) 
    \\
    \alphapush_{p', e, p} & \eqdef &\htb \jpts \sent \hpts p \hunion h \hunion
    \garb \aand \htb \spts \histS \aand \llist(p, l, h) &  \rightsquigarrow 
    \\
    & & \htb \jpts \sent \hpts p' \hunion (p' \hpts (e, p) \hunion h) \hunion
    \garb ~ \aand \\
    & & \htb \spts \histS \hunion \tfr{\histS \hunion \histO} \hpts
    (l, e::l) 
    \end{array}
}\]
In $\taupop$, the sentinel pointer is swapped from used-to-be head $p$
to its next one, $p'$, whereas $(p \hpts -)$ logically joins the
garbage. The transition $\alphapush$ describes how a heap of the shape
$p' \hpts (e, p)$, describing the node to be pushed, is acquired and
placed at the top of the stack. It's an external transition, which
means it only fires when entangled with a concurroid from which the
heap $p' \hpts (e, p)$ can be taken away. In our case, that will be
the concurroid $\privcon$ for private state. Importantly, $\tbcon$
doesn't have a release transition; once a memory chunk is in the joint
state, it never leaves, capturing that $\tbcon$ doesn't allow
deallocation.  
%
%Another important stability property, derivable from the above
%transitions, is that no pointer except $\sent$ in the joint part of
%$\tbcon$ ever changes its value.

\paragraph{Method specs.}
We give the following history-based specs.
{\small
\begin{gather*}
\hspace{-5pt}
\begin{array}{r@{\ }c@{\ }l}
\spec{\!\!\!\!
  \begin{array}{c}
    \hpriv \spts \hempty ~\sep \\
    \histso{\htb}{\hempty, -, \hist}
  \end{array}
\!\!\!\!} &\esc{push}(e)& 
\spec{\!\!\!
  \begin{array}{c}
    \exists t~l\ldot \hpriv \spts \hempty ~\sep\\
   \histso{\htb}{t \!\hpts \! (l,e\!::\!l), -, \hist} \aand \hist < t
  \end{array}
\!\!\!}@\privcon \entangle \tbcon
\end{array}
\\[-5pt] 
\tag{\normalsize \arabic{tags}}\refstepcounter{tags}\label{eq:stack-spec}
\\[-5pt]
\begin{array}{c}
\spec{\!\!\!\!
  \begin{array}{l}
    \histso{\htb}{\hempty, -, \hist}
  \end{array}
\!\!\!\!} \\
\esc{pop}()
\\
\spec{\!\!\!
  \begin{array}{c}
    \exists e~t~l\ldot \res = \Some e \aand 
    \histso{\htb}{t \hpts (e\!::\!l,l), -, \hist} \aand
    \hist < t \oor \hbox{}\\
    \exists \histO~t\ldot \res = \None \aand \histso{\htb}{\hempty, \histO, \hist} \aand \tani{\nil}{\histO}{t}\\
  \end{array}
\!\!\!}@\tbcon
\end{array}
\end{gather*}
}
\hspace{-5pt}
\code{push} runs with empty private heap and history, thus by framing,
it can run with any private heap and history. After termination, the
self history is incremented by a singleton exposing that a push event
has been executed at a time stamp $t$; $\hist < t$ indicates that the
push event appeared strictly after the events preceding the call. The
spec for \code{pop} is slightly more complicated as \code{pop} checks
for stack emptiness, but ultimately proceeds in the similar
manner. \code{push} works over the entangled concurroid $\privcon
\entangle\tbcon$, as it needs to allocate memory; \code{pop} works
over $\tbcon$ only, as it doesn't deallocate.

In Figure~\ref{fig:push-proof} we present the proof outline for
\code{push}.\footnote{The proof for \code{pop} can be found in the Coq
  files.} It's mostly self-explanatory, so we only
point out a few technicalities.  First, the atomic actions
\code{alloc} and \code{write} are specific to the $\privcon$
concurroid and have the following specs.
\[
{\small
\tag{\arabic{tags}}\refstepcounter{tags}\label{eq:alloc-spec}
\begin{array}{r@{\ }c@{\ }l}
\sspec{~
    \hpriv \spts \hempty
~} &\act{alloc}()& 
\sspec{~
    \hpriv \spts  \res \hpts -
~}@\privcon
\\[2pt]
\sspec{~\hpriv \spts x \hpts - ~}
&
\act{write}(x, e)
&
\sspec{~\hpriv \spts x \hpts e~}@\privcon
\end{array}
}
\]
Thus, in Figure~\ref{fig:push-proof}, they have to be explicitly
injected into $\privcon \entangle \tbcon$, by means of the coercion
$[-]$ introduced in Section~\ref{sec:background}. Similarly for
\code{readSentinel}, whose concurroid is $\tbcon$. Somewhat
surprisingly, the call to \code{readSentinel} in line 6 is irrelevant
for the (partial) correctness of \code{tryPush}; thus line 7 doesn't
say anything about \esc{p1}.\footnote{Though, taking a random \esc{p1}
  here will affect liveness, as \code{push} will keep looping until it
  finds the chosen \esc{p1} at the top of the stack.} The
\code{tryPush} action appears in the proof outline with its precise
specification; that is, line 9 contains its precondition, and 11
contains the postcondition, describing that a successful outcome of
\code{tryPush} removed a heap from $\privcon$, moved it to the joint
heap of $\tbcon$, and updated the history to reflect the move,
following the $\alphapush$ transition.

\begin{figure}
\centering
\[
{\small
\hspace{-5pt}
\begin{array}{r@{\ \ }ll}
  \Num{1}& \sspec{~~\hpriv \spts \hempty ~\sep~ \histso{\htb}{\hempty, -, \hist}~~} & 
  \\
  \Num{2} &  ~\esc{p <- [\act{alloc}()];}~ & 
  \\
  \Num{3} & \sspec{~~\hpriv \spts \esc{p} \hpts - ~\sep~ \histso{\htb}{\hempty, -, \hist}~~}
%  & \text{by \textsc{Seq}, \eqref{eq:alloc-spec} and \textsc{Inject}}
  \\
  \Num{4} &  ~\esc{\textbf{fix} loop() \{}~ & 
  \\
  \Num{5}& \sspec{~~\hpriv \spts \esc{p} \hpts - ~\sep~ \histso{\htb}{\hempty, -, \hist}~~}
%  & \text{loop precondition}
  \\
  \Num{6} &  ~\esc{p1 <- [\act{readSentinel}()];}~ & 
  \\
  \Num{7}& \sspec{~~
     \hpriv \spts \esc{p} \hpts - ~\sep~
     \histso{\htb}{\hempty, -, \hist}
     % ~ \aand \\
     % (~\esc{p1} = \Null \oor \htb \jpts (\esc{p1} \hpts -) \hunion - ~)
  ~~} 
% \text{by \textsc{Act}, \textsc{Seq} and \textsc{Inject}}
  \\
  \Num{8} &  ~\esc{[\act{write}(p, (e, p1))];}~ & 
  \\
  \Num{9}& \sspec{~~
     \hpriv \spts \esc{p} \hpts (\esc{e}, \esc{p1}) ~\sep~
     \histso{\htb}{\hempty, -, \hist} 
     % ~ \aand \\
     % (~\esc{p1} = \Null \oor \htb \jpts (\esc{p1} \hpts -) \hunion - ~)
  ~~} 
%  \text{by~\eqref{eq:write-spec} and \textsc{Inject}}
  \\
  \Num{10} &  ~\esc{ok <- \act{tryPush}(p1, p);}~ & 
  \\
  \Num{11}& \spec{\!\!
    \begin{array}{l}
     \esc{ok} = \True \aand \exists t\ l\ldot \hpriv \spts \hempty ~\sep~
     \histso{\htb}{t \hpts (l,e\!::\!l), -, \hist}\aand 
     \hist < t\\
     \esc{ok} = \False \aand
     \hpriv \spts \esc{p} \hpts (\esc{e}, \esc{p1}) ~\sep~ \histso{\htb}{\hempty, -, \hist}
    \end{array}
  \!\!} 
%  \text{by~\textsc{Act} and \textsc{Inject}}
  \\
  \Num{12} &  ~\esc{\textbf{if} ok \textbf{then return} ();}~ & 
  \\
  \Num{13}&\sspec{~~\exists t\ l\ldot
     \hpriv \spts \hempty  ~\sep~\histso{\htb}{t \hpts (l,e\!::\!l), -, \hist} \aand \hist < t ~~}
%   & \text{by~\textsc{If}}
  \\
  \Num{14} &  ~\esc{\textbf{else}}~ & 
  \\
  \Num{15} &  
  \sspec{~~\hpriv \spts \esc{p} \hpts - ~\sep~ \histso{\htb}{\hempty, -, \hist}~~}& 
  \\
  \Num{16} &  ~\esc{loop();\}();}~ & 
  \\
  \Num{17}&\sspec{~~\exists t\ l\ldot
     \hpriv \spts \hempty  ~\sep~ \histso{\htb}{t \hpts (l,e\!::\!l), -, \hist} \aand \hist < t ~~}
  % & \text{by~\textsc{If}, \textsc{Fix} and \textsc{App}}
%
\end{array}
}
\]
\caption{A proof outline of Treiber's \code{push} method. The proof
  rule for \code{fix} allows assuming the spec of a procedure in the
  proof of the body, and is presented in
  \ifext{Appendix~\ref{sec:rules}.}{the Appendix of the extended
    version.}}
\label{fig:push-proof}
\vspace{-7pt}
\end{figure}

\paragraph{Recovering sequential specifications.}\label{sec:rest-seq}
We next show that the subjective spec~\eqref{eq:stack-spec} is a
generalization of the canonical sequential
spec~\eqref{eq:seqstack-spec}. In particular, if there's no
interference from other threads, \eqref{eq:stack-spec} can be reduced
to~\eqref{eq:seqstack-spec}. The mechanism for achieving the reduction
relies on the self/other dichotomy, thus substantiating our point that
the dichotomy is important for precise reasoning with histories.

To this end, we use the $\mathsf{hide}$ constructor from
Section~\ref{sec:background}. $\mathsf{Hide}$ introduces a concurroid
in a delimited scope, and prohibits the environment threads from
interfering on it. The heap for the introduced concurroid is
appropriated from the private heap. In the case of \code{push}, we
will appropriate a heap storing the sentinel and the linked list of
the stack, install the $\tbcon$ concurroid over this heap, perform
\code{push} with interference disabled, then return the heap back to
private heaps. We will derive the following specification, which is
essentially an elaborated version of~\eqref{eq:seqstack-spec}, modulo
the memory leak inherent to Treiber stack (hence $\garb$ in the postcondition).
\vspace{-3pt}
\[
\tag{\arabic{tags}}\refstepcounter{tags}\label{eq:seq}
{\small
\begin{array}{c}
\sspec{~\exists p~h\ldot \hpriv \spts (\sent \hpts p \hunion h) \aand
  \llist(p, l, h)~} 
\\[3pt]
\mathsf{hide}_{\Phi, \hempty}\ \esc{\{}~\code{push}(e);~\esc{\}}
\\[3pt]
\sspec{~\exists p~h~\garb\ldot \hpriv \spts (\sent \hpts p \hunion h \hunion \garb) \aand \llist(p, e::l, h)~} @\privcon
\end{array}
}
\]
The self/other dichotomy affords explicit access to other-owned
histories, so that we can define the following predicate $\Phi$
stating that other-histories remain empty within the scope of
$\mathsf{hide}$.
\[
\tag{\arabic{tags}}\refstepcounter{tags}\label{eq:phi}
{\small
\hspace{-13pt}
\begin{array}{c}
\Phi(\hist) \eqdef \exists l\ldot \htb \spts ((0 \hpts (l, l)) \hunion
\hist) \aand \htb \opts \hempty \aand W_{\tbcon}
\end{array}
}
\]
Inside hide, the stack is initialized (the history contains the
singleton $0 \hpts (l, l)$), there's no interference ($\htb \opts
\hempty$), and the state is a valid one for $\tbcon$ (\ie, it is
captured by the definition~\eqref{eq:tb-states}).

One can prove that if the histories are erased from any state in
$\Phi(\hist)$, the remaining concrete heap consists of $\sent$ and the
stack. Moreover, the contents of the stack is the last entry of
$\hist$ (or $l$ if $\hist$ is empty). In other words, using
$\Psi$~\eqref{eq:hide-rule}, defined in Section~\ref{sec:background}:
\[
\tag{\arabic{tags}}\refstepcounter{tags}\label{eq:phieq}
{\small
\begin{array}{c}
\Psi~\hist~\hempty \iff \exists p~h\ldot \hpriv \spts (\sent \hpts p \hunion h \hunion -) \aand \llist(p, l', h)  
\end{array}
}
\]
where $l' = \hist[\mathsf{last}(\hist)]$ (or $l' = l$ if $\hist$ is
empty).

The derivation is in Figure~\ref{fig:seq-proof}, and we comment on the
main points. In line~2, the right conjunct uses the property inherent
in $\Psi$, that $\Phi(\hempty)$ erases to the heap storing $l$. Thus,
this is the $l$ that appears in the consequent of $\wand$. In line~7,
the right conjunct implies that the history $\hist$, whose existence
obtains from the rule for hiding~\eqref{eq:hide-rule}, must be the
self-history returned by \code{push}. Hence, it's equal to $0 \hpts
(l, l) \hunion t \hpts (l', e::l')$ for some $t$ and $l'$. But, we
also know that $\hist$ must be complete (no gaps between timestamps)
and continuous. Hence $t = 1$ and $l' = l$ in line~9, which then
derives the postcondition by~\eqref{eq:phieq}.

\paragraph{A stack client.}
We next illustrate how the specs (\ref{eq:stack-spec}) are exploited
by the \emph{concurrent} clients of Treiber stack to abstract from the
fine-grained nature of Treiber's implementation.
%
% and view each effect as if it appeared atomically.
%
The example code in Figure~\ref{fig:client} presents two procedures,
\code{produce} and \code{consume}, that communicate via a common
Treiber stack~$\htb$. \code{produce} pushes onto the stack the
elements of its array \code{ap} in order, whereas \code{consume} pops
from the stack, to fill its array \code{ac}. Both arrays are of equal
size $n$. The procedure \code{exchange} runs \code{produce} and
\code{consume} concurrently. Our goal is to prove that after
\code{exchange} terminates, \code{ap} has been copied to \code{ac},
modulo element permutation. The inference will only use the specs
(\ref{eq:stack-spec}) but not the code of Treiber methods, thus
obtaining a coarse-grained view of effects inherent in the histories.
% and a program \code{exchange} that moves the contents
%of a dedicated array \code{ap} to an array \code{ac}. This is done by
%running \code{produce} in parallel with \code{consume} and making them
%communicating through a {shared} Treiber stack labelled $\htb$ in a
%way that the producer only \code{push}es to the stack, whereas the
%consumer only \code{pop}s from it.
%
\begin{figure}
\centering
\[
{\small
\hspace{-5pt}
\begin{array}{r@{\ \ }l}
  \Num{1}& \sspec{~\exists p~h\ldot \hpriv \spts (\sent \hpts p \hunion h) \aand \llist(p, l, h)~} 
  \\
  \Num{2} & \sspec{~\Psi\ \hempty\ \hempty~\sep~\left(\Phi(\hempty)
      \wand \histso{\htb}{0 \hpts (l, l), -, -} \right)  ~} \ \mbox{by~\eqref{eq:phieq}}

  \\[2.5pt]
  \Num{3} &  ~\mathsf{hide}_{\Phi, \hempty}~\esc{\{} ~  
  \\
  \Num{4} & 
  \sspec{~\hpriv \spts \hempty ~\sep~ 
    \histso{\htb}{0 \hpts (l, l), -, -}  ~} 
  \\[2.5pt]
  \Num{5} &  ~\esc{push}(e); ~  
  \\
  \Num{6} &
  \sspec{~\exists t~l'\ldot \hpriv \spts \hempty ~\sep~ 
    \histso{\htb}{0 \hpts (l, l) \hunion t \hpts (l', e::l'), -, -}
    ~} ~~~~~~~\esc{\}} 
  \\[2.5pt]
  \Num{7} &
  \sspec{~\exists \hist\ldot \Psi~\hist~\hempty~\sep~ 
    (\Phi(\hist) \wand \exists t~l'\ldot \histso{\htb}{0 \!\hpts \! (l, l) \hunion t \!\hpts \! (l', e::l'), -, -}  ~} 
  \\[2.5pt]
  \Num{8} &
  \spec{\!\!\begin{array}{l}
          \exists t~l'~\hist\ldot\hist = 0 \hpts (l, l) \hunion t \hpts (l', e::l') \aand \hbox{}\\
          \mathsf{complete}(\hist) \aand \mathsf{continuous}(\hist) \aand \Psi~\hist~\hempty
         \end{array}\!\!}
       \\[2.5pt]
  \Num{9} &
  \sspec{~\exists \hist\ldot \hist = 0 \hpts (l, l) \hunion 1 \hpts (l, e::l) \aand \Psi~\hist~\hempty}\\
  \Num{10}& \sspec{~\exists p'~h\ldot \hpriv \spts (\sent \hpts p' \hunion h \hunion -) \aand \llist(p', e::l, h)~} \ \mbox{by~\eqref{eq:phieq}}
\end{array}
}
\]
\vspace{-2mm}
\caption{Proof outline for sequential specification for \code{push}.}
\label{fig:seq-proof}
\end{figure}

We use several auxiliary predicates. First, $\Array{a}{n}{l}{h}$
defines an array of size $n$ as a sequence of consecutive pointers in
the heap $h$, starting from pointer $a$, and storing elements of the
list $l$:
\[
\tag{\arabic{tags}}\refstepcounter{tags}\label{eq:array}
{\small
\begin{array}{rcl}
\Array{a}{n}{l}{h} & \eqdef & \size{l} = n \aand h = \Hunion{i < n} (a + i)
\hpts l(i)
\end{array}
}
\]
Next, the predicates $\mathsf{Pushed}$ and $\mathsf{Popped}$ extract
the lists of pushed and popped elements from a stack history $\hist$.
{\small
\begin{align*}
\Pushed{\hist}{l} & \eqdef  l \eqm \mset{e ~|~
  \exists t~l\ldot t \!\hpts\! (l, e::l) \in \hist 
  \oor 0\!\hpts\! (l, l) \in \hist \aand e \in l}
\\
\tag{\normalsize \arabic{tags}}\refstepcounter{tags}\label{eq:pp}
\Popped{\hist}{l} & \eqdef  l \eqm \mset{e ~|~ \exists t~ l\ldot t \hpts
  (e::l, l) \in \hist}
\end{align*}
}
\hspace{-4pt}
The notation $\mset{-}$ stands for multisets, and $\eqm$ is multiset
equality, which we conflate with list equality modulo permutation.
We can now ascribe the following specs to \code{produce} and
\code{consume}:
{\small
\begin{align*}
\hspace{-5pt}
\spec{\!\!
\begin{array}{c}
\Prod(h_p, l_{<i}) \aand 
  \Array{\esc{ap}}{n}{l}{h_p} 
\end{array}
\!\!} 
& ~\esc{produce}(n, i)~
\spec{\!\!
\begin{array}{c}
\Prod(h_p, l) \aand 
  \Array{\esc{ap}}{n}{l}{h_p} 
\end{array}
\!\!}%@\privcon\entangle(\tbcon \apart \acon)
\\[7pt]
\tag{\normalsize \arabic{tags}}\refstepcounter{tags}\label{eq:pcspecs}
\spec{\!\!\!
\begin{array}{c}
\exists h_c~l\ldot \Cons(h_c, l_{<i}) ~\aand \\
  \Array{\esc{ac}}{n}{l}{h_c} 
\end{array}
\!\!\!} 
& ~\esc{consume}(n, i)~  
\spec{\!\!\!
\begin{array}{c}
\exists h_c~l\ldot \Cons(h_c, l) ~\aand \\
  \Array{\esc{ac}}{n}{l}{h_c} 
\end{array}
\!\!\!}%@\privcon\entangle\tbcon
\end{align*}}
\hspace{-5pt}
both over the $\privcon \entangle \tbcon$ concurroid. $\Prod$ and
$\Cons$ are defined as follows:
\vspace{-3pt}
\[
%\tag{\arabic{tags}}\refstepcounter{tags}\label{eq:pcpreds}
{\small
\def\arraystretch{1.3}
\begin{array}{r@{\ }c@{\ }l}
\Prod(h_p, l)&\eqdef&
   \hpriv \spts h_p ~\sep~ \htb \spts \histS \aand 
   \Pushed{\histS}{l}  \aand  \Popped{\histS}{\nil}
\\
\Cons(h_c, l)&\eqdef&
   \hpriv \spts h_c ~\sep~ \htb \spts \histS \aand
   \Pushed{\histS}{\nil}  \aand  \Popped{\histS}{l},
\end{array}
}
\]
so they essentially describe the producer/consumer loop invariants;
$l_{<i}$ is a prefix of $l$ for elements with indices less than $i$.
The specs~\eqref{eq:pcspecs} show that \code{produce} pushes all the
elements from \code{ap}, and \code{consume} fills \code{ac} with
elements of some sequence of the length $n$. The proofs of both specs
derive easily from (\ref{eq:stack-spec}) after these are framed to
allow running in arbitrary initial self heap and history. We omit the
proofs here, but provide them in the Coq files.

\begin{figure}
{\centering
{\small
\begin{tabular}{l@{\ \ }l}
%
% Ap: \(\Array A\), 
\begin{minipage}[l]{4.0cm}
\begin{alltt}
\num{1} produce(n: \(\nat\), i: \(\nat\)) \{
\num{2}  \textbf{if} i == n
\num{3}  \textbf{then return} ();
\num{4}  \textbf{else} \{
\num{5}   e <- ap[i];
\num{6}   push\(\sb{\htb}\)(e);
\num{7}   produce(i + 1);
\num{8}  \}
\num{9} \}

\end{alltt} 
\end{minipage}
& 
\begin{minipage}[l]{4.0cm}
\begin{alltt}
\num{ 1} consume(n: \(\nat\), i: \(\nat\)) \{
\num{ 2}  \textbf{if} i == n
\num{ 3}  \textbf{then return} ();
\num{ 4}  \textbf{else} \{
\num{ 5}   r <- pop\(\sb{\htb}\)();
\num{ 6}   \textbf{if} r == \textsf{Some} e
\num{ 7}   \textbf{then} \{
\num{ 8}    ac[i] := e;
\num{ 9}    consume(i + 1);\}
\num{10}   \textbf{else} consume(i);\}\}
\end{alltt} 
\end{minipage}
\\\\[-3pt]
\hline 
\\[-3pt]
\multicolumn{2}{c}{
\begin{minipage}[l]{8.0cm}
\begin{alltt}
~~~~~~~\num{1} exchange(n: \(\nat\))\!: \Unit~\{ \textsf{hide}\(\sb{\Phi, \hempty}\) \{ \\
\phantom{aaaaaaa}\num{2} ~~produce(n, 0); \(||\) consume(n, 0); \\
\phantom{aaaaaaa}\num{3} \}\}
\end{alltt} 
\end{minipage}
}
\end{tabular}
}
}
\vspace{1mm}
\caption{A parallel stack-based producer/consumer program.}
\label{fig:client}
\end{figure}

The interesting part of the example is proving \code{exchange}, where
we compose \code{produce} and \code{consume} in parallel, and then use
hiding to infer that the \code{ap} and \code{ac} arrays in the end
contain the same elements, modulo permutation. The proof outline is in
Figure~\ref{fig:pc-proof}, and it relies on the following important
lemmas about histories.
%
%The following two lemmas hold for histories wrt.  $\mathsf{Pushed}$
%and $\mathsf{Popped}$.
%
\vspace{1mm}
\begin{lemma}[Combining \textsf{Pushed} and \textsf{Popped} histories]
\label{lm:pushpop1}
\[{\small
\begin{array}{l}
\Pushed{\hist_1}{l_1} \aand \Popped{\hist_1}{\nil} \aand 
\Popped{\hist_2}{l_2} \aand \Pushed{\hist_2}{\nil} \implies \\
\Pushed{\hist_1 \hunion \hist_2}{l_1} \aand \Popped{\hist_1 \hunion
  \hist_2}{l_2}    
\end{array}
}
\]
\end{lemma}
\begin{lemma}
\label{lm:pushpop2}
If $\hist$ is $\complete$ and $\cont$, then 
\[{\small
  \begin{array}{c}
    \Pushed{\hist}{l_1} \aand \Popped{\hist}{l_2} \aand |l_1| =
|l_2| \implies l_1 \eqm l_2.
  \end{array}
}
\]
\end{lemma}
\vspace{-1mm}
The proof outline in Figure~\ref{fig:pc-proof} starts in the
concurroid $\privcon$, which extends to $\privcon\entangle\tbcon$ in
the scope of $\mathsf{hide}$. The invariant $\Phi$ of $\mathsf{hide}$
is the one we already used, defined in~\eqref{eq:phi}. It introduces a
Treiber stack structure with an initial history $0 \hpts (\nil,
\nil)$. Also, the heaplet $\sent \hpts \Null$ with the sentinel
pointer has been donated to the state space of the Treiber stack, so
it is removed from the private heap.
Next, the self-heap and history are split via $\ssep$; the parts are
given to \code{produce} and \code{consume}, respectively, according to
the parallel composition rule~\eqref{eq:parcom}. Next, we reason out
of specifications~\eqref{eq:pcspecs} for producer/consumer and combine
the subjective views back via $\ssep$ upon joining of the parallel
threads: we thus derive that the contents of \code{ap} and
\code{ac}, are $l$ and $l'$ respectively. By unfolding the definitions of $\mathsf{Pr}$
and $\mathsf{Cn}$, and using Lemma~\ref{lm:pushpop1}, we derive
$\Pushed{\histS}{l} \aand \Popped{\histS}{l'}$, where $\histS$ is the
combined history of \code{produce} and \code{consume}.
Finally, $\histS$ is complete and stack-like (since other-history is
provably $\hempty$ thanks to hiding). Moreover, both $l$ and $l'$ have size 
$n$, as ensured by the assertion $\AAArray_n$ constraining both of them.
Thus, in the last assertion, we can use Lemma~\ref{lm:pushpop2} to obtain the 
desired equality of $l$ and $l'$ modulo permutation. 
Note also that the sentinel pointer is returned
back to the private heap, along with the garbage heap (existentially
abstracted by $-$).

\begin{figure}
{\small
\begin{gather*}
\spec{
\begin{array}{c}
\hpriv \spts h_p \hunion h_c \hunion \sent \hpts \Null \aand 
  \Array{\esc{ap}}{n}{l}{h_p}  \aand
  \Array{\esc{ac}}{n}{-}{h_c}
   \end{array}
}
%@\privcon \entangle \acon
\\[-3pt]
\!\!\!\!\!\!\!\!\!\!\!\!\mathsf{hide}_{\Phi, \hempty}~\esc{\{} 
\\[-3pt] 
\spec{
\begin{array}{c}
\hpriv \spts h_p \hunion h_c \aand \Array{\esc{ap}}{n}{l}{h_p}  \aand
\Array{\esc{ac}}{n}{-}{h_c} ~\sep~\\
\htb \spts 0 \hpts (\nil, \nil) \aand \htb \opts \hempty
\end{array}}
%@\privcon \entangle \acon \entangle \tbcon
%& \text{by \textsc{Hide}}
\\
\!\!\!\!\!\!\!\!\!\!\!\!\!\!\!
\spec{\left(\!\!\!
\begin{array}{c}
\hpriv \spts h_p \aand \Array{\esc{ap}}{n}{l}{h_p} \\
~\sep~ \htb \spts 0 \hpts (\nil, \nil)      
\end{array}
\!\!\!\right) \ssep 
\left(\!\!\!
\begin{array}{c}
  \hpriv \spts h_c \aand
  \Array{\esc{ac}}{n}{-}{h_c} \\
~\sep~ \htb \spts \hempty      
\end{array}
\!\!\!\right)}
% & \text{by def. } \ssep
\\
\begin{array}{r||l}
\spec{\Prod(h_p, l_{<0}) \aand
  \Array{\esc{ap}}{n}{l}{h_p}}
&
\spec{\exists l'\ldot~\Cons(h_c, l'_{<0}) \aand
  \Array{\esc{ac}}{n}{l'}{h_c}}
\\
\esc{produce}(n, 0); & \esc{consume}(n, 0); 
\\
\spec{\Prod(h_p, l) \aand
  \Array{\esc{ap}}{n}{l}{h_p}}
&
\spec{\exists h'_c~l'\ldot~\Cons(h_c, l') \aand
  \Array{\esc{ac}}{n}{l'}{h'_c}}
\end{array}
\\
~~~~\spec{\left(\!\!\!
\begin{array}{c}
\Prod(h_p, l) \aand \Array{\esc{ap}}{n}{l}{h_p}
\end{array}
\!\!\!\right) \ssep 
\left(\!\!\!
\begin{array}{c}
\exists h'_c~l'\ldot~\Cons(h_c, l') \aand \Array{\esc{ac}}{n}{l'}{h'_c}
\end{array}
\!\!\!\right)}
\\
\spec{\!\!\!
\begin{array}{c}
\exists h'_c~l'\ldot 
\hpriv \spts h_p \hunion h_c % \hunion \sent \hpts - \hunion -) \aand
\aand
\Array{\esc{ap}}{n}{l}{h_p}
\aand \Array{\esc{ac}}{n}{l'}{h'_c} \\
~\sep~\exists \histS, \htb \spts \histS \aand \Pushed{\histS}{l} \aand \Popped{\histS}{l'}   
\aand \htb \opts \hempty
\end{array}
\!\!\!}~~~\esc{\}}
%\\
%\!\!\!\!\!\!\!\!\!\!\!\!\}
\\[-4pt]
\spec{
\begin{array}{c}
\exists h'_c~l'\ldot
  \hpriv \spts h_p \hunion h'_c \hunion (\sent \hpts -) \hunion -
  \aand \hbox{}\\
  \Array{\esc{ap}}{n}{l}{h_p}
  \aand \Array{\esc{ac}}{n}{l'}{h'_c} \aand l \eqm l'
\end{array}
}
%@\privcon \entangle \acon
\end{gather*}}
%\vspace{-4mm}
\caption{Proof outline for producer/consumer.}
\label{fig:pc-proof}
\end{figure}

\section{Flat combining}
\label{sec:flatco}
This section shows how PCMs in general, and histories in particular,
can formalize the concurrent algorithm design pattern of helping,
whereby one concurrent thread may execute code on behalf of
another. We use Hendler \etal's flat combining algorithm as an
example~\cite{Hendler-al:SPAA10}.  Unlike other proofs of this
algorithm~\cite{Cerone-al:ICALP14,Turon-al:ICFP13}, we don't require
any additional logical infrastructure aside from ordinary auxiliary
state, represented by a
PCM~\cite{LeyWild-Nanevski:POPL13,Nanevski-al:ESOP14}.
%
%This algorithm has been verified
%before~\cite{Cerone-al:ICALP14,Turon-al:ICFP13}, but we illustrate
%that the proof can be developed in a
%
We verify the algorithm \wrt~a generic PCM, and then instantiate with
the PCM of histories. Thus, our proof is usable even in examples where
the specs don't rely on histories. 
%(\eg, the incrementor~\cite{Nanevski-al:ESOP14}).

The flat combiner structure (FC) generalizes a coarse-grained
lock~\cite{Owicki-Gries:CACM76,Nanevski-al:ESOP14,OHearn:TCS07} as
follows. In the case of a lock, threads acquire exclusive access to
the shared resource protected by the lock, \emph{in succession}. With
the flat combiner, threads register the work that they want to perform
over the shared resource. The lock-acquiring thread (aka.~the
\emph{combiner}) then executes all the registered work, so the other
threads don't need to compete for the lock anymore. This reduces the
contention on the lock, and improves performance. The
higher-order \code{flatCombine} procedure (Figure~\ref{fig:flatco}) works as
follows.\footnote{For simplicity, we consider a modified version of
  the original algorithm. In particular, (a) we use an array rather
  than a priority queue for registration of help requests, and (b) we
  don't expunge help requests that haven't been served for
  sufficiently long time.} It takes as input a \emph{sequential}
function $f$ and argument $x$, and registers the invoking thread for
help with executing $f\ x$ over the shared resource. It does so by
storing $\Req{\eesc{f}}{\eesc{x}}$ into the shared \emph{publication}
array, at index \code{tid} (line~2), where \code{tid} is the id of the
invoking thread.
It next enters the main loop (line~3) and tries to acquire the lock to
the shared heap (line~4).
The acquiring thread becomes a combiner (line~5); it traverses the
publication array, checking for help requests (lines 6--11). For each
request found (which can arrive even while the combiner holds the
lock), the combiner executes the appropriate function with the
provided arguments (line~9) over the shared heap. It informs the
requesting thread \code{i} of the result \code{w}, by writing
$\Resp{\eesc{w}}$ into the slot \code{i} of the publication array
(line~10).
After the traversal, the combiner releases the lock (line~12).
Finally, the thread (combiner or otherwise), checks the publication
array to see if it has been helped (line~13). If so, it extracts the
result \code{w} from its slot in the publication array, and fills the
slot with $\Done$ (all line~13). The result of the help, if one
exists, is returned in line~15. Otherwise, the thread loops for help
again.

To supply the intuition behind the proof, we first review how ordinary
locks work with auxiliary state, in the subjective setting of
FCSL~\cite{Nanevski-al:ESOP14}. As in CSL~\cite{OHearn:TCS07}, and the
Owicki-Gries method~\cite{Owicki-Gries:CACM76}, a lock comes with a
resource invariant $I$ which relates the auxiliary state to the heap
of the shared resource. When the lock is not taken, the shared heap
satisfies $I$. When the lock is taken, the heap is in the exclusive
possession of the acquiring thread, which can invalidate $I$, but has
to restore it before releasing the lock. The subjective setting is
similar, except the values of the auxiliary state are drawn from a PCM
$\pcmS$, and specs keep track of two values $\gS$ and $\gO$,
describing how much the thread (\emph{self}) and its environment
(\emph{other}) have contributed to the resource, respectively. When
the lock is free, the heap of the shared resource satisfies $I(\gS
\join \gO)$. When the lock is released by a thread, the thread may
update its $\gS$ by some value $\gd$, reflecting that its contribution
to the resource changed. Thus, if before locking, the resource
satisfied $I(\gS \join \gO)$, after unlocking it will satisfy $I(\gS
\join \gd \join \gO)$.

The setup of the flat combiner is similar, but in addition to $\gS$
and $\gO$, FC also keeps an array $\gp$ storing a $\pcmS$-value
for each thread. The entry $\gp[i]$ signifies how much the thread $i$
has been helped by the combiner. If $\gp[i]=\gd$ is non-unit, $i$ can
collect the help by joining $\gd$ to its own $\gS$, and setting
$\gp[i]$ to the unit $\pcmU$ of $\pcmS$, after which it can ask
for help again. Thus, the overall relation between the auxiliary state
and the heap of the shared resource, when the lock is free, is
captured by the invariant $I~(\jjoin{i}\gp[i] \join \gS \join \gO)$.

\begin{figure}[t]
\centering 
{\small
\hspace{-11mm}
\begin{tabular}{l@{\!\!\!\!\!\!\!\!\!\!\!\!\!\!\!\!\!\!\!\!\!}l}
\begin{minipage}[l]{5.3cm}
\begin{alltt}
\num{ 1} flatCombine(f: \(A{}\to{}B\), x: \(A\)): \(B\) \{
\num{ 2}  \act{reqHelp}(tid, f, x);          
\num{ 3}  \textbf{fix} loop() \{       
\num{ 4}   locked <- \act{tryLock}(); 
\num{ 5}   \textbf{if} locked \textbf{then} \{
\num{ 6}    \textbf{for} i\(\in\!\set{0, \ldots, n-1}\) \{
\num{ 7}     req <- \act{readReq}(i);
\num{ 8}     \textbf{if} req == \textsf{Req} fi xi \textbf{then} \{         
\num{ 9}      w <- fi(xi);    
\num{10}      \act{doHelp}(i, w);
\num{11}     \}\}
\num{12}    \act{unlock}();\}
\num{13}  rc <- \act{tryCollect}(tid);
\num{14}  \textbf{if} rc == \textsf{Some} w 
\num{15}  \textbf{then} \textbf{return} w;
\num{16}  \textbf{else return} loop();\}();\}
\end{alltt} 
\end{minipage}
&
\begin{minipage}[l]{3cm}
\vspace{3.5mm}
\begin{alltt}
\emph{// Request help for myself}
\emph{// Start looping for help}
\emph{// Try to become a combiner}
\emph{// Now I'm a combiner}
\emph{// Helping loop}



\emph{// Notify i of helping}
\emph{// Finish the helping loop}
\emph{// Release the lock}
\emph{// Try to collect my help}
\emph{// I have been helped}
\emph{// Return the result}
\emph{// Try again}
\end{alltt} 
\end{minipage}
\end{tabular}
}
\caption{Code of the flat combining algorithm. $n$ is a global variable bounding the number of threads.}
\vspace{-3mm}
\label{fig:flatco}
\end{figure}

\subsection{Flat combiner state and specs}\label{sec:fc-conc-spec}

The states of the FC concurroid $\fccon$ are described by the assertion:
% with a label $\hfc$:
%
\[
%\tag{\arabic{tags}}\refstepcounter{tags}\label{eq:fc-state}
\hspace{-5pt}
{\small
\begin{array}{c}
  W_{\fccon} \!\eqdef \!
  \hfc \!\spts\! (\tS, \mS, \gS) \aand 
  \hfc \!\opts\! (\tO, \mO, \gO) \!\!\!\begin{array}[t]{l}
    \hbox{}\aand \hfc \!\jpts\! {\angled{\lk \hpts b \hunion h_p \hunion h_r, \gp}} \\
    \hbox{}\aand \exists l_p\ldot \Array{\ap}{n}{l_p}{h_p} 
\end{array}
\end{array}
}
\]
The auxiliary state in the self/other components consists of the
following. $\tS$ and $\tO$ are sets of thread ids, which form a PCM
under disjoint union.\footnote{One thread may hold many thread id's,
  which it distributes between its children upon forking.}
$\mS$ and $\mO$ are elements of the \emph{mutual exclusion} set $O =
\{\lockNown,
\lockOwn\}$~\cite{LeyWild-Nanevski:POPL13,Nanevski-al:ESOP14} and
record whether the lock $\lk$ is owned by the thread, or the
environment. $O$ is a PCM under the operation defined as $x
\join \lockNown = \lockNown \join x = x$, with $\lockOwn \join
\lockOwn$ undefined. The unit element is $\lockNown$, and the
undefinedness of $\lockOwn \join \lockOwn$ means that two threads
can't simultaneously own the lock.
$\gS$ and $\gO$ are elements of a generic PCM $\pcmS$, as
described above.
The self/other triples form a PCM with component-wise lifted joins
and~units.
 
The joint component of $\fccon$ contains a concrete heap, and the
auxiliary array $\gp$.
The concrete heap keeps the pointer $\lk \hpts b$, which stands for
the lock, with the boolean $b$ representing the lock status. It also
stores the publication array with the origin pointer $\ap$ into the
heaplet $h_p$ (see notation~\eqref{eq:array}). The array stores
elements of type $\mathsf{Stat} \eqdef \Done~|~\Req{f}{x}~|~\Resp{w}$,
as already apparent from Figure~\ref{fig:flatco}.  We abuse the
notation and refer to the array represented by $h_p$ as $\ap$.
The heap $h_r$ is the resource protected by the FC lock. Upon locking
it moves to the exclusive ownership of the combiner. 

%The array $\gp$ stores the auxiliary state obtained by helping. When
%the combiner executes \code{fi} on behalf of $i$, the execution should
%influence not only the locked heap, but also the auxiliary $\gS$ of
%the thread $i$. However, the combiner can't change the self components
%of other threads. Thus, it records into the $i$-th entry of the joint
%auxiliary array $\gp$ a value $\gd$ obtained as an auxiliary value
%after the call to $\code{fi}$. The thread $i$ eventually collects this
%value, and joins it with its own $\gS$.

%The FC concurroid comes with a resource invariant $I$, similar to
%CSL~\cite{OHearn:TCS07}. $I$ is predicate relating the resource heap
%$h_r$, with the auxiliaries $\gS$ of the various threads, thus
%allowing threads to reason about the resource heap abstractly in terms
%of $\gS$'s. 
We further assume the following properties of $W_{\fccon}$:
\begin{itemize}
\item [$(i)$] for any $\tid$, if $\gp[\tid] \neq \pcmU$, then $\ap[\tid] =
  \Resp{w}$ for some $w$;
\item [$(ii)$] if $b$ is $\True$ then $\h_r = \hempty$ and $\mS \join
  \mO = \lockOwn$; otherwise $\mS \join \mO = \lockNown$ and
  $I~(\jjoin{i}\gp[i] \join \gS \join \gO )~h_r$.
\end{itemize}
Property $(i)$ ensures that the auxiliary array $\gp$ holds a pending
contribution in a cell $\tid$ only if the corresponding entry in the
publication array $\ap$ points to the response with some (uncollected)
result.
Property $(ii)$ formally relates the auxiliary state to the resource
heap $h_r$, as already described.
%
%important difference: it connects the resource $h_r$ via the invariant
%$I$ not only to the sum of self/other, but also to the sum of
%client-specific pending contributions $\gp[i]$: $i$~in the last join
%ranges of the set off all thread ids $\set{0, \dots, n-1}$.

Now we can provide a spec for \code{flatCombine} in terms of the
concurroid $\fccon$. We assume $f : A \to B$, $x : A$, and $f$ comes
with the following spec over concurroid~$\privcon$ for private
heaps.\footnote{Thus, we don't require $f$ to be sequential, but every
  sequential function can be given a spec in $\privcon$.}
\[
\tag{\normalsize \arabic{tags}}\refstepcounter{tags}\label{eq:fspec}
{\small
\hspace{-5pt}
\begin{array}{c}
\spec{
\exists h\ldot \hpriv \spts h \aand I~\g~h
} 
f(x)
\spec{
\exists h'~\gd\ldot\hpriv \spts h' \aand I~(\g \join \gd)~h' \aand
\fspec{f}~x~\res~\g~\gd
}
\end{array}
}\]
The spec allows the input heap $h$ to change to $h'$. The resource
invariant $I$ has to be preserved, up to a change of the auxiliary
state, from $\g$ to $\g \join \gd$. $\fspec{f}$ is a client-supplied
predicate which specifies $f$. We call it \emph{validity predicate};
it's functional with respect to~$\gd$, and relates the input value
$v$, the result value~$\res$, the initial auxiliary state~$\g$ and the
``auxiliary delta'' $\gd$ resulting from the invocation of~$f$.
For instance, if $f$ were a sequential push operation on stacks, with
$\g$ and $\gd$ being set to histories $\hist$ and $\histd$, we might
choose
\[
\tag{\normalsize \arabic{tags}}\refstepcounter{tags}\label{eq:push-spec}
{\small
%\hspace{-4.1mm}
\begin{array}{r@{\ }c@{\ }l}
  \fspec{\push}~x~\res~\hist~\histd & \eqdef & \res = () \aand \histd =
  \tfr{\hist} \hpts (l, x::l),
\end{array}
}\] 
where $l = \hist[\last{\hist}]$. That is, $\fspec{\push}$ fixes the
result of $\push$ to be unit and its effect to be the singleton
history describing the action of pushing.

%
%We require that $f$ preserves the invariant $I$. 
%The  operates over the private heaps only, it can be
%considered as sequential. 
%spec~\eqref{eq:fspec} is indeed sequential, as it captures only
%elements of the privately-owned heap and can possibly make use of the
%unconstrained allocator.

For the spec of \code{flatCombine} we need two auxiliary predicates.
$\NoReqs$ indicates that the thread $\tid$ currently requests no
help. $\histso{\cdot}{\cdot}$, generalizes~\eqref{eq:histso} from
histories to PCM $\pcmS$.
{\small{
\begin{align*}
\noreqs{\tid} & \eqdef \hfc \spts (\set{\tid}, \lockNown, -) \aand \ap[\tid] = \Done    
\tag{\normalsize \arabic{tags}}\refstepcounter{tags}\label{eq:noreqs}
\\
\histso{\hfc}{\gS, \gO, \g} & \eqdef \hfc \!\spts \! (-, -, \gS) \aand
\hfc \!\opts\! (-, -, \gO) \aand
\g \pre \jjoin{i}\gp[i] \!\join\! \gS \!\join\! \gO    
\end{align*}
}}
\hspace{-1.5mm}
Here, the partial order $\pre$ on PCM elements is defined as $\g_1
\pre \g_2 \eqdef \exists \g, \g_2 = \g_1 \join \g$. It generalizes the
relation $\pre$ from histories to the PCM $\pcmS$, and in the
specs captures that the value $\g_1$ was ``current'' before $\g_2$.

The spec for \code{flatCombine} is given \wrt a specific thread id
$\tid$. % in a concurroid $\privcon\entangle\fccon$, which we omit for brevity:
{\small
{\begin{gather*}
\spec{\!\!
\begin{array}{c}
  \hpriv \spts \hempty ~\sep~
  \histso{\hfc}{\pcmU, -, \g} \aand
  \noreqs{\tid} 
\end{array}
\!\!} 
\\
\tag{\normalsize \arabic{tags}}\refstepcounter{tags}\label{eq:fc-spec}
\esc{flatCombine}(f, x): B
\\
\spec{\!\!
\begin{array}{c}
\exists \g'~\gd\ldot \hpriv \spts \hempty ~\sep~
\histso{\hfc}{\gd, -, \g'} ~\aand \\
\noreqs{\tid} \aand 
\g \pre \g' \aand
\fspec{f}~x~\res~\g'~\gd 
\end{array}
\!\!}@\privcon\entangle\fccon
%@(\privcon\acon)\entangle\fccon
\end{gather*}}
}
\hspace{-2.2mm}
\code{flatCombine} starts and ends in a state in which the thread
$\tid$ doesn't request the help ($\NoReqs$), and in which $\g$ names
the sum total of the contributions. It doesn't change the
privately-owned heap, but increases self-contribution by amount of an
auxiliary delta $\gd$. The mediating value $\g'$ is a sum-total of the
contributions at the moment when the thread received help; thus,
$\fspec{f}~x~\res~\g'~\gd$. As $\g'$ is current sometime after the
initial $\g$, the spec postulates $\g \pre \g'$.
%
%relating the input and result values with $\gd$ using $\fspec{f}$:
%$\g'$ captures the overall auxiliaries combined at \emph{some moment}
%during the execution of \code{flatCombine}. Using $\g$ for this
%purpose would make the postcondition non-stable.

\subsection{Flat combiner transitions}
\label{sec:fc-transitions}
External transitions intuitively correspond to locking/unlocking the
heap $h_r$, thus moving it from the joint to private state, and
vice-versa. We don't present them formally, as they are similar to the
transitions in CSL~\cite{Nanevski-al:ESOP14}.
The internal transitions $\reqtrans$, $\helptrans$ and $\colltrans$
synchronously change the contents of $\ap$ and $\gp$ for a particular
thread id $i$ (one at a time) as the following diagram illustrates.
\begin{center}
\includegraphics[width=0.40\textwidth]{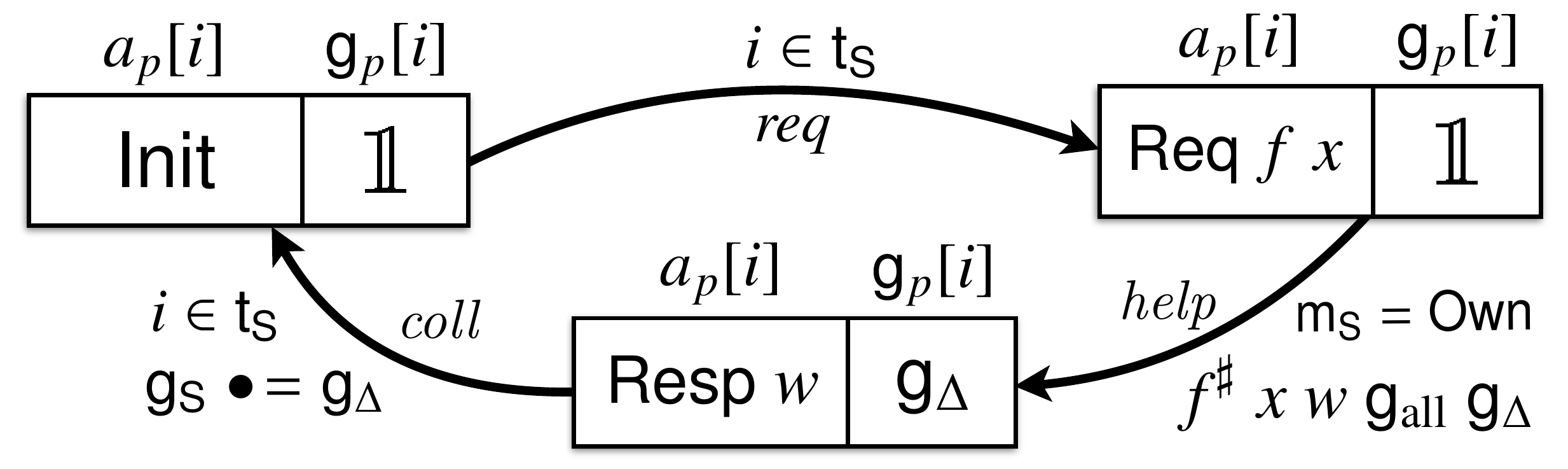}    
\end{center}
The transition $\reqtrans$ can be taken only by a thread holding the
thread id $i$; it changes the value of $\ap[i]$ from $\Done$ to
$\Req{f}{x}$ for some $f$ and $x$.
The transition $\helptrans$ can be performed by any thread that owns
the lock (not necessarily the one with the id $i$); it replaces the
contents of $\ap[i]$ and $\gp[i]$ with an appropriate result $w$ and
an auxiliary delta $\gd$, respectively. The two are valid \wrt~the
input $x$ and the cumulative auxiliary $\gall$, as ensured by the
constraint $\fspec{f}$.
Finally, $\colltrans$ is invoked by the thread with id $i$; it flushes
the contents of $\gp[i]$, into the self-contribution $\gs$ and puts
$\Done$ into $\ap[i]$.

\subsection{Verifying the flat combiner}
\label{sec:verifying-fc}

\begin{figure}
\centering 
\[
\hspace{-2mm}
{\small
\begin{array}{r@{\ \ \ }l@{\ }l}
  \Num{1} & \sspec{~\hpriv \spts \hempty ~\sep~
    \histso{\hfc}{\pcmU, -, \g} \aand \noreqs{\tid} ~} & 
  \\[1pt]
  \Num{2} &  \esc{[}\act{reqHelp}\esc{(}tid, f, x\esc{)]}; & 
  \\[1pt]
  \Num{3} & \sspec{~\hpriv \spts \hempty ~\sep~ 
    \hfc \spts (\set{\tid}, \lockNown, \pcmU) \aand \requested{\tid, f, x, \g} ~} & 
  \\[1.5pt]
  \Num{4} &  \esc{\textbf{fix}~loop() \{}        & 
  \\
  \Num{5} & \sspec{~\hpriv \spts \hempty ~\sep~
    \hfc \spts (\set{\tid}, \lockNown, \pcmU) \aand \requested{\tid, f, x, \g} ~} & 
  \\[1.5pt]
  \Num{6} &  \esc{\textbf{if} } \act{tryLock}\esc{()} \esc{ \textbf{then} \{}        & 
  \\
  \Num{7} & 
  \sspec{~\exists h_r~\gall\ldot \hpriv \spts h_r  \aand I~\gall~h_r 
    ~\sep~ \lhr{\tid, f, x, \g, \gall}~} & 
  \\[1.5pt]
  \Num{8} &  \esc{\textbf{for} }i \in \set{0, \ldots, n-1}~\esc{\{}        & 
  \\
  \Num{9} &   \sspec{~\exists h_r~\gall\ldot \hpriv \spts h_r  \aand I~\gall~h_r 
    ~\sep~  \lhr{\tid, f, x, \g, \gall} ~} & 
  \\[1.5pt]
  \Num{10} &  \esc{\textbf{if} [}\act{readReq}\esc{(}i\esc{)] == \textsf{Req}
  }f_i~x_i\esc{ \textbf{then} \{}        & 
  \\
  \Num{11} & \sspec{~\exists h_r~\gall\ldot \hpriv \spts h_r  \aand I~\gall~h_r ~\sep~
    \ap[i] = \Req{f_i}{x_i} \aand 
    \lhr{\tid, f, x, \g, \gall}~} & 
  \\
  \Num{12} &  w\esc{ <- [}f_i(x_i)\esc{];}        & 
  \\
  \Num{13} & \spec{\!\!\!
    \begin{array}{r@{\ }l}
     \exists h_r~\gd~\gall\ldot& \hpriv \spts h_r  \aand
    I~(\gall \join \gd)~h_r \aand
    \fspec{f_i}~x_i~w~\gall~\gd ~\sep~ \\[1.5pt]
    &     \ap[i] = \Req{f_i}{x_i} \aand 
    \lhr{\tid, f, x, \g, \gall} 
    \end{array}
  \!\!\!} & 
  \\[1pt]
  \Num{14} &  \esc{[}\act{doHelp}\esc{(}i, w\esc{)];}        & 
  \\
  \Num{15} & \sspec{~
     \exists h_r~\gd~\gall\ldot \hpriv \spts h_r  \aand
    I~(\gall \join \gd)~h_r ~\sep~
    \lhr{\tid, f, x, \g, \gall \join \gd}~} & 
  \\[1pt] 
  \Num{16} &  \esc{\}\}}        & 
  \\
  \Num{17} &   \sspec{~\exists h_r~\gall\ldot \hpriv \spts h_r  \aand I~\gall~h_r 
    ~\sep~ \lhr{\tid, f, x, \g, \gall}~} & 
  \\[1.5pt]
  \Num{18} &  \act{unlock}\esc{();\}}        & 
  \\
  \Num{19} & \sspec{~\hpriv \spts \hempty ~\sep~
    \hfc \spts (\set{\tid}, \lockNown, \pcmU) \aand \requested{\tid, f, x, \g} ~} & 
  \\[1.5pt]
  \Num{20} &  rc\esc{ <- [}\act{tryCollect} \esc{(}tid\esc{)];}        & 
  \\
  \Num{21} & \sspec{~\hpriv \spts \hempty ~\sep~ 
    \ack{\tid, f, x, \g, {rc}} ~} & 
  \\[1.5pt]
  \Num{22} &  \esc{\textbf{if} }rc\esc{ == \textsc{Some} }w\esc{
    \textbf{then} \textbf{return} } w\esc{;}        & 
  \\
  % \Num{23} & \sspec{~\exists \g'~\gd\ldot \hpriv \spts \hempty \!\aand\!
  %   \histso{\hfc}{\gd, -, \g'} \!\aand\! 
  %   \noreqs{\tid} \!\aand\! 
  %   \g \pre \g' \!\aand\!
  %   \fspec{{f}}~{x}~\res~\g'~\gd 
  %   ~}
  \Num{23} & \sspec{~\text{postcondition}~(\ref{eq:fc-spec})~}
  \\[1.5pt]
  \Num{24} &  \esc{\textbf{else}}        & 
  \\
  \Num{25} & \sspec{~\hpriv \spts \hempty ~\sep~
    \hfc \spts (\set{\tid}, \lockNown, \pcmU) \aand \requested{\tid, f, x, \g} ~} & 
  \\[1.5pt]
  \Num{26} &  \esc{\textbf{return} loop();\}();}        & 
  \\
%   \Num{27} & \sspec{~\exists \g'~\gd\ldot \hpriv \spts \hempty \!\aand\!
% \histso{\hfc}{\gd, -, \g'} \!\aand\! 
% \noreqs{\tid} \!\aand\! 
% \g \pre \g' \!\aand\!
% \fspec{{f}}~{x}~\res~\g'~\gd 
% ~} & 
  \Num{27} & \sspec{~\text{postcondition}~(\ref{eq:fc-spec})~} & 
\end{array}
}
\]
\caption{Proof outline for \code{flatCombine}.}
\label{fig:flatco-proof}
\end{figure}

Figure~\ref{fig:flatco-proof} presents the proof outline for
\code{flatCombine}. We go over it in detail, providing specs for the
employed atomic operations and auxiliary predicates as we go.
The procedure starts by a call to $\act{reqHelp}(\tid, f, x)$ in
line~2, which requests help for running $f$ with argument $x$.  The
action $\act{reqHelp}$ has the following spec:
%is supported by an internal transition
%$\reqtrans$ and has the following stable specification:
%
\[
\tag{\normalsize \arabic{tags}}\refstepcounter{tags}\label{eq:reqhelp}
{\small
\hspace{-10pt}
\begin{array}{r@{\ }c@{\ }l}
\spec{\!\!
\begin{array}{c}
\histso{\hfc}{\pcmU, -, \g} \\
\aand~ \noreqs{\tid}
\end{array}
\!\!}
&
\act{reqHelp}(\tid, f, x)
&
\spec{\!\!
\begin{array}{c}
\hfc \spts (\set{\tid}, \lockNown, \pcmU) \\
\aand ~ \requested{\tid, f, x, \g}    
\end{array}
\!\!}@\fccon
\end{array}
}
\]
where the auxiliary predicate $\Requested$ is defined as follows:
\[
%\tag{\normalsize \arabic{tags}}\refstepcounter{tags}\label{eq:hasreq}
{\small
\hspace{-1mm}
\begin{array}{l}
  \requested{\tid, f, x, \g} ~ \eqdef \\
  ~~~~~~\exists \gO\ldot \ap[\tid] = \Req{f}{x} \aand \hfc \opts (-,-,\gO) \aand \g \pre
  \jjoin{i}\gp[i] \join \gO ~\oor \\
  ~~~~~~ \exists w~\g'~\gO\ldot \ap[\tid] = \Resp{w} \aand \g' \pre
  \jjoin{i}\gp[i] \join \gO ~\aand \\
  ~~~~~~ ~~~~~~~~~~~~~~~~~\gp[\tid] = \gd \aand \g \pre \g' \aand \fspec{f}~x~w~\g'~\gd 
\end{array}
}\]
$\Requested$ indicates that once help is requested by a thread $\tid$,
it can remain unanswered. But if it's answered, than it's answered
appropriately. That is, the result \code{w} and the auxiliary
$\gp[\tid]$ are obtained by a call to $f$, and are related by
$\fspec{f}$. 

The assertion in line~3 serves as a loop invariant for lines
4--26. Right after entering the loop, the thread tries to acquire the
shared resource by calling $\act{tryLock}\esc{()}$ in
line~6. $\act{tryLock}$ transfers the ownership of the heap $h_r$ from
$\fccon$ to $\privcon$'s self-part (hence, its concurroid is
$\privcon\entangle\fccon$) along with establishing the assertion
$\Locked$ and invariant $I~\gall~h_r$.
In the spec of $\act{tryLock}$ below, $\gall$ is a cumulative
auxiliary value of $\fccon$. Notice that this value is stable under
interference.
%
% Notice that this value is stable throughout the execution of the
% combiner.
%
The environment threads may collect their entries from $\gp$, and move
them to their self components, but they can't change the sum total
$\gall$.
\[
%\tag{\normalsize \arabic{tags}}\refstepcounter{tags}\label{eq:trylock}
{\small
\hspace{-8pt}
\begin{array}{c}
\spec{
\begin{array}{c}
\hpriv \spts \hempty ~\sep~ 
\hfc \spts (\set{\tid}, \lockNown, \pcmU) \aand \requested{\tid, f, x, \g}
\end{array}
}
\\[4pt]
\act{tryLock}()
\\[2pt]
\spec{\!\!\!
\begin{array}{c}
  \res = \True \aand  \exists h_r~\gall\ldot \hpriv \spts h_r  \aand I~\gall~h_r
  ~\sep~ \lhr{\tid, f, x, \g, \gall} \\
  \oor~ \res = \False \aand \hpriv\!\spts\!\hempty ~\sep~ \hfc \spts
  (\set{\tid}, \lockNown, \pcmU) \aand \requested{\tid, f, x, \g}    
\end{array}
\!\!\!}
\end{array}
}
\]
\vspace{-5pt}
\[
%\tag{\normalsize \arabic{tags}}\refstepcounter{tags}\label{eq:lhr}
{\small
\hspace{-32pt}
\begin{array}{r@{\ }c@{\ }l}
  \lhr{\tid, f, x, \g, \gall} & \eqdef & \locked{\tid, \gall} \aand \requested{\tid, f, x, \g}
\end{array}
}\]
\vspace{-9pt}
\[
%\tag{\normalsize \arabic{tags}}\refstepcounter{tags}\label{eq:locked}
{\small
%\hspace{-4mm}
\begin{array}{l}
  \locked{\tid, \gall} ~~~~~~ \eqdef \\
  ~~~~~\exists \gO\ldot \hfc \spts (\set{\tid}, \lockOwn, \pcmU) \aand \hfc \opts (-,-,\gO) \aand \gall
  = \jjoin{i}\gp[i] \join \gO
\end{array}
}\]
The assertion on line~7 serves as a loop invariant for the ``combiner
loop'' of lines~8--18.
The action $\act{readReq}(i)$ in line~10 returns the contents of
$\ap[i]$. The assertion in line~11 is stable since only the combiner
can change the requests in $\ap$, by replacing them with responses.
The call $f_i(x_i)$ in line~12 changes the assertion according to the
spec~\eqref{eq:fspec}, producing the result value $w$ and an auxiliary
delta $\gd$.
Calling $\act{doHelp}(i, w)$ changes the contents of $\ap[i]$ from
$\Req{f_i}{x_i}$ to $\Resp{w}$ and sets $\gp[i]$ to be $\gd$,
following the transition $\helptrans$. This changes the cumulative
value of $\fccon$'s auxiliaries from $\gall$ to $\gall \join \gd$,
however, the invariant is preserved. Any assertion about $i$'s status
isn't stable at this point (as nothing prevents $\ap[i]$ and $\gp[i]$
to be modified according to the transitions of $\fccon$), so we don't
mention it on line~15.
The combiner loop invariant on line~17 implies the precondition of the
$\act{unlock}$ action invoked on line~18, which releases the lock and
transfers the ownership of $h_r$ from $\privcon$'s self back
to~$\fccon$:
\[
%\tag{\normalsize \arabic{tags}}\refstepcounter{tags}\label{eq:unlock}
{\small
\hspace{-6pt}
\begin{array}{c}
\spec{\!\!\!
\begin{array}{c}
  \exists h_r~\gall\ldot \hpriv \spts h_r \aand I~\gall~h_r   
  ~\sep~ \lhr{\tid, f, x, \g, \gall}
\end{array}
\!\!\!}
\\[4pt]
\act{unlock}()
\\[1pt]
\spec{\!\!\!
\begin{array}{c}
\hpriv \spts \hempty ~\sep~ \hfc \spts (\set{\tid}, \lockNown, \pcmU)
\aand \requested{\tid, f, x, \g}
\end{array}
\!\!\!}@\privcon\entangle\fccon
\end{array}
}
\]
Regardless of whether the thread managed to be a combiner
(lines~6--18) or not, it tries to collect its result and the
contribution on line~20 by calling $\act{tryCollect}$ action:
\[
%\tag{\normalsize \arabic{tags}}\refstepcounter{tags}\label{eq:unlock}
{\small
\begin{array}{r@{\ }c@{\ }l}
\spec{
  \begin{array}{c}
    \hfc \spts (\set{\tid}, \lockNown, \pcmU) ~\aand \\
    \requested{\tid, f, x,\g}
  \end{array}
}
&
\act{tryCollect}(\tid)
&
\spec{~
\ack{\tid, f, x, \g, \res}
~}@\fccon
\end{array}
}
\]
\vspace{-2mm}
\[
\tag{\normalsize \arabic{tags}}\refstepcounter{tags}\label{eq:ack}
{\small
%\hspace{-4.1mm}
\begin{array}{l}
  \ack{\tid, f, x, \g, r} ~ \eqdef \\
  ~~~~~r = \None \aand  \hfc \spts (\set{\tid}, \lockNown, \pcmU)
  \aand \requested{\tid, f, x, \g}~\oor 
  \\
  ~~~~~ \exists w~\g'~\gd\ldot 
  r = \Some{w} \aand \noreqs{\tid} \aand   \g \pre \g' ~\aand \\
  ~~~~~ ~~~~~~~~~~~~~~~~\histso{\hfc}{\gd, -, \g'} \aand \fspec{f}~x~w~\g'~\gd
\end{array}
}\]
Operationally, if the content of $\ap[\tid]$ was $\Resp{w}$,
$\act{tryCollect}$ replaces it by $\Done$ and simultaneously flushes
the content $\gd$ of $\gp[i]$ into the self-component, returning
$\Some{w}$ as its result; otherwise it returns $\None$ without
changing anything. The predicate $\Ack$ describes these two possible
outcomes.
The rest of the proof goes by branching on the result of
$\act{tryCollect}$ (line~22), selecting the appropriate disjunct from
$\Ack$~\eqref{eq:ack}, and restarting the \code{loop} if $\None$ was
returned (line~26).

\subsection{Instantiating the flat combiner for stacks}
\label{sec:instantiating-fc}
To illustrate that the abstract spec for the flat combiner follows the
expected intuition, we consider an instance where $\gS, \gO, \gp$ are
histories, and $f$ is the sequential \code{push} method for stacks,
satisfying~the generic sequential spec~\eqref{eq:fspec} with the
validity predicate $\fspec{\push}$ defined by~\eqref{eq:push-spec} and
the stack invariant~\eqref{eq:tb-states}.
%
%A sequential implementation of $\push$ would come with the following
%accompanying validity predicate $\fspec{\push}$:
%
%
%\[
%\tag{\normalsize \arabic{tags}}\refstepcounter{tags}\label{eq:push-spec}
%{\small
%%\hspace{-4.1mm}
%\begin{array}{r@{\ }c@{\ }l}
%  \fspec{\push}~x~w~\hist~\histd & \eqdef & w = () \aand \histd =
%  \tfr{\hist} \hpts (l, x::l),
%\end{array}
%}\]
%%
%where $l = \hist[\last{\hist}]$. That is, $\fspec{\push}$ fixes the
%result of $\push$ to be unit and its effect to be the singleton
%history describing the ``push''. 
%
So by instantiating~\eqref{eq:fc-spec}, after some simplification, we
obtain:
{\small {\begin{gather*} \spec{\!\!
\begin{array}{c}
  \hpriv \spts \hempty ~\sep~
  \histso{\hfc}{\hempty, -, \hist} \aand
  \noreqs{\tid} 
\end{array}
\!\!} 
\\
\tag{\normalsize \arabic{tags}}\refstepcounter{tags}\label{eq:fc-push-spec}
\esc{flatCombine}(\push, e): \mathsf{Unit}
\\
\spec{\!\!
\begin{array}{c}
\exists t~l\ldot \hpriv \spts \hempty ~\sep~
\histso{\hfc}{t \hpts (l, e::l), -, \hist} \aand
\hist < t \aand
\noreqs{\tid}
\end{array}
\!\!}
%@(\privcon\acon)\entangle\fccon
\end{gather*}}
}
\hspace{-7.5pt}
Note that~\eqref{eq:fc-push-spec} is very similar to the spec
\eqref{eq:stack-spec} for Treiber \code{push}; the only difference is
in the FC-specific components such as thread id's, the $\NoReqs$
predicate, and the lock status views used in the definition of
$\NoReqs$. Thus, the spec~\eqref{eq:fc-spec} is adequate.

Strictly speaking, instantiating \eqref{eq:fc-spec} yields the
postcondition:
\[{\small
\spec{\!\!
\begin{array}{c}
  \exists \hist'~\histd\ldot \hpriv \spts \hempty ~\sep~
  \histso{\hfc}{\histd, -, \hist'} \aand
  \hist \pre \hist' \aand
  \fspec{\push}~e~()~\hist'~\histd
  \aand \ldots
\end{array}
\!\!}
}\]
but this can be easily weakened into~\eqref{eq:fc-push-spec}. The main
difficulty is in deriving the assertion $\hist < t$
in~\eqref{eq:fc-push-spec}'s postcondition. Intuitively, the assertion
holds because $t$, such that $\histd = t \hpts (l, e::l)$, has been
taken to be \emph{fresh} \wrt $\hist'$ by definition of
$\fspec{\push}$~\eqref{eq:push-spec}. Thus, $\hist' < t$, so the
result follows from $\hist \pre \hist'$.
A similar derivation can be done for an FC-specification of $\pop$.

%
%definition~\eqref{eq:noreqs} of~$\Histso$), and therefore may be used
%in exactly the same way for client-side reasoning as it has been
%demonstrated in Section~\ref{sec:examples}.

\section{Related and future work}
\label{sec:related}
%% History-based semantics for concurrency
Histories are a recurring idea in the semantics of shared-memory
concurrency, in one form or another. For example, the classical
Brookes' semantics~\cite{Brookes:TCS07} uses \emph{traces} to give a
model for CSL.
%
%Arguing for grainless semantics, Reynolds~\cite{Reynolds:FSTTCS04}
%suggests traces with , similar
%to our self/other combining.
%
Traces are similar to histories, but don't contain time stamps. The
explicit time-stamping makes it straightforward to define a merge
(\ie, join) for histories, and endows them with PCM structure. While
Brookes uses traces in the semantics, we use histories in the specs.

%% Hindsight

Temporal reasoning about shared-memory concurrent programs has also
been employed before. For example, O'Hearn
\etal~\cite{OHearn-al:PODC10} advocate \emph{hindsight lemmas} to
directly and elegantly capture the intuition about linearizability of
a class of concurrent data structures and algorithms. In this paper,
we put histories to use in ordinary Hoare-style specs. This avoids the
relational reasoning about permuting traces of \emph{two} programs, as
required by linearizability, but is strong enough to provide Hoare
logic specs that are expressive, and capable of abstracting
granularity. In our Coq formalization, we discovered that deriving
stability of history-based specs very much resembles reasoning by
hindsight.

%% Logics with temporal reasoning

HLRG by Fu \etal is a Hoare logic for concurrency which admits
history-based assertions~\cite{Fu-al:CONCUR10}. However, their
histories are hard-coded into the logic. In contrast, our histories
are just a specific PCM, that one can use to instantiate the general
framework of FCSL. This affords greater flexibility: if history-based
specifications are not needed (\eg, the incrementation
example~\cite{Nanevski-al:ESOP14}), they don't have to be used.  HLRG
defines separating conjunction~$\sep$ over histories as follows:
conjoined histories must have equal length, and their corresponding
entry heaps are merged via disjoint union.  In contrast, our histories
are not required to have heaps in the codomain. One can choose an
arbitrary datatype to capture what is important for an example at
hand.

Gotsman \etal use temporal reasoning to verify several concurrent
memory reclamation algorithms using the notion of \emph{grace
  period}~\cite{Gotsman-al:ESOP13}. Their logic extends
RGSep~\cite{Vafeiadis-Parkinson:CONCUR07} with a very specific notion
of histories, which live in the shared state. In contrast, we use
histories not as shared, but as private auxiliary state, following the
self/other dichotomy. This enables us to directly reuse the frame
rule and other logical infrastructure from the separation logic FCSL,
without any extensions.

%with no possibility of framing with
%respect to them.
%%
%Even though we believe that the notion grace period can be captured
%using our formulation of the histories, in our experience it is not
%significantly more difficult to specify and verify correct memory
%reclamation in FCSL without relying on any sort of temporal reasoning.
%%%
%% \footnote{Our paper-and-pencil proof of the Michael's
%% stack~\cite{Michael:TPDS04} employs heap ownership transfer to
%% specify memory reclamation and can be carried out solely basing on a
%% formulation of the concurroid, and do not require to make statements
%% about histories. Since we did not implement this proof in Coq by the
%% time of submission, we do not present it here, leaving it for future
%% work.}
%%
%Unlike our host logic FCSL, none of the above mentioned
%logics~\cite{Fu-al:CONCUR10,Gotsman-al:ESOP13} has a notion of
%subjective contributions, which prevents them from giving principal
%elementary Hoare-style specifications as well as restricting
%other-interference by means of hiding.

Several recent approaches, such as Turon \etal's
CaReSL~\cite{Turon-al:ICFP13} (which also verifies the flat combiner),
and the logic of Liang and Feng (L\&F)~\cite{Liang-Feng:PLDI13}
support granularity abstraction by unifying Hoare-style reasoning with
linearizability and contextual refinement. 
%
%\ab{Drop LF. It has other connotations.}
%
In contrast, in this paper, we argue that a form of granularity
abstraction can already be obtained without relying on
linearizability. Instead, by using histories, one obtains Hoare-style
specs which hide the fine-grained nature of the underlying
programs. This can be done in a simple Hoare logic (and we reuse FCSL
off the shelf), whereas CaReSL and L\&F require significant additional
logical infrastructure~\cite{Turon-al:POPL13,Liang-al:POPL12,Liang-Feng:PLDI13}, as
linearizability is a stronger property than our specs. One example of
the additional infrastructure has to do with helping (\eg, in the flat
combiner), where these logics consider the refined effectful commands
as resources, and make them subject to ownership
transfer~\cite{Turon-al:ICFP13}.
While on the surface there's a similarity between
commands-as-resources and histories-as-resources, there are also
significant differences. Commands-as-resources are about executing
specification-level programs (and an effectful abstract program, once
executed, can't be ``re-executed'', since it has reached a value),
while histories are about what has transpired. Unlike
commands-as-resources, histories also contain information about the
order in which something happened in the form of timestamps, thus
enabling temporal reasoning by
hindsight~\cite{OHearn-al:PODC10}. Histories have a PCM structure,
whereas commands-as-resources don't. Hence, histories in FCSL are
subject to the same set of inference rules as heaps, in contrast to
commands-as-resources which requires a number of dedicated inference
rules.

% Liang and Feng presented a methodology to verify linearizability by
% means of establishing linearization points in via Hoare-style
% reasoning~\cite{Liang-Feng:PLDI13} by extending a general-purpose
% concurrency logic LRG~\cite{Feng:POPL09}.

Many of our history-based proofs are very close in spirit to proofs of
linearizability (\eg, the proofs of Treiber stack in
Section~\ref{sec:examples} compared to the proofs in
L\&F~\cite{Liang-Feng:PLDI13}), since adding an entry to a self-history
can be seen as linearizing an effectful operation.
However, we obtain some simplification in the proofs of pure methods
such as \code{readPair}. In particular, L\&F and related logics require
\emph{prophecy variables}~\cite{Qadeer-al:TR09} (or, equivalently,
\emph{speculations}~\cite{Liang-Feng:PLDI13,Turon-al:POPL13}) in their
proofs of \code{readPair}, but we don't.
We do expect, however, that prophecy variables will be required in
examples where the shape of the event to be inserted into the history
can't be fully determined at the moment when it logically takes place
(\eg, Harris \etal's MCAS~\cite{Harris-al:DISC02,Vafeiadis:PhD}). We
plan to address such examples in the future work, by choosing another
history-based PCM; that of branching-time histories, in contrast to the
linear-time ones used here.
 
% Alternatively, an operational approach to prophecies using
% oracle semantics~\cite{Zhang-al:TAMC12} might be adopted as a minor
% extension to FCSL's model.
%
% Similarly to Liang and Feng's logic, CaReSL by Turon
% \etal~\cite{Turon-al:ICFP13} allows to establish the contextual
% refinement in a Hoare-style setting. While is expressive enough to
% prove the refinement for the flat combining, CaReSL does not provide
% it with a suitable elementary spec, which makes its contribution
% orthogonal to our development.

In this work, we argued for the abstraction of atomicity via the
singleton histories of the form $t \hpts (s_1, s_2)$, which describe
the atomic changes in the abstract state.
A different approach to express atomicity abstraction is suggested by
da Rocha Pinto \etal's logic TaDA~\cite{ArrozPincho-al:ECOOP14} (a
successor of the Concurrent Abstract Predicates framework
(CAP)~\cite{DinsdaleYoung-al:ECOOP10}) using the notion of an ``atomic
Hoare triple'' of the form $\drspec{p}~c~\drspec{q}$, where the
precondition $p$ is required to be stable, whereas $q$ is not. Such
triples can be explicitly stabilized to obtain specs similar
to~\eqref{eq:stack-weak}.
TaDA proposes a \emph{make\_atomic} command and a number of related
inference rules, which allow one to specify synchronized changes of
auxiliary resources across several shared regions. The changes
themselves don't have to be physically atomic; it's sufficient that
they appear atomic from the point of view of specs. TaDA's assertions
range over \emph{atomic tracking} resources, similar to the
operations-as-resources in the linearizability
proofs~\cite{Liang-Feng:PLDI13,Turon-al:ICFP13}. Unlike histories,
these resources don't have the PCM structure, and thus require special
treatment in TaDA's metatheory. The atomic tracking resources aren't
subject of ownership transfer, which is why TaDA currently doesn't
support reasoning about helping.

Yet another view of atomicity abstraction and canonical concurrent
specifications, which also bypasses linearizability, is advocated by
Svendsen~\etal in a series of papers on Higher-Order and Impredicative
Concurrent Abstract
Predicates~\cite{Svendsen-al:ESOP13,Svendsen-Birkedal:ESOP14}.
Both HOCAP and iCAP leverage the idea, originated by Jacobs and
Piessens~\cite{Jacobs-Piessens:POPL11}, of parametrizing specs of
concurrent data types by a user-provided auxiliary code.
Such auxiliary code can be seen as a callback, which, when invoked at
some point during the execution of a specified method, changes the
values of auxiliary resources in several regions simultaneously.
Thus, when proving a parametrized spec, one should locate a right
moment to invoke the provided auxiliary code, so its precondition
would be ensured and the postcondition handled properly, a reasoning
similar to locating a linearization point.
The use of the first-class auxiliary code can introduce circularity in
the domain underlying the logic---the issue tackled in HOCAP by means
of indirection via ``region types'' and resolved in iCAP by providing
a (non-elementary) model in the topos of trees, which enables
reasoning about helping. 
%
%HOCAP and iCAP don't observe that PCM structure on histories can
%provide a simple and unified way to abstracting granularity.

One difference between iCAP and TaDA is that \emph{make\_atomic} in
TaDA presents a more localized view of atomicity, whereas the specs
in iCAP have to predict the uses of the data structure, and provide
hooks for callbacks. The hooks lead to somewhat indirect specs, and
pollute the reasoning about the structure with client-side
information. We haven't considered either of these two ways of
exploiting abstract atomicity in the current paper, but plan to add
\emph{make\_atomic} to FCSL in the future work. The challenge will be
to generalize \emph{make\_atomic} to work with different notions of
histories (\eg, branching-time histories may be useful, as mentioned
above). We believe that the PCM approach (together with subjectivity),
neither of which is exploited by TaDA and iCAP, will be beneficial in
that respect. In particular, we plan to use PCMs to generalize the
notion of logical atomicity afforded by histories, that we explored in
this paper. Given a PCM $\pcmS$, the element $x \in \pcmS$ is
\emph{prime} if it can't be represented as $x = x_1 \bullet x_2$, for
non-unit $x_1$, $x_2$. For example, in the PCM of heaps, the prime
elements are the singleton heaps. In the PCM of natural numbers with
multiplication, the prime elements are the prime numbers. In the PCM
of histories, the prime elements are the singleton histories $t \hpts
a$. A program can be considered logically atomic if it augments the
self-owned portion of its state by a prime element, or by a
unit. According to this definition, all the examples presented in this
paper are atomic. We expect it should be possible to soundly apply
\emph{make\_atomic} to programs that are atomic in this logical sense.

\section{Conclusion}
\label{sec:conclusion}

In this work we proposed using specifications over auxiliary state in
the form of histories as means of providing general specs for
fine-grained concurrent data structures in a separation style logic.

We relied on singleton time-stamped histories $t \hpts a$, to specify
that a program at time $t$ performs an action $a$. The action is
viewed as \emph{logically} atomic, even though the program may
implement it in a fine-grained manner. Client programs that reason with
this spec can treat the program as if it were coarse-grained. Thus, in
the context of Hoare logic, history-based specs can achieve one of the
main goals behind linearizability.

Histories satisfy the algebraic properties of PCMs, and thus can
directly reuse the underlying infrastructure from an employed
separation logic, such as the assertion logic and the frame rule.
Furthermore, as we illustrated with the proof of the flat combiner
algorithm in Section~\ref{sec:flatco}, the concept of ownership
transfer from separation logic, when specialized to the PCM of
histories, directly formalizes the design pattern of helping.  
%
%whereby a thread can execute code on behalf of another.
%

In addition to the flat combiner, we have verified a number of
benchmark fine-grained structures, such as the pair snapshot
structure, and the Treiber stack. The interesting and novel point
about the specs and the proofs is that they all rely in an essential
way on the subjective dichotomy between self and other auxiliary
state, in order to directly relate the result of a program execution
with the interference of other threads.
Such explicit dichotomy provides for what we consider very concise
proofs. We substantiate this observation by mechanizing all the
reasoning in Coq.

%
%
%\is{Points to stress:}
%
%\begin{itemize}
%
%\item Common approaches for FG specifications are ...
%
%\item We focused on a differen thing - histories
%
%\item By doing so, we got the following insights and benefits (e.g.,
%  the reasoning is similar in spirit to hindsight)
%
%\item The Coq proofs we got are proportional in the size to programs
%  and are conceptually simple
%
%\item Exciting future work is to formalise more patterns
%
%\item Therefore we contributed to general understanding to the ideas
%  behind design and use of concurrent algorithms
%
%\end{itemize}

% The main contribution of our work is an observation that the
% time-stamped histories provide a simple and expressive abstraction to
% capture the behavior of concurrently used data structures. The idea to
% treat such histories as ``strictly growing'' resources allowed us to
% make an effective use of FCSL in order to provide intuitive specs for
% a series of concurrent resources.
% %
% Moreover, the idea of a subjective split of resources to self/other,
% originating from the work on SCSL, made it possible to argue for the
% \emph{principality} of the specs we give as well as gave a very
% straightforward way to reason about other-interference, allowing one,
% in particular to derive sequential specifications out of concurrent
% ones.

% Importantly, we required \emph{no} extensions to the metatheory of
% FCSL in order to construct the reasoning with state histories.

%\bibliographystyle{plainnat}
\bibliographystyle{abbrvnat}

\ifdefined\extflag 
\setlength{\bibsep}{2.2pt}
\else
\setlength{\bibsep}{2.2pt}
\fi

\bibliography{bibmacros,references,proceedings-short}

\providecommand{\noopsort}[1]{}
\begin{thebibliography}{34}
\providecommand{\natexlab}[1]{#1}
\providecommand{\url}[1]{\texttt{#1}}
\expandafter\ifx\csname urlstyle\endcsname\relax
  \providecommand{\doi}[1]{doi: #1}\else
  \providecommand{\doi}{doi: \begingroup \urlstyle{rm}\Url}\fi

\bibitem[Brookes(2007)]{Brookes:TCS07}
S.~Brookes.
\newblock A semantics for concurrent separation logic.
\newblock \emph{Th. Comp. Sci.}, 375\penalty0 (1-3), 2007.

\bibitem[Calcagno et~al.(2007)Calcagno, O'Hearn, and Yang]{Calcagno-al:LICS07}
C.~Calcagno, P.~W. O'Hearn, and H.~Yang.
\newblock Local action and abstract separation logic.
\newblock In \emph{LICS}, 2007.

\bibitem[Cerone et~al.(2014)Cerone, Gotsman, and Yang]{Cerone-al:ICALP14}
A.~Cerone, A.~Gotsman, and H.~Yang.
\newblock {Parameterised Linearisability}.
\newblock In \emph{ICALP}, 2014.

\bibitem[da~Rocha~Pinto et~al.(2014)da~Rocha~Pinto, Dinsdale-Young, and
  Gardner]{ArrozPincho-al:ECOOP14}
P.~da~Rocha~Pinto, T.~Dinsdale-Young, and P.~Gardner.
\newblock {TaDA: A Logic for Time and Data Abstraction}.
\newblock In \emph{ECOOP}, 2014.

\bibitem[Dinsdale-Young et~al.(2010)Dinsdale-Young, Dodds, Gardner, Parkinson,
  and Vafeiadis]{DinsdaleYoung-al:ECOOP10}
T.~Dinsdale-Young, M.~Dodds, P.~Gardner, M.~J. Parkinson, and V.~Vafeiadis.
\newblock {Concurrent Abstract Predicates}.
\newblock In \emph{ECOOP}, 2010.

\bibitem[Elmas et~al.(2010)Elmas, Qadeer, Sezgin, Subasi, and
  Tasiran]{Elmas-al:TACAS10}
T.~Elmas, S.~Qadeer, A.~Sezgin, O.~Subasi, and S.~Tasiran.
\newblock Simplifying linearizability proofs with reduction and abstraction.
\newblock In \emph{TACAS}, 2010.

\bibitem[Feng(2009)]{Feng:POPL09}
X.~Feng.
\newblock Local rely-guarantee reasoning.
\newblock In \emph{POPL}, 2009.

\bibitem[Feng et~al.(2007)Feng, Ferreira, and Shao]{Feng-al:ESOP07}
X.~Feng, R.~Ferreira, and Z.~Shao.
\newblock On the relationship between concurrent separation logic and
  assume-guarantee reasoning.
\newblock In \emph{ESOP}, 2007.

\bibitem[Filipovic et~al.(2010)Filipovic, O'Hearn, Rinetzky, and
  Yang]{Filipovic-al:TCS10}
I.~Filipovic, P.~W. O'Hearn, N.~Rinetzky, and H.~Yang.
\newblock Abstraction for concurrent objects.
\newblock \emph{Theor. Comput. Sci.}, 411\penalty0 (51-52), 2010.

\bibitem[Fu et~al.(2010)Fu, Li, Feng, Shao, and Zhang]{Fu-al:CONCUR10}
M.~Fu, Y.~Li, X.~Feng, Z.~Shao, and Y.~Zhang.
\newblock Reasoning about optimistic concurrency using a program logic for
  history.
\newblock In \emph{CONCUR}, 2010.

\bibitem[Gotsman and Yang(2012)]{Gotsman-Yang:CONCUR12}
A.~Gotsman and H.~Yang.
\newblock {Linearizability with Ownership Transfer}.
\newblock In \emph{CONCUR}, 2012.

\bibitem[Gotsman et~al.(2013)Gotsman, Rinetzky, and Yang]{Gotsman-al:ESOP13}
A.~Gotsman, N.~Rinetzky, and H.~Yang.
\newblock Verifying concurrent memory reclamation algorithms with grace.
\newblock In \emph{ESOP}, 2013.

\bibitem[Harris et~al.(2002)Harris, Fraser, and Pratt]{Harris-al:DISC02}
T.~L. Harris, K.~Fraser, and I.~A. Pratt.
\newblock A practical multi-word compare-and-swap operation.
\newblock In \emph{DISC}, 2002.

\bibitem[Hendler et~al.(2010)Hendler, Incze, Shavit, and
  Tzafrir]{Hendler-al:SPAA10}
D.~Hendler, I.~Incze, N.~Shavit, and M.~Tzafrir.
\newblock Flat combining and the synchronization-parallelism tradeoff.
\newblock In \emph{SPAA}, 2010.

\bibitem[Herlihy and Shavit(2008)]{Herlihy-Shavit:08}
M.~Herlihy and N.~Shavit.
\newblock \emph{The art of multiprocessor programming}.
\newblock M. Kaufmann, 2008.

\bibitem[Herlihy and Wing(1990)]{Herlihy-WingTOPLAS90}
M.~Herlihy and J.~M. Wing.
\newblock Linearizability: A correctness condition for concurrent objects.
\newblock \emph{ACM Trans. Prog. Lang. Syst.}, 12\penalty0 (3), 1990.

\bibitem[Jacobs and Piessens(2011)]{Jacobs-Piessens:POPL11}
B.~Jacobs and F.~Piessens.
\newblock Expressive modular fine-grained concurrency specification.
\newblock In \emph{POPL}, 2011.

\bibitem[Jones(1983)]{Jones:IFIP83}
C.~B. Jones.
\newblock Specification and design of (parallel) programs.
\newblock In \emph{IFIP Congress}, pages 321--332, 1983.

\bibitem[Ley-Wild and Nanevski(2013)]{LeyWild-Nanevski:POPL13}
R.~Ley-Wild and A.~Nanevski.
\newblock Subjective auxiliary state for coarse-grained concurrency.
\newblock In \emph{POPL}, 2013.

\bibitem[Liang and Feng(2013)]{Liang-Feng:PLDI13}
H.~Liang and X.~Feng.
\newblock Modular verification of linearizability with non-fixed linearization
  points.
\newblock In \emph{PLDI}, 2013.

\bibitem[Liang et~al.(2012)Liang, Feng, and Fu]{Liang-al:POPL12}
H.~Liang, X.~Feng, and M.~Fu.
\newblock A rely-guarantee-based simulation for verifying concurrent program
  transformations.
\newblock In \emph{POPL}, 2012.

\bibitem[Nanevski et~al.(2014)Nanevski, Ley-Wild, Sergey, and
  Delbianco]{Nanevski-al:ESOP14}
A.~Nanevski, R.~Ley-Wild, I.~Sergey, and G.~A. Delbianco.
\newblock {Communicating State Transition Systems for Fine-Grained Concurrent
  Resources}.
\newblock In \emph{ESOP}, 2014.

\bibitem[O'Hearn(2007)]{OHearn:TCS07}
P.~W. O'Hearn.
\newblock Resources, concurrency, and local reasoning.
\newblock \emph{Th. Comp. Sci.}, 375\penalty0 (1-3), 2007.

\bibitem[O'Hearn et~al.(2010)O'Hearn, Rinetzky, Vechev, Yahav, and
  Yorsh]{OHearn-al:PODC10}
P.~W. O'Hearn, N.~Rinetzky, M.~T. Vechev, E.~Yahav, and G.~Yorsh.
\newblock Verifying linearizability with hindsight.
\newblock In \emph{PODC}, 2010.

\bibitem[Owicki and Gries(1976)]{Owicki-Gries:CACM76}
S.~S. Owicki and D.~Gries.
\newblock Verifying properties of parallel programs: An axiomatic approach.
\newblock \emph{Commun. ACM}, 19\penalty0 (5), 1976.

\bibitem[Qadeer et~al.(2009)Qadeer, Sezgin, and Tasiran]{Qadeer-al:TR09}
S.~Qadeer, A.~Sezgin, and S.~Tasiran.
\newblock Back and forth: Prophecy variables for static verification of
  concurrent programs.
\newblock Technical Report MSR-TR-2009-142, Microsoft Research, 2009.

\bibitem[Svendsen and Birkedal(2014)]{Svendsen-Birkedal:ESOP14}
K.~Svendsen and L.~Birkedal.
\newblock {Impredicative Concurrent Abstract Predicates}.
\newblock In \emph{ESOP}, 2014.

\bibitem[Svendsen et~al.(2013)Svendsen, Birkedal, and
  Parkinson]{Svendsen-al:ESOP13}
K.~Svendsen, L.~Birkedal, and M.~J. Parkinson.
\newblock Modular reasoning about separation of concurrent data structures.
\newblock In \emph{ESOP}, 2013.

\bibitem[Treiber(1986)]{Treiber:TR}
R.~K. Treiber.
\newblock Systems programming: coping with parallelism.
\newblock Technical Report RJ 5118, IBM Almaden Research Center, 1986.

\bibitem[Turon et~al.(2013{\natexlab{a}})Turon, Dreyer, and
  Birkedal]{Turon-al:ICFP13}
A.~Turon, D.~Dreyer, and L.~Birkedal.
\newblock Unifying refinement and {H}oare-style reasoning in a logic for
  higher-order concurrency.
\newblock In \emph{ICFP}, 2013{\natexlab{a}}.

\bibitem[Turon et~al.(2013{\natexlab{b}})Turon, Thamsborg, Ahmed, Birkedal, and
  Dreyer]{Turon-al:POPL13}
A.~J. Turon, J.~Thamsborg, A.~Ahmed, L.~Birkedal, and D.~Dreyer.
\newblock Logical relations for fine-grained concurrency.
\newblock In \emph{POPL}, 2013{\natexlab{b}}.

\bibitem[Vafeiadis(2007)]{Vafeiadis:PhD}
V.~Vafeiadis.
\newblock \emph{Modular fine-grained concurrency verification}.
\newblock PhD thesis, University of Cambridge, 2007.

\bibitem[Vafeiadis and Parkinson(2007)]{Vafeiadis-Parkinson:CONCUR07}
V.~Vafeiadis and M.~J. Parkinson.
\newblock A marriage of rely/guarantee and separation logic.
\newblock In \emph{CONCUR}, 2007.

\bibitem[Vafeiadis et~al.(2006)Vafeiadis, Herlihy, Hoare, and
  Shapiro]{Vafeiadis-al:PPOPP06}
V.~Vafeiadis, M.~Herlihy, T.~Hoare, and M.~Shapiro.
\newblock Proving correctness of highly-concurrent linearisable objects.
\newblock In \emph{PPOPP}, 2006.

\end{thebibliography}

\ifext{ 
\appendix
%\softraggedright 

\setlength{\parindent}{0.0in}
\setlength{\parskip}{5pt}
\titlespacing*{\section}{0pt}{*1}{*1}
\titlespacing*{\subsection}{0pt}{*0.7}{*0.5}
\titlespacing*{\paragraph}{0pt}{*0.5}{*0.5}

\section*{Optional appendices}

In the optional appendices we provide detailed overview of main
concepts of Fine-grained Concurrent Separation Logic~(FCSL), necessary
for the formal reasoning. These include semantics of the logical
assertions as well as inference rules. We address the curious reader
to the original paper on FCSL~\cite{Nanevski-al:ESOP14} and its
extended version (or the Coq development accompanying this manuscript)
for the details of FCSL's denotational semantics and the soundness
proof.
\textbf{Appendix~\ref{sec:broccoli}} provides the formal semantics of
the FCSL assertions.
\textbf{Appendix~\ref{app:conc}} formally presents concurroids and
entanglement, along with several examples.
\textbf{Appendix~\ref{sec:appactions}} describes properties of
atomic actions of FCSL concurroids.
Finally, \textbf{Appendix~\ref{sec:rules}} provides the rules of FCSL,
explaining some of them in detail.

\section{Semantics of FCSL assertions}
\label{sec:broccoli}

State in FCSL is divided along two different axes. The first axis is
labels (isomorphic to $\mathsf{nat}$). Labels identify concurroids,
\ie data structures that are stored in the state, with specific
restrictions on their evolution. The second axis is ownership. Each
label contains self, other and joint component, describing how much of
each concurroid is owned privately by the specified thread, privately
by that thread's environment, and how much is shared, respectively.

To formally define the concept, we introduce the notion of PCM-map and
type-maps.  A PCM-map is a finite map from labels to a dependent
product $\Sigma_{{\pcmS}{:}\textrm{pcm}} \pcmS$, where
$\mathbb{U}$ is a PCM, and $v \in \pcmS$. A type map is similar,
except we don't require the range to be a PCM; it can be an arbitrary
type.

PCM-maps are composed by means of two operations. Disjoint union $m_1
\hunion m_2$ collects the labels from $m_1$ and $m_2$, ensuring that
there's no overlap. This operation applies to type-maps as well.
However, PCM-maps have another operation which doesn't apply to
type-maps: $m_1 \zip m_2$ joins the values of individual labels, \ie,
$\hempty \zip \hempty = \hempty$, and $((\hlabel \hpts_{\pcmS}
v_1) \hunion m'_1) \zip ((\hlabel \hpts_{\pcmS} v_2) \hunion m'_2)
= (\hlabel \hpts_{\pcmS} v_1 \join v_2) \hunion (m'_1 \zip m'_2)$,
and undefined otherwise.

State, ranged over by $w$, is a triple $\state{s}{j}{o}$, where~$s$
and~~$o$ are PCM-maps, and $j$ is a type map. We refer to them as
\emph{self}, \emph{other}, and \emph{joint} components of $w$.  In
specifications, the three components signify different state
ownership: $s$ is the state owned by the specified thread, and is
inaccessible to the environment; $o$ is the state owned by the
environment, and is inaccessible to the specified thread; $j$ is the
shared (or joint) state, accessible to every thread.
Notice that unlike $s$ and $o$ which are PCM-maps, $j$ is a
type-map. In other words, the joint component is not subject to
PCM-laws, as we don't shuffle its components upon forking, joining,
and framing, as we do in the cases of $s$ and $o$.

The state $w = \state{s}{j}{o}$ is valid iff:
\begin{itemize}
\item[$(i)$] the components $s$, $j$ and $o$ contain the same labels.
 
\item[$(ii)$] $s\ {\zip}\ o$ is defined, \ie, equals labels in $s$ and $o$
  contain equal PCMs. Notice that the labels in $j$ are independent,
  and may contain elements of other types;
\item[$(iii)$] the heaps that may be stored in the labels of $s$, $j$,
  $o$ are disjoint.
\end{itemize}

Figure~\ref{fig:broccoli} collects the definitions the main assertions
of FCSL in terms of the two operations on PCM-maps.

\begin{figure}[t]
\[
{\small
\begin{array}[t]{l@{\,}l}
  w \models \top & \mbox{iff always}\\
  w \models \hlabel \spts v & \mbox{iff valid $w$, and $w = w_1 \hunion
    w_2$, and $w_1.\mathself = \hlabel \hpts v$}\\ 
  w \models \hlabel \jpts h & \mbox{iff valid $w$, and $w = w_1 \hunion
    w_2$, and $w_1.\mathjoint = \hlabel \hpts v$}\\ 
  w \models \hlabel \opts v & \mbox{iff valid $w$, and $w = w_1 \hunion
    w_2$, and $w_1.\mathother = \hlabel \hpts v$}\\ 
  w \models p \aand q & \mbox{iff $w \models p$ and $w \models q$}\\
  w \models p \lsep q & \mbox{iff valid $w$, and $w = w_1 \hunion w_2$, and $w_1 \models p$ and $w_2 \models q$}\\
  w \models p \wand q & \mbox{iff for every $w_1$, valid $w \hunion w_1$, $w_1 \models p$ implies $w \hunion w_1 \models q$}\\
  w \models p \ssep q & \mbox{iff valid $w$, and $w.\mathself= \mathself_1 \hunion \mathself_2$, and}\\
  & \hphantom{\mbox{iff}}\ \mbox{$\state{\mathself_1}{w.\mathjoint}{{\mathself_2} \zip {w.\mathother}}
    \models p$ and $\state{\mathself_2}{w.\mathjoint}{{\mathself_1} \zip
      {w.\mathother}} \models q$}\\
  w \models \mathsf{this}\ w' & \mbox{if $w = w'$}\\
  \hphantom{w} \models p \downarrow h & \mbox{iff for every valid $w$, $w
    \models p$ implies $\flatten w = h$}
  \\
  \\
  \mbox{valid $w$} & \mbox{iff $w=\state{\mathself}{\mathjoint}{\mathother}$, 
                           $\mathsf{dom}\,\mathself = \mathsf{dom}\,\mathjoint = \mathsf{dom}\,\mathother$,}\\
  & \mbox{ $\mathself\,{\zip}\,\mathother$ is defined, 
           and the heaps in $\mathself$, $\mathjoint$, $\mathother$
           are disjoint}
         \\\\
  \flatten w & \eqdef \mbox{ disjoint union of all the heaps in $w$}
  \\\\
  w_1 \hunion w_2 & \eqdef \mbox{pairwise disjoint union of $w_{1,2}$'s
    PCM-components}
  \\\\
  \hlabel \hpts \state{{v_s}}{{v_j}}{{v_o}} & \eqdef \mbox{ $\state{{\hlabel \hpts v_s}}{{\hlabel \hpts v_j}}{{\hlabel \hpts v_o}}$}
\end{array}}
\]
\caption{Notation and semantics of main FCSL assertions.}
\label{fig:broccoli}
\end{figure}

\section{Concurroids: properties and examples}
\label{app:conc}

A concurroid is a 4-tuple $\ucon = (L, W, I, {E})$ where:
(1) $L$ is a set of labels, where a label is a nat; (2) $W$
is the \emph{set of states}, each state $w \in W$ having the
structure described in Section~\ref{sec:broccoli}; (3) $I$ is the
set of \emph{internal transition}, which are relations on $W$
and one of which is always an identity
relation $\id$; (4) $ E$ is a 
set of pairs $(\alpha, \rho)$, where $\alpha$ and $\rho$ are
\emph{external transitions} of $\ucon$. An external transition is a
function, mapping a heap $h$ into a relation on $W$. The
components must satisfy a further set of requirements, discussed next.

%\vspace{5pt}

\paragraph{State properties.}

Every state $w \in W$ is $\mathsf{valid}$ as defined in
Figure~\ref{fig:broccoli}, and its label footprint is $L$, \ie
$\mathsf{dom}\ (w.\mathself) = \mathsf{dom}\ (w.\mathjoint) =
\mathsf{dom}\ (w.\mathother) = {L}$. Additionally, $W$
satisfies the property:
\begin{mathpar}
{\small
\begin{array}{ll}
\textit{Fork-join closure:} & \forall t{:}\textrm{PCM-map}\ldot w \zig t \in {W} \iff w \zag t \in {W}, \\
& \mbox{where}\ w \zig t = \state{t \zip
  w.\mathself}{w.\mathjoint}{w.\mathother}, 
\\
&\mbox{and}\ w \zag t = \state{w.\mathself}{w.\mathjoint}{t \zip w.\mathother}
\end{array}
}
\end{mathpar}
The property requires that $W$ is closed under the realignment of
\self and \other components, when they exchange a PCM-map $t$ between
them. Such realignment is part of the definition of $\ssep$, and thus
appears in proofs whenever the rule \textsc{Par}~\eqref{eq:parcom} is
used, \ie whenever threads fork or join. Fork-join closure ensures
that if a parent thread forks in a state from $W$, then the child
threads are supplied with states which also are in $W$, and
dually for joining.

%\vspace{5pt}

\paragraph{Transition properties.}

A concurroid transition $\gamma$ is a relation on $W$ satisfying:
\begin{mathpar}
{\small
\begin{array}{ll}
\textit{Guarantee:} & (w, w') \in \gamma \implies w.\mathother =
w'.\mathother
\\[5pt]
\textit{Locality:} & \forall t{:}\textrm{PCM-map}\ldot 
  w.\mathother = w'.\mathother \implies\\
&  (w \zag t, w' \zag t) \in \gamma \implies (w \zig t, w' \zig t) \in \gamma
\end{array}
}
\end{mathpar}
Guarantee restricts $\gamma$ to only modify the \self and \joint
components. Therefore, $\gamma$ describes the behavior of a viewing
thread in the subjective setting, but not of the thread's
environment. In the terminology of Rely-Guarantee
logics~\cite{Feng-al:ESOP07,Feng:POPL09,Vafeiadis-Parkinson:CONCUR07},
$\gamma$ is a \emph{guarantee} relation. To describe the behavior of
the thread's environment, \ie, obtain a \emph{rely} relation, we
merely \emph{transpose} the self and other components of~$\gamma$.
\[
\tag{\arabic{tags}}\refstepcounter{tags}\label{eq:transp}
{\small
\gamma^\top = \{(w_1^\top, w_2^\top) \mid (w_1, w_2) \in \gamma\},\ 
\mbox{where $w^\top = \state{w.\mathother}{w.\mathjoint}{w.\mathself}$}
}
\]
In this sense, FCSL transitions always encode \emph{both} guarantee
and rely relations.

Locality ensures that if $\gamma$ relates states with a certain \self
components, then $\gamma$ also relates states in which the \self
components have been simultaneously \emph{framed} by a PCM-map $t$,
\ie, enlarged according to $t$. It thus generalizes the notion of
locality from separation logic, with a notable difference. In
separation logic, the frame $t$ materializes out of nowhere, whereas
in FCSL, $t$ has to be appropriated from \other; that is, taken out
from the ownership of the environment.

An \emph{internal} transition $\iota$ is a transition which preserves
heap footprints. An \emph{acquire} transition $\alpha$, and a
\emph{release} transition $\rho$ are functions mapping heaps to
transitions which extend and reduce heap footprints, respectively, as
show below.  An external transition is either an acquire or a release
transition. If $(\alpha, \rho) \in  E$, then $\alpha$ is an
acquire transition, and $\rho$ is a release transition.
\begin{mathpar}
{\small
\begin{array}{lcl}
\textit{Footprint preservation} & : & 
 (w, w') \in \iota \implies \mathsf{dom}\ \flatten{w} = \mathsf{dom}\
 \flatten{w'}
\\[5pt]
\textit{Footprint extension} & : &
 \forall h{:}\mathrm{heap}\ldot (w, w') \in \alpha(h) \implies
\\ 
&& \mathsf{dom}\ (\flatten{w} \hunion h) = \mathsf{dom}\ \flatten{w'}
\\[5pt]
\textit{Footprint reduction} & : &
 \forall h{:}\mathrm{heap}\ldot (w, w') \in \rho(h) \implies
 \\
 &&\mathsf{dom}\ (\flatten{w'} \hunion h) = \mathsf{dom}\ \flatten{w}
\end{array}
}
\end{mathpar}
The set of Internal transitions always includes at least the identity
transition $\id$ (\ie, transition from a state to itself). Footprint
preservation requires internal transitions to preserve the domains of
heaps obtained by state flattening. Internal transitions may exchange
the ownership of subheaps between the \self and \joint components, or
change the contents of individual heap pointers, or change the values
of non-heap (\ie, auxiliary) state, which flattening erases. However,
they cannot add new pointers to a state or remove old ones, which is
the task of external transitions, as formalized by Footprint extension
and reduction.

\subsection{The concurroid of private heaps}
\label{sec:conc-priv-heaps}

The private heap concurroid is defined as follows.
\[
\tag{\arabic{tags}}\refstepcounter{tags}\label{eq:privcon}
\privcon = (\{\hpriv\}, {W}_{\privcon}, \set{\iota_{\privcon},
  \id}, \{(\alpha_{\privcon}, \rho_{\privcon})\})
\]
It is identified by a \emph{fixed} dedicated label $\hpriv$
and directly captures the notion of heap \emph{ownership}, as presented in
CSL~\cite{OHearn:TCS07}.
Its state-space ${W}_{\privcon}$ is defined as a set of states of the
shape
\[
\hpriv \hpts \state{\hL}{\hempty}{\hE},
\]
where $\hS$ and $\hO$ are disjoint heaps (which are known to form a
PCM). The concurroid's \emph{internal} transitions $\iota_{\privcon}$
allow the values in the codomain of the heap $\hL$, privately-owned by
\emph{self}, to be changed arbitrarily. There is only one channel of
acquire/release transitions $\alpha_{\privcon}$ and $\rho_{\privcon}$
that account for the addition/removal of a heap chunk to/from~$\hL$
correspondingly, given that the state validity is
preserved. Transitions of~$\privcon$ can be formally defined using the
notation from Figure~\ref{fig:broccoli} as follows:
\[ 
\tag{\arabic{tags}}\refstepcounter{tags}\label{eq:priv-trans}
{\small
  \begin{array}{lcr@{\ \ }c@{\ \ }l}
    \iota_\privcon & \eqdef & \hpriv \spts (x \hpts v \hunion \hL) & {\rightsquigarrow} &
    \hpriv \spts {(x \hpts w \hunion \hL)}
    \\
    \alpha_\privcon(h) & \eqdef & \hpriv \spts \hL & {\rightsquigarrow} &
    \hpriv \spts {(\hL \hunion h)}
    \\
    \rho_\privcon(h) & \eqdef & \hpriv \spts (\hL \hunion h) & {\rightsquigarrow} & \hpriv \spts {\hL}
  \end{array}
}\]
Importantly, as demonstrated by the rule fo
hiding~\eqref{eq:hide-rule}, the concurroid $\privcon$ serves as the
primary one in FCSL: all other concurroids are it in a scoped manner
via the \emph{hiding} mechanism (see Appendix~\ref{sec:rules}).
In order to describe allocation/deallocation, the private heap
concurroid is typically being entangled with an allocator concurroid
$\acon$, which we have implemented in Coq as an instance of a
spin-lock with a specific resource invariant (see
Section~\ref{sec:concurroid-spin-lock}), but omitted from the
presentation. The entangled concurroid $\privcon \entangle \acon$ is
referred to as simply $\privcon$ in the main body of the paper.

\subsection{The concurroid for a spin-lock}
\label{sec:concurroid-spin-lock}

A simple CAS-based spin-lock is defined by the concurroid
\[
\lcon_{\hlock,lk,\Inv} = (\{\hlock\}, {W}_L, \set{\id}, \{(\alpha_{\lcon},
\rho_{\lcon})\})
\] 
with ${W}_{\lcon} = \{~w \mid w
\models~\mbox{assertion}~\eqref{lock}~\}$, where
\[
%\hspace{-5pt}
{\small
\begin{array}{l}
\hlock \spts (\mL, \gL) \aand \hlock \opts (\mE, \gE) \aand \hlock \jpts ((lk \hpts b) \hunion h) \aand \hbox{}\tag{\arabic{tags}}\refstepcounter{tags}\label{lock}\\
~~~~~~~~\mathsf{if}\ b\ \mathsf{then}\ h = \hempty \aand \mL \join \mE =
\lockOwn\ \\
~~~~~~~~\mathsf{else}\ \Inv\ (\gL \join \gE)\ h \aand \mL \join \mE = \lockNown
\end{array}
}\]
The assertion states that if the lock is taken ($b = \mathsf{true}$)
then the heap $h$ is given away, otherwise it satisfies the
resource~invariant $\Inv$. In either case, the thread-relative views
$\mL$, $\mE$, $\gL$ and $\gE$ are consistent with the resource's views
of $lk$ and $h$. Indeed, notice how $\mL$, $\mE$ and $\gL, \gE$ are
first $\join$-joined (by the $\join$-operations of $O = \{\lockNown,
\lockOwn\}$, defined in Section~\ref{sec:flatco}, and a
client-provided PCM $\pcmS$, respectively) and then related to $b$ and
$h$; the former implicitly by the conditional, the latter explicitly,
by the resource invariant $\Inv$, which is now parametrized by $\gL
\join \gE$.

The external transitions of the lock are defined as follows (assuming
$w.\mathother = w'.\mathother$ everywhere):
\begin{mathpar}
{\small
\begin{array}{l@{\ }c@{\ }l}
(w, w') \in \alpha_{\lcon}(h) & \iff & 
\begin{array}[t]{ll}
w.\mathself & = \hlock \hpts (\lockOwn, \gL), 
\\
w.\mathjoint & = \hlock \hpts (lk \hpts \mathsf{true}), 
\\
w'.\mathself & = \hlock \hpts (\lockNown, \gL'), 
\\
w'.\mathjoint & = \hlock \hpts ((lk \hpts \mathsf{false}) \hunion h)
\\
\end{array}
\\\\
(w, w') \in \rho_{\lcon}(h) & \iff & 
\begin{array}[t]{ll}
w.\mathself & = \hlock \hpts (\lockNown, \gL), \\
 w.\mathjoint & = \hlock \hpts ((lk \hpts \mathsf{false}) \hunion h), \\
w'.\mathself & = \hlock \hpts (\lockOwn, \gL), \\
 w'.\mathjoint & = \hlock \hpts (lk \hpts \mathsf{true})
\end{array}
\end{array}
}
\end{mathpar}
The internal transition admits no changes to the state $w$. The
$\alpha_{\lcon}$ transition corresponds to unlocking, and hence to the
acquisition of the heap $h$. It flips the ownership bit from
$\lockOwn$ to $\lockNown$, the contents of the $lk$ pointer from
$\mathsf{true}$ to $\mathsf{false}$, and adds the heap $h$ to the
resource state. The $\rho_{\lcon}$ transition corresponds to locking, and is
dual to $\alpha_{\lcon}$. When locking, the $\rho_{\lcon}$ transition keeps the
auxiliary view $\gL$ unchanged. Thus, the resource ``remembers'' the
auxiliary view at the point of the last lock. Upon unlocking, the
$\alpha_{\lcon}$ transition changes this view into $\gL'$, where $\gL'$ is
some value that is coherent with the acquired heap $h$, \ie, which
makes the resource invariant $\Inv~(\gL \join \gE)~h$ hold, and thus,
the whole state belongs to ${W}_{\lcon}$.
 
\subsection{Entanglement}

Let ${\ucon} = ({L}_{\ucon}, {W}_{\ucon}, I_{\ucon}, {
  E}_{\ucon})$ and ${\vcon} = ({L}_{\vcon}, {W}_{\vcon}, I_{\vcon}, {
  E}_{\vcon})$, be concurroids. The entanglement ${\ucon} \entangle {\vcon}$ is a
concurroid with the label component ${L}_{{\ucon} \entangle {\vcon}} =
{L}_{\ucon} \cup {L}_{\vcon}$.
The state set component combines the individual states of ${\ucon}$
and ${\vcon}$ by taking a union of their labels, while ensuring that
the labels contain only non-overlapping heaps.
\begin{mathpar}
{\small
{W}_{{\ucon} \entangle {\vcon}} = \{w \hunion w' \mid w \in {W}_{\ucon},
w' \in {W}_{\vcon}, \mbox{and $\flatten {w}$ disjoint from $\flatten{w'}$}\}
}
\end{mathpar}
To define the transition components of ${\ucon} \entangle {\vcon}$, we first need
the auxiliary concept of transition interconnection. Given transitions
$\gamma_{\ucon}$ and $\gamma_{\vcon}$ over ${W}_{\ucon}$ and ${W}_{\vcon}$,
respectively, the interconnection $\gamma_1 \relentangle \gamma_2$ is
a transition on ${W}_{{\ucon} \entangle {\vcon}}$ which behaves as $\gamma_{\ucon}$
(resp. $\gamma_{\vcon}$) on the part of the states labeled by ${\ucon}$
(resp. ${\vcon}$).  
\[
{\small
\begin{array}{r@{\ }c@{\ }l}
\gamma_1 \relentangle \gamma_2 &= &
\left\{
({w_1} \hunion {w_2}, {w'_1} \hunion {w'_2})
\left| 
  \begin{array}{l}
    (w_i, w'_i) \in \gamma_i, w_1 \hunion w_2, w'_1 \hunion \\
    w'_2 \in {W}_{{\ucon} \entangle {\vcon}}
  \end{array}
\right.\right\}.
\end{array}
} 
\]
The internal transition of ${\ucon} \entangle {\vcon}$ is defined as follows,
where $\mathsf{id}_{\ucon}$ is the diagonal of ${W}_{\ucon}$.
\[
{\small
\hspace{-5pt}
\begin{array}{r@{\ }c@{\ }l}
I_{{\ucon} \entangle {\vcon}} & = & \set{\iota_{\ucon} \relentangle \mathsf{id}_{\vcon}} \cup
\set{\mathsf{id}_{\ucon} \relentangle \iota_{\vcon}} ~\cup
\\[3pt]
&& \bigcup_{\scriptsize{\begin{array}{c}h, (\alpha_{\ucon}, \rho_{\ucon})\in{
        E}_{\ucon}, (\alpha_{\vcon}, \rho_{\vcon})\in{E}_{\vcon}\end{array}}}\!\!\!(\alpha_{\ucon}\ h
\relentangle \rho_{\vcon}\ h) \cup (\alpha_{\vcon}\ h \relentangle \rho_{\ucon}\ h)
    
\end{array}
}
\]
Thus, ${\ucon}
\entangle {\vcon}$ steps internally whenever ${\ucon}$ steps and ${\vcon}$ stays idle,
or when ${\vcon}$ steps and ${\ucon}$ stays idle, or when there exists a heap $h$
which ${\ucon}$ and ${\vcon}$ exchange ownership over by synchronizing their
external transitions.

\vspace{5pt}

\begin{example}
  We have already presented the transitions $\alpha_\privcon$ of
  $\privcon$ and $\rho_\lcon$ of $\lcon_{\hlock,lk,\Inv}$ in
  Sections~\ref{sec:conc-priv-heaps}
  and~\ref{sec:concurroid-spin-lock}.

  The following display~(\ref{trans}) presents the interconnection
  $\alpha_\privcon\ h \relentangle \rho_\lcon\ h$, which moves $h$ from
  $\lcon_{\hlock,lk,\Inv}$ to $\privcon$, and is part of the definition of $I_{\privcon
    \entangle \lcon_{\hlock,lk,\Inv}}$. The latter further allows moving $h$
  in the opposite direction ($\alpha_\lcon\ h \relentangle \rho_\privcon\ h)$,
  independent stepping of $\privcon$ ($\iota_\privcon \relentangle \mathsf{id}_\lcon$)
  and of $\lcon_{\hlock,lk,\Inv}$ ($\mathsf{id}_\privcon \relentangle \id$).
\[
{\small
\hspace{-7pt}
\begin{array}{l@{\ \lsep\ }l@{\ \aand\ }l}
\hpriv \spts \hL & (\hlock \spts (\lockNown, \gL) & \hlock \jpts ((lk \hpts \mathsf{false}) \hunion h)) \rightsquigarrow \hbox{}
\tag{\arabic{tags}}\refstepcounter{tags}\label{trans} \\ 
\hpriv \spts {(\hL \hunion h)} & (\hlock \spts (\lockOwn, \gL) & \hlock \jpts (lk \hpts \mathsf{true}))
\end{array}
}\]
\end{example}

The external transitions of ${\ucon} \entangle {\vcon}$ are those of ${\ucon}$, framed
\wrt~the labels of ${\vcon}$.
\begin{mathpar}
{\small
{E}_{{\ucon} \entangle {\vcon}} = \{(\lambda h\ldot (\alpha_{\ucon}\ h) \relentangle \mathsf{id}_{\vcon},
\lambda h\ldot (\rho_{\ucon}\ h) \relentangle \mathsf{id}_{\vcon}) \mid (\alpha_{\ucon}, \rho_{\ucon}) \in {E}_{\ucon}\}
}
\end{mathpar}
We note that ${E}_{{\ucon} \entangle {\vcon}}$ somewhat arbitrarily chooses
to frame on the transitions of ${\ucon}$ rather than those of ${\vcon}$. In this
sense, the definition interconnects the external transitions of ${\ucon}$
and ${\vcon}$, but it keeps those of ${\ucon}$ ``open'' in the entanglement, while
it ``shuts down'' those of ${\vcon}$. The notation ${\ucon} \entangle {\vcon}$ is meant
to symbolize this asymmetry. The asymmetry is important for our
example of encoding CSL resources, as it enables us to iterate the
(non-associative) addition of new resources as $((\privcon \entangle
\lcon_{\hlock_1, lk_1, \Inv_1}) \entangle \lcon_{\hlock_2, lk_2, \Inv_2}) \entangle
\cdots $ while keeping the external transitions of $\privcon$ open to
exchange heaps with new resources.

Clearly, many ways exist to interconnect transitions of two
concurroids and select which transitions to keep open. In our
implementation, we have identified several operators implementing
common interconnection choices, and proved a number of equations and
properties about them (\eg, all of them validate an instance of the
\textsc{Inject} rule).

\vspace{5pt}
\begin{lemma} 
${\ucon} \entangle {\vcon}$ is a concurroid. 

\end{lemma}

We can also reorder the iterated addition of lock concurroids.

\vspace{5pt}
\begin{lemma}[Exchange law] $({\ucon} \entangle {\vcon}) \entangle W = ({\ucon}
  \entangle W) \entangle {\vcon}$.\end{lemma}

\subsection{The empty concurroid}
\label{sec:empty-concurroid}

We close the section with the definition of the \emph{empty}
concurroid $\econ$ which is the right unit of the entanglement
operator $\entangle$. $\econ$ is defined as $\econ = (\emptyset,
{W}_E, \set{id}, \emptyset)$, where ${W}_\econ$ contains only the
empty state (\ie,~the state with no labels).

\begin{figure*}
\centering
{\small
\begin{mathpar}
\inferrule*[Right={\scriptsize{Seq}}]
 {\Gamma \vdash \stconcTy{p}{c_1}{B}{q}{\ucon} \\ 
  \Gamma, x:B \vdash \stconcTy{[x/\result]q}{c_2}{A}{r}{\ucon} \\
  x \not\in \mathsf{FV}(r)}
 {\Gamma \vdash \stconcTy{p}{x \leftarrow c_1; c_2}{A}{r}{\ucon}}
\and
\inferrule*[Right={\scriptsize{Par}}]
  {\Gamma \vdash \stconcTy{p_1}{c_1}{A_1}{q_1}{\ucon} \\
   \Gamma \vdash \stconcTy{p_2}{c_2}{A_2}{q_2}{\ucon}}
  {\Gamma \vdash \stconcTy{p_1 \ssep p_2}{c_1 \parallel c_2}{A_1 \times A_2}{[\pi_1\,{\result}/\result]q_1 \ssep [\pi_2\,{\result}/\result]q_2}{\ucon}}
\and
\inferrule*[Right={\scriptsize{Hyp}}]
  {\forall x{:}B\ldot \stconcTy{p}{f(x)}{A}{q}{\ucon} \in \Gamma}
  {\Gamma \vdash \forall x{:}B\ldot \stconcTy{p}{f(x)}{A}{q}{\ucon}}
\and
\inferrule*[Right={\scriptsize{Conseq}}]
  {\Gamma \vdash \stconcTy{p_1}{c}{A}{q_1}{\ucon} \\
   \Gamma \vdash (p_1, q_1) \sqsubseteq (p_2, q_2)}
  {\Gamma \vdash \stconcTy{p_2}{c}{A}{q_2}{\ucon}}
\and
\inferrule*[Right={\scriptsize{Frame}}]
  {\Gamma \vdash \stconcTy{p}{c}{A}{q}{\ucon} \\
  \mbox{$r$ stable under $\ucon$}}
  {\Gamma \vdash \stconcTy{p\ssep r}{c}{A}{q \ssep r}{\ucon}}
\and
\inferrule*[Right={\scriptsize{If}}]
  {\Gamma \vdash \stconcTy{e = \mathsf{true}\aand p}{c_1}{A}{q}{\ucon}\\
   \Gamma \vdash \stconcTy{e = \mathsf{false}\aand p}{c_2}{A}{q}{\ucon}}
  {\Gamma \vdash \stconcTy{p}{\mathsf{if}\ e\ \mathsf{then}\ c_1\ \mathsf{else}\ c_2}{A}{q}{\ucon}}
\and
\inferrule*[Right={\scriptsize{Conj}}]
  {\Gamma \vdash \stconcTy{p_1}{c}{A}{q_1}{\ucon}\\
   \Gamma \vdash \stconcTy{p_2}{c}{A}{q_2}{\ucon}}
  {\Gamma \vdash \stconcTy{p_1 \aand p_2}{c}{A}{q_1 \aand q_2}{\ucon}}
\and
\inferrule*[Right={\scriptsize{Exist}}]
  {\Gamma \vdash \stconcTy{p}{c}{A}{q}{\ucon}\\ 
   \alpha \not\in\mathsf{dom}\ \Gamma}
  {\Gamma \vdash \stconcTy{\exists \alpha{:}B\ldot p}{c}{A}{\exists \alpha{:}B\ldot q}{\ucon}}
\and
\inferrule*[Right={\scriptsize{Ret}}]
  {\Gamma \vdash e: A \\ \mbox{$p$ stable under $\ucon$}}
  {\Gamma \vdash \stconcTy{p}{\mathsf{return}~e}{A}{p \aand \result = e}{\ucon}}
\and
\inferrule*[Right={\scriptsize{Fix}}]
  {\Gamma, \forall x{:}B\ldot\stconcTy{p}{f(x)}{A}{q}{\ucon}, x{:}B \vdash \stconcTy{p}{c}{A}{q}{\ucon}}
  {\Gamma \vdash \forall x{:}B\ldot\stconcTy{p}{(\mathsf{fix}\ f\ldot x\ldot c)(x)}{A}{q}{\ucon}}
\and
\inferrule*[Right={\scriptsize{App}}]
  {\Gamma \vdash \forall x{:}B\ldot \stconcTy{p}{F(x)}{A}{q}{\ucon}\\
   \Gamma \vdash e : B}
  {\Gamma \vdash \stconcTy{[e/x]p}{F(e)}{A}{[e/x]q}{\ucon}}
\and
\inferrule*[Right={\scriptsize{Inject}}]
  {\Gamma \vdash \stconcTy{p}{c}{A}{q}{\ucon}\\
    \mbox{$r \subseteq W_\vcon$ stable under $\vcon$}}
  {\Gamma \vdash \stconcTy{p \lsep r}{[c]}{A}{q \lsep
      r}{\ucon \entangle \vcon}}
\and
\inferrule*[Right={\scriptsize{Action}}]
  {a = (\ucon, A, \sigma, \mu)\ \mbox{is an atomic action}\\
   \Gamma \vdash (\sigma \aand \mathsf{this}\ w, \lambda w'\ldot (w, w', \result) \in \mu) \sqsubseteq (p, q)\\
   \mbox{$p, q$ stable under $\ucon$}}
  {\Gamma \vdash \stconcTy{p}{\mathsf{act}\ a}{A}{q}{\ucon}}
\and
\inferrule*[Right={\scriptsize{Hide}}]
{\Gamma \vdash \stconc{\hpriv\spts h \lsep p}{c}{\hpriv\spts h' \lsep q}{(\privcon \entangle \ucon) \entangle \vcon}
 \and \text{$\privcon$, $\ucon$ and $\vcon$ have disjoint sets of labels}}
{\Gamma \vdash \stconc{\Psi\ g\ h \lsep (\Phi\,(g) \wand p)}{\mathsf{hide}_{\Phi, g}\ c}{\exists g'. \Psi\ g'\ h' \lsep (\Phi\,(g') \wand q)}{\privcon \entangle \ucon}}
\and
\mbox{where}\ \Psi\ g\ h = \exists k{:}\mathsf{heap}.\, \hpriv \spts {h \hunion k} \aand \Phi\,(g) \downarrow k
\end{mathpar}}
\caption{FCSL inference rules.}\label{fig:rules}
\end{figure*}

\section{Atomic actions}
\label{sec:appactions}

A concurroid $\ucon$'s transitions, described in
Section~\ref{app:conc}, specify all possibles ``degrees of freedom''
along which a state (auxiliary or real) governed by $\ucon$ can
evolve. To tie these specifications to actual programming primitives
(\ie, machine commands like \textsf{read}, \textsf{write},
\textsf{skip} or various read-modify-write operations), FCSL
introduces a notion of an \emph{atomic action}.

An atomic action is a 4-tuple $a = (\ucon, A, \sigma, \mu)$, where (1)
$\ucon$ is a concurroid, whose \emph{internal} transitions an action
respects; (2) $A$ is a return type of the action; (3) $\sigma$
describes states of $\ucon$, which $a$ can be run from; and (4) the
$\mu$ relates the initial and final states, and the result $\res$ of
the action. FCSL imposes a soft requirement that, if all ghost
information is erased from an action's definition (\eg, manipulating
with histories), it becomes operationally equivalent to a mere
heap-manipulating machine command.

\vspace{5pt}

\begin{definition}[Action erasure] Given an atomic action $a$,
  the erasures $\flatten \sigma$ and $\flatten \mu$ of $a$'s safety
  predicate and stepping relation are relations on heaps defined as
  follows.
{\small
\[
\begin{array}{lcl}
\flatten {w} \in \flatten{\sigma} & \iff & w \in \sigma\\
(\flatten{w}, \flatten{w'}, r) \in \flatten{\mu} & \iff & (w, w', r) \in \mu
\end{array}
\]}
\end{definition}

An \emph{atomic} is a triple $\alpha = (A, \sigma, \mu)$. It's a
special kind of actions, but over concrete heaps, rather than over
states. States differ from heaps in that they are decorated with
additional information such as auxiliary state and partitioning
between \self, \joint and \other.  As with actions, $A$ is the
return type, $\sigma$ is the safety predicate and $\mu$ is the
stepping relation, but they all range over heaps.

We consider four different (parametrized classes of) atomics,
corresponding to the four (parametrized) primitive memory operations
that we consider.

\vspace{5pt}

\begin{definition}[Primitive atomic actions]
\label{def:actions}
{\small
\[
\begin{array}{lcl}
  \mathsf{Read}^A_x & = & (A, (x \hpts_A -) \hunion h, (x \hpts v) \hunion h \rightsquigarrow (x \hpts v) \hunion h \aand \result = v)\\
  \mathsf{Write}\ x\ v & = & (\mathsf{unit}, (x \hpts -) \hunion h, (x
  \hpts -) \hunion h \rightsquigarrow (x \hpts v) \hunion h)\\
  \mathsf{Skip} & = & (\mathsf{unit}, h, h \rightsquigarrow h) 
  \\
  \mathsf{RMW}^{A~B}_{x~f~g} & = & (B, (x \hpts_A -) \hunion h, (x
  \hpts v) \hunion h \rightsquigarrow \\
  && (x \hpts f(v)) \hunion h \aand \result = g(v))
\end{array}
\label{eq:actions}
\]}
  
\end{definition}

The last class $\mathsf{RMW}^{A~B}_{x~f~g}$ corresponds to the family
of \emph{Read-Modify-Write} operations: they all atomically replace
the current register value $v$ with $f(v)$ for some pure function $f$,
and return the result according to the function
$g$~\cite[\S5.6]{Herlihy-Shavit:08}. One particular representative of
this family is the CAS operation, which instantiates the parameters of
$\mathsf{RMW}$ as follows:
{\small
\[
\begin{array}{rcl}
\text{CAS}_{A~x~v_1~v_2} & \eqdef &
\mathsf{RMW}^{A~\mathsf{bool}}_{x~f({v_1},{v_2})~g({v_1},{v_2})}, \text{where}   
\\ \\
f({v_1},{v_2})(v) & = & \mathsf{if}~ (v = v_1) ~\mathsf{then}~ v_2
~\mathsf{else}~ v_1 \\
g({v_1},{v_2})(v) & = & (v = v_1)
\end{array}
\]}

\vspace{5pt}

\begin{definition}[Operational actions] 
\label{def:opact}
An action $a$ is \emph{operational} if its erasure corresponds to one
of the atomics, \ie, if there exists $b \in \set{\mathsf{Read}^A_x, \mathsf{Write}\ x\ v, \mathsf{Skip},
\mathsf{RMW}^{A~B}_{x~f~g}}$ such that
{\small
\[
\flatten{\sigma_a} \subseteq \sigma_b \aand 
\forall h \in \flatten{\sigma_a}\ h'\ r\ldot (h, h', r) \in \flatten{\mu_a} \implies (h, h', r) \in \mu_b
\]}
\end{definition}
In our examples we only considered operational actions, though the
inference rules and the implementation in Coq don't currently enforce
this requirement (the operationality of actions in the examples has
been proved by hand).

\subsection{Properties of atomic actions}
\label{sec:prop-atom-acti}

Let $\ucon = (L, W, I, {E})$. The action $a = (\ucon, A,
\sigma, \mu)$ is required to satisfy the following properties.

\[
{\small
\begin{array}{rcl}
\textit{Coherence} & : & w \in \sigma \implies w \in {W}\\[3pt]
\textit{Safety monotonicity} & : & w \zag t \in \sigma \implies w \zig t \in \sigma\\[3pt]
\textit{Step safety} & : & (w, w', r) \in \mu \implies w \in \sigma\\[3pt]
\textit{Internal stepping} & : & (w, w', r) \in \mu \implies (w, w') \in I\\[3pt]
\textit{Framing} & : & w \zag t \in \sigma \implies (w \zig t, w', r) \in \mu \implies \hbox{}\\
& & \quad \exists w''\ldot w' = w'' \zig t \aand  (w \zag t, w'' \zag
t, v) \in \mu
\\[3pt]
\textit{Erasure} & : &  \mathsf{defined} (\flatten w \hunion h) \implies \flatten w \hunion h = \flatten{w'} \hunion h' \implies \hbox{}\\
& &  (w, w_1, r) \in \mu \implies (w', w'_1, r') \in \mu \implies \hbox{}\\
& &  r = r' \aand \flatten w_1 \hunion h = \flatten{w'_1} \hunion h'
\\[3pt]
\textit{Totality} & : & \forall w\ldot w \in \sigma \implies \exists w'\ v\ldot (w, w', v) \in \mu
\end{array}
}
\]

The properties of Coherence, Step safety and Internal stepping are
straightforward.  
Safety monotonicity states that if the action is safe in a state with
a smaller \self component (because the other component is enlarged by
$t$), the action is also safe if we increase the \self component by
$t$.
%The property is analogous to the equally named property in
%abstract separation logic.
%

Framing property says that if $a$ steps in a state with a large \self
component $w \zig t$, but is already safe to step in a state with a
smaller \self component $w \zag t$, then the result state and value
obtained by stepping in $w \zig t$ can be obtained by stepping in $w
\zag t$, and moving $t$ afterwards.

The Erasure property shows that the behavior of the action on the
concrete input state obtained after erasing the auxiliary fields and
the logical partition, doesn't depend on the erased auxiliary fields
and the logical partition. In other words, if the input state have
\emph{compatible} erasures (that is, erasures which are sub-heaps of a
common heap), then executing the action in the two states results in
equal values, and final states that also have compatible
erasures. This is a standard property proved in concurrency logics
that deal with auxiliary state and
code~\cite{Owicki-Gries:CACM76,Brookes:TCS07}.

The Totality property shows that an action whose safety predicate is
satisfied always produces a result state and value. It doesn't loop
forever, and more importantly, it doesn't crash. We will use this
property of actions in the semantics of programs to establish that if
the program's precondition is satisfied, then all of the
approximations in the program's denotation are either done stepping,
or can actually make a step (\ie, they make progress).

Usually, the actions are defined in a so-called \emph{large footprint}
style.
To enable writing various actions in a \emph{small footprint} style,
we also enforce the property 
\[
{\small
\begin{array}[t]{c}
\textit{Locality} ~:~ w.\mathother = w'.\mathother \implies (w \zag t, w' \zag t, v) \in \mu \implies (w \zig t, w' \zig t, v) \in \mu
\end{array}
}\]
%
%This property corresponds to what we have taken up calling
%\emph{locality}, though it's not at all equivalent to locality in the
%The property is \emph{not} provable from the rest, unless we require
%that each action is deterministic (which right now seems
%unnecessary). Thus, if we want it, it has to be required as primitive,
%which is what we have done.
Curiously, if the default use of the logic is in a large footprint
notation, then this property is not necessary as it is not used in any
proofs.

\subsection{Example: pair snapshot reading and writing actions}
\label{sec:ops}

In the pair snapshot concurroid (Section~\ref{sec:concurroids}), the
reading from~$x$ can be implemented by means of an atomic action
\[
\mathit{readX} = (\pscon, (A \times \nat), \sigma_{\mathit{rx}},
\mu_{\mathit{rx}}),\] 
where
\[
\tag{\arabic{tags}}\refstepcounter{tags}\label{eq:readx-act}
 {\small
\hspace{-5pt}
\begin{array}{lcl}
  \sigma_{\mathit{rx}}(w) & \eqdef & w \in W_{\pscon} \\
  \mu_{\mathit{rx}}(w, w', \res) & \eqdef &  w = w' \aand w.\mathjoint = (x
  \hpts (c_x, v_x) \hunion y \hpts -)~\aand \\
 && \res = (c_x, v_x).
\end{array}
}
\]
Similarly, writing into $x$ and updating its version simultaneously is
implemented via the action
\[
\mathit{writeAndIncX}(v) = (\pscon, \mathsf{Unit},
\sigma_{\mathit{wx}}, \mu_{\mathit{wx}}(v)),
\]
such that
\[
\tag{\arabic{tags}}\refstepcounter{tags}\label{eq:writex-act}
 {\small
\begin{array}{lcl}
  \sigma_{\mathit{wx}}(w) & \eqdef & w \in W_{\pscon} \\
  \mu_{\mathit{wx}}(v)(w, w', \res) & \eqdef & \res = \mathsf{unit} \aand
  \iota_{\pscon}^x(w, w')|_{c'_x~=~v}
\end{array}
}
\]
where by $wr_x (w, w')|_{c'_x~=~\res}$ we mean a restricted version of
the relation induced by the transition $wr_x$ defined
in~\eqref{eq:pair-trans}, such that $c'_x$ is taken to be the action
argument $v$, which is being written as a new value $c'_x$ to the
snapshot cell $x$.
It is not difficult to check that \emph{readX} corresponds to the
$\id$ transition of $\pscon$, whereas \emph{writeAndIncX} naturally
corresponds to the internal transition~$wr_x$~\eqref{eq:pair-trans}.

\section{Language and logic inference rules}
\label{sec:rules}

Program specifications in FCSL take the form of Hoare 4-tuple
$\stconc{p}{c}{q}{\ucon}$ expressing that the thread $c$ has a
precondition $p$, postcondition $q$, in a state space and under
transitions defined by the concurroid $\ucon$, which in FCSL plays
both the role of a resource context from CSL and the role of
Rely/Guarantee. The Hoare 4-tuple $\stconcTy{p}{c}{A}{q}{\ucon}$ is
satisfied by a command $c$ if $c$'s effect is approximated by the
\emph{internal} transition of the concurroid $\ucon$, $c$ is
\emph{memory-safe} when executed from a state satisfying $p$, and
concurrently with any environment that respects the transitions
(internal and external) of $\ucon$; if $c$ terminates, it returns a
value of type $A$ in a state satisfying $q$. A dedicated variable
$\res$ of type $A$ is used to name the return result in $q$.
In FCSL, the first-order looping commands are represented by recursive
procedures implemented using the fixpoint operator. In the case of
recursive procedures, $p$ and $q$ in the procedure tuple correspond to
a loop invariant, which is supposed provided by the programmer.
Judgments in FCSL are formed under hypotheses from a context $\Gamma$
that maps \emph{program variables} $x$ to their types and
\emph{procedure variables} $f$ to their specifications. $\Gamma$ is
omitted in most of the examples, as it is clear from the context. The
scope of logical variables is limited to the Hoare tuples in which
they appear.
Figure~\ref{fig:rules} lists FCSL rules.

The rule \textsc{Fix} requires proving a Hoare tuple for the procedure
body, under a hypothesis that the recursive calls satisfy the same
tuple. The procedure \textsc{App}lication rule uses the typing
judgment for expressions $\Gamma \vdash e : A$, which is the customary
one from a typed $\lambda$-calculus, so we omit its rules; in our
formalization in Coq, this judgment will correspond to the CiC's
typing judgment.

\subsection{Definition of Hoare ordering %
  $(p_1, q_1) \sqsubseteq (p_2, q_2)$}
\label{sec:ordering}

The \textsc{Action} and \textsc{Conseq} rules use the judgment $\Gamma
\vdash (p_1, q_1) \sqsubseteq (p_2, q_2)$, which generalizes the
customary side conditions $p_2\,{\implies}\,p_1$ for strengthening the
precondition and $q_1\,{\implies}\,q_2$ for weakening the
postcondition, to deal with the local scope of logical variables

The generalization is required in FCSL because of the local scope of
logical variable. In first order Hoare logics, the logical variables
have global scope, so the above implications over $p_1, p_2$ and $q_1,
q_2$ suffice. In FCSL, the logical variables have scope locally over
Hoare triples, and this scope has to be reflected in the semantic
definition of $\sqsubseteq$ by introducing quantifiers.
\[
{\small
\begin{array}{l}
(p_1, q_1) \sqsubseteq (p_2, q_2) \iff \hbox{}\\
\qquad \forall w\ w'\ldot \begin{array}[t]{l}
(w \models \exists \bar{v}_2\ldot p_2 \implies w \models \exists \bar{v}_1\ldot p_1) \aand \hbox{}\\
((\forall \bar{v}_1\ \result\ldot w \models p_1 \implies w' \models
q_1) \implies \\
~~~~(\forall \bar{v}_2\ \result\ldot w \models p_2 \implies w' \models q_2))
\end{array}
\end{array}
}\]
where $\bar{v}_i = \mathsf{FLV}(p_i, q_i)$ are the free logical
variables.
The definition makes it apparent that the Hoare triple
$\stconc{p}{c}{q}{\ucon}$ is essentially a syntactic sugar for a different
kind of Hoare triple, which may be written as:
\[
{\small
\stconc{w\ldot\exists \bar{v}\ldot w \models p}{c}{\result\ w\ w'\ldot \forall \bar{v}\ldot w \models p \implies w' \models q}{\ucon}
}\]
where $\bar{v} = \mathsf{FLV}(p,q)$. In this alternative Hoare triple,
the postconditions are predicates ranging over input and output states
$w$ and $w'$ (they are thus called binary postconditions). The
advantage of the alternative Hoare triple is that the logical
variables are explicitly bound, making their scoping explicit. In our
Coq implementation we use this alternative formulation of Hoare
triples.

\subsection{Turning atomic actions into commands}
\label{sec:stable-spec-atom}
Since all pre- and postconditions in FCSL are stable under the
interference of the corresponding concurroid, the use of an atomic
action requires explicit stabilization of its specification $\mu$,
as captured by the rule \textsc{Action}. This rule has been implicitly
used in most of the examples in the paper body in order to obtain
stable specifications for methods like
$\act{readX}$~\eqref{eq:readx-spec}, $\act{tryCollect}$~\eqref{eq:ack}
\etc.

To demonstrate the use of the \textsc{Action} rule, let us consider
one of the most commonly used commands: writing into a privately owned
heap, to which we gave the spec~\eqref{eq:alloc-spec}. As one may
expect, such command ``lives'' in a concurroid of private heaps
$\privcon$, supported by its internalt transition $\iota_{\privcon}$,
and has the following obviously stable specification (given in a
\emph{large footprint} with explicit universally-quantified
\emph{self}-owned heap $\hL$):
\[
\tag{\arabic{tags}}\refstepcounter{tags}\label{eq:write-spec}
{\small
\begin{array}{r@{\ }c@{\ }l}
\spec{\hpriv \spts (x \hpts -) \hunion \hL}
&
\act{write}(x, e)
&
\spec{\hpriv \spts (x \hpts e) \hunion \hL}@\privcon    
\end{array}
}
\]
The specification~\eqref{eq:alloc-spec}, used in the paper body, can
be obtained from~\eqref{eq:write-spec} by taking $\hL = \hempty$.

Another example of a command obtained from an atomic action a method
for reading from $\pscon$'s pointer $x$ from
Section~\ref{sec:overview}.  It is easy to make sure that the
spec~\eqref{eq:readx-spec}, which was used for verification of the
\code{readPair} procedure, can be obtained by stabilization of the
assertions defining $\mu_{\mathit{rx}}$~\eqref{eq:readx-act} of the
corresponding atomic action $\mathit{readX}$ in Section~\ref{sec:ops}.

\subsection{Properties of $\Phi$ functions from the hiding rule}
\label{sec:phi-properties}

The abstraction function $\Phi$ is a user-specified annotation on the
hide command (see rule \textsc{Hide} in Figure~\ref{fig:rules} or
display~\eqref{eq:hide-rule}). It maps values $g : \pcmS$ (where
$\pcmS$ is a user-specified PCM) to assertions, that is, predicates
over states (equivalently, sets of states) of a
concurroid~$\vcon$. For the soundness of the hiding rule, $\Phi$ is
required to satisfy the following properties.
\[
{\small
\begin{array}[t]{l@{\ }c@{\ }l}
\textit{Coherence} & : & w \in \Phi(g) \implies w \in W_\vcon
\\[3pt]
\textit{Injectivity} & : & w \in \Phi(g_1) \implies w \in \Phi(g_2)
\implies g_1 = g_2
\\[3pt]
%& : & i \in \mathsf{coh}\ W \arrow \exists g, i \in \Phi(g)\\
\textit{Surjectivity} & : & w_1 \in \Phi(g_1) \implies w_2 \in W_\wcon
\implies w_1.\mathother = w_2.\mathother \implies \\
&&\exists g_2\ldot w_2 \in \Phi(g_2)
\\[3pt]
\textit{Guarantee} & : & w_1 \in \Phi(g_1) \implies w_2 \in \Phi(g_2)
\implies w_1.\mathother = w_2.\mathother
\\[3pt]
\textit{Precision} & : & w_1 \in \Phi(g) \implies w_2 \in \Phi(g)
\implies \\
&& \flatten {w_1} \hunion h_1 = \flatten {w_2} \hunion h_2 \implies w_1 = w_2
\end{array}
}
\]
Coherence and Injectivity are obvious. Surjectivity states that for
every state $w_2$ of the concurroid $\wcon$ one can find an image $g$,
under the condition that the \other component of $w_2$ is well-formed
according to $\Phi$ (typically, that the \other component is equal to
the unit of the PCM-map monoid for $\wcon$). Guarantee formalizes that
environment of $\mathsf{hide}$ can't interference on $\vcon$, as
$\vcon$ is installed locally. Thus, whatever the environment does, it
can't influence the \other component of the states $w$ described by
$\Phi$.

Precision is a technical property common to separation-style logics,
though here it has a somewhat different flavor. Precision ensures that
for every value $g$, $\Phi(g)$ precisely describes the underlying
heaps of its circumscribed states; that is, each state $\Phi(g)$ is
uniquely determined by its heap erasure.

}{} 

\end{document}